\documentclass[nofootinbib,showkeys,floatfix,onecolumn,times]{revtex4-2}
\usepackage{bm,morefloats}
\usepackage{dcolumn}
\usepackage[latin1]{inputenc}
\usepackage[T2A,T1]{fontenc} 
\usepackage{fonttable}
\usepackage{graphicx}
\usepackage{balance}
\usepackage[latin1]{inputenc}
\usepackage[T1]{fontenc}
        \usepackage{cellspace}
\usepackage{latexsym,color}
\usepackage{centernot}
\usepackage{mathtools}
\usepackage{stmaryrd}
\usepackage{amssymb}
\usepackage{natbib}
\usepackage{amsmath,amssymb,amsthm,mathrsfs,amsfonts,dsfont}

\usepackage{color}

\usepackage{rotating}
\usepackage[dvipsnames]{xcolor}
\definecolor{LinkBlue}{RGB}{6,69,173}
\definecolor{DarkBlue}{RGB}{11,0,128}
\definecolor{red}{rgb}{1,0.,0.}
\usepackage[colorlinks=true,linkcolor=LinkBlue,urlcolor=magenta,
	citecolor=green,hyperfootnotes=true,urlcolor=red]{hyperref}

\usepackage{enumitem}
\count\footins = 1000

\newcommand{\mnras}{MNRAS}

\newcommand{\apjj}{ApJ}
\newcommand{\prdd}{Phys. Rev. \textbf{D}}
\begin{document}

\title{On photons  and matter  inversion spheres\\ from complex super-spinars accretion structures  
 }
\author{D. Pugliese\&Z. Stuchl\'{\i}k}
\email{daniela.pugliese@physics.slu.cz}
\affiliation{
Research Centre for Theoretical Physics and Astrophysics, Institute of Physics,
  Silesian University in Opava,
 Bezru\v{c}ovo n\'{a}m\v{e}st\'{i} 13, CZ-74601 Opava, Czech Republic
}

\begin{abstract}Our analysis focus on the    dragging   effects on the accretion flows and jet emission in  Kerr super-spinars.
These attractors are characterized by peculiar accretion structures   as double tori, or special dragged tori in the ergoregion, produced by the balance of the hydrodynamic and centrifugal  forces  and also effects of super-spinars repulsive gravity.  We investigate  the accretion flows, constituted by particles and photons, from   toroids orbiting a central Kerr super-spinar. As results of our analysis, in both accretion  and jet flows, properties characterizing these geometries, that constitute possible strong observational signatures or these attractors, are highlighted. We found that the  flow is  characterized by  closed  surfaces, defining    {inversion coronas} (spherical shell), with null  the  particles flow toroidal  velocity ($u^{\phi}=0$) embedding  the central singularity. We proved that this  region distinguishes  proto-jets and accretion driven flows, co-rotating and counter-rotating flows.
Therefore in both cases the flow carries information about the accretion structures around the central attractor,  demonstrating  that  inversion points can constitute an observational aspect capable of distinguishing the super-spinars.
\end{abstract}
\date{\today}

\maketitle

\def\be{\begin{equation}}
\def\ee{\end{equation}}
\def\bea{\begin{eqnarray}}
\def\eea{\end{eqnarray}}
\newcommand{\tb}[1]{\textbf{\texttt{#1}}}
\newcommand{\actaa}{Acta Astronomica}
\newcommand{\laa}{\mathcal{L}}
\newcommand{\ba}{\mathcal{B}}
\newcommand{\Sie}{\mathcal{S}}
\newcommand{\Mie}{\mathcal{M}}
\newcommand{\La}{\mathcal{L}}
\newcommand{\Em}{\mathcal{E}}

\newcommand{\mso}{\mathrm{mso}}
\newcommand{\mbo}{\mathrm{mbo}}

\newcommand{\rtb}[1]{\textcolor[rgb]{1.00,0.00,0.00}{\tb{#1}}}
\newcommand{\gtb}[1]{\textcolor[rgb]{0.17,0.72,0.40}{\tb{#1}}}
\newcommand{\ptb}[1]{\textcolor[rgb]{0.77,0.04,0.95}{\tb{#1}}}
\newcommand{\btb}[1]{\textcolor[rgb]{0.00,0.00,1.00}{\textbf{#1}}}
\newcommand{\otb}[1]{\textcolor[rgb]{1.00,0.50,0.25}{\tb{#1}}}
\newcommand{\non}[1]{{\LARGE{\not}}{#1}}

\newcommand{\cc}{\mathrm{C}}

\newcommand{\il}{~}
\newcommand{\la}{\mathcal{A}}
  \newcommand{\Qa}{\mathcal{Q}}
\newcommand{\Sa}{\mathcal{\mathbf{S}}}
\newcommand{\Ta}{{\mbox{\scriptsize  \textbf{\textsf{T}}}}}
\newcommand{\Ca}{\mathcal{\mathbf{C}}}

\section{Introduction}

The study  of  possible  astrophysical signatures   for
Kerr super-spinars   constitutes an extensive  and debated literature on a  huge variety
of astrophysical phenomena--see for example \cite{z01,z04,z05,z06,z07,z08,z09,z010,z011,z013,1980BAICz..31..129S,1981BAICz..32...68S,z015,z017}.

In the present analysis, we  discuss   newly introduced  observationally relevant astrophysical
phenomena which can expose distinctions between
Kerr Naked Singularities (\textbf{NSs})  and Black holes (\textbf{BHs}). We implement this  analysis in the accretion and jet emission  frame   investigating    the   accretion tori flow  and  jet emission  in the super-spinnar gravitational  field.

Inversion points  investigated here are points of vanishing  toroidal  velocity of the particles and photons   as related to distant static observers, therefore defined by the condition   $u^{\phi}=0$ (or equivalently $\Omega=0$) on the flow particles and  photons velocities (relativistic angular velocity).
This condition defines a spherical surface, the \emph{inversion sphere}, embedding the central attractor, where $\Omega=0$. The inversion  sphere is  independent from the normalization  condition on the flow components at  the inversion point, therefore describes equivalently photons and timelike particles.
Jet emission and  accretion flow from the orbiting structures define a closed  region surrounding the central attractors  bounded by an inner and outer inversion sphere, and    fixed  only by the spin--mass ratio of the attractor. This closed region  (spherical shell), defined    \emph{inversion coronas},  filled with null    particles flow toroidal  velocity ($u^{\phi}=0$)  distinguishes  proto-jets and accretion driven flows, co-rotating and counter-rotating flows. The location and topology of this sphere
carries information about details of the  accretion structures around the central attractor.
Therefore an inversion corona   is  as a valid astrophysical tracer distinguishing Kerr super-spinar solutions
from Kerr \textbf{BH} solutions, providing also  possible strong
observational signature of the \textbf{NSs}.

This  analysis follows paper \cite{published}, where the case of Kerr \textbf{BH} attractors has been considered, while in \cite{submitted}
a comparison between \textbf{BH} and \textbf{NS}  inversion sphere is discussed.

The inversion
points in the super-spinner geometries are related to a combination of the repulsive gravity effects and the frame
dragging effects, typical of the \textbf{NSs} ergoregion.
 Toroidal  configurations around the central  super-spinning attractor have been  thoroughly  characterized in a several studies, here we discuss the accretion flow     velocity components, energy and  momentum at the inversion points taking into account  the peculiar  \textbf{NS}   causal structure and the frame dragging effects from the \textbf{NSs} ergoregions.

We detail, the fluid inversion points  driven by the orbiting toroidal structures.
To fix the  discussion we   considered geometrically thick general relativistic hydrodynamic (GRHD) disks,  centered on the equatorial plane of the central attractor, which is also the tori symmetry plane. It should be noted  that  the inversion points  analysis is    however independent from an eventual tilted angle,  and the assumption of coplanar tori is in fact  not necessary.
More precisely   we consider barotropic, radiation pressure supported accretion tori, cooled by advection, with low viscosity and
opaque,  with  super-Eddington  luminosity (high  matter accretion rates)
\cite{Jaroszynski(1980),Pac-Wii,Koz-Jar-Abr:1978:ASTRA:,abrafra}.
This model  (Polish doughnut)  is    widely used in  literature as the initial conditions  of the  GRMHD (magnetohydrodynamic)  evolution of accretion   disks \cite{Shafee,Fragile:2007dk,DeVilliers,arXiv:0910.3184,Porth:2016rfi}.
In the  GRHD  barotropic models,   the time scale of the dynamical processes $\tau_{dyn}$ (driven by gravitational and inertial forces) is much lower than the time scale of the thermal ones $\tau_{therm}$   (including heating and cooling processes, radiation) which  is lower than the time scale of the viscous processes $\tau_{\nu}$. In other words,
the  strong gravitational field of the background  is dominant with respect to the  dissipative forces,  in determining  the  tori unstable phases \cite{F-D-02,abrafra,pugtot,Pac-Wii,Hawley1990,Fragile:2007dk,DeVilliers,Hawley1987,Hawley1991,Hawley1984,Fon03,Lei:2008ui}.
Within  this approximation, during the  dynamical processes, the functional form of the angular
momentum and entropy distribution in the fluid, depends on the system initial conditions and not on the
the details of the dissipative processes.

 The  accretion  process  turns therefore to be  regulated by the  mass  loss in  the  outer   toroid Roche  lobe  overflow,  associated to the occurrence of  the surface   cusp. The instability   process  follows   the un-balance of the gravitational, inertial forces, and the pressure gradients in the fluid.  This is a   important  self-regulated process, which turns out to
locally  stabilize   the accreting torus from the   thermal  and  viscous   instabilities    and it   globally  stabilizes the torus from the Papaloizou\&Pringle instability\cite{Abramowicz-nATURE,Blaes,Blaes1987,abrafra,Pac-Wii,Koz-Jar-Abr:1978:ASTRA:}.).

As for the \textbf{BH} case, analyzed in \cite{published}, the  flow inversion points  in the  \textbf{NS} spacetimes,  define the  inversion sphere as a closed and regular surface determined by the constant fluid specific angular momentum $\ell$. Whereas the turning corona is defined by the \emph{range} of location of the  disk inner edge or the proto-jets cusps for accretion driven or proto-jets driven flows.

Therefore, although the inversion points do not depend on the details of the accretion processes, or the precise location of the tori  inner edge, in our analysis the flow  of photons and   free-falling particles   are driven from the inner edge (the toroidal surface cusp) of the accretion disk or proto-jets (open, cusped configuration).

In general, the concept of  accretion  disks inner edge    depends  on   the details of the accretion process.
However analytic models  of accretion tori  fix a certain  radius $r_\times$, inner edge, which is usually an unstable geodesic circular orbit, the cusp, as a  theoretical radius  distinguishing the accretion flows in  (free falling) particles   from  the region $r<r_\times$, and at $r>r_{\times}$, defines  the accretion torus up to an outer edge $r_{outer}$ \cite{Capellupo:2017qpt,McClintock:2006xd,Daly:2008zk} \citep{Krolik:2002ae,BMP98,2010A&A...521A..15A,Agol:1999dn,Paczynski:2000tz}.
For a stellar attractor,  the inner edge is generally located near the star surface, for  \textbf{BHs} and the \textbf{NSs} the inner  edge lies between the marginally bound circular orbit  and the marginally stable circular orbit.
  The proto-jets have a cusp in the orbital range defined by the marginally circular orbit (usually a photon orbit) and marginally bound circular orbit\footnote{The inner edge of the accreting tori has also be framed in the  jet emission mechanism, with several indications of  a jet emission-accretion disk correlation \cite{Krolik:2002ae,BMP98,2010A&A...521A..15A,Agol:1999dn,Paczynski:2000tz}.}.   Therefore, the fluid specific angular momentum at these orbits defines the accretion driven and defines the proto-jets  driven inversion coronas respectively.
 The morphological structure of the jets and particularly  the velocity components on the central axis of rotation constitute an important  topic of jet emission collimation, thus for the proto-jets analysis we  consider  in particular the presence of flow  inversion  point  in the direction of  the rotational axis of the central attractors (vertical direction).

\medskip

More in details,  this article plan is as follows:

In Sec.\il(\ref{Sec:quaconsta}) we introduce the spacetime metric. The toroidal structures are discussed  in Sec.\il(\ref{Sec:accretion-Tori}).

The inversion points are analyzed in Sec.\il(\ref{Sec:contain-inner-torus}). In  Sec.\il(\ref{Sec:FIP})  we define the flow inversion points and  in Sec.\il(\ref{Sec:fromaccretionflows})  flow inversion points are related  to the orbiting structures defining accretion driven and proto-jets driven inversion coronas.

In Sec.\il(\ref{Sec:double-tori}) we address the special case of double tori with equal specific angular momentum.
Tori and inversion points for  $\ell=\pm
a$ are studied  in Sec.\il(\ref{Sec:l=pma}).
Excretion driven inversion points and torus outer edges are focused in Sec.\il(\ref{Sec:outer-edge}).
The location of inversion points in relation to the ergoregion is discussed in
Sec.\il(\ref{Sec:ergo-inversion-in-out}).
\textbf{NSs} are distinguished from the \textbf{BHs} by the presence of  double inversion points on planes  different from the equatorial, we discuss this aspect in Sec.\il(\ref{Sec:double-inversion-points}).
Proto-jets driven inversion points and inversion points verticality are focused in Sec.\il(\ref{Sec:inversionpoint-proto-jets}):
Some notes on the inversion coronas thickness and slow counter-rotating  inversion spheres
are in Sec.\il(\ref{Sec:slow-Invers}).
We conclude  in Sec.\il(\ref{Sec:deta-tori-integr}) with further notes on the flow inversion points from orbiting tori.
Sec.\il(\ref{Sec:discussion}) contains discussion and final remarks.
Three  appendix sections follow:
some more in depth notes on the tori models  are in Sec.\il(\ref{Sec:fluid-effe-pot}),
relative location of the inversion points in accreting flows is the focus of Sec.\il(\ref{Sec:relative-location}).
Further notes on the co-rotating   flows inversion points are in Sec.\il(\ref{Sec:co-space-like}).
\section{The Kerr spacetime metric}\label{Sec:quaconsta}
The   Kerr  spacetime metric  reads
%
\bea \label{alai}&& ds^2=-\left(1-\frac{2Mr}{\Sigma}\right)dt^2+\frac{\Sigma}{\Delta}dr^2+\Sigma
d\theta^2+\left[(r^2+a^2)+\frac{2M r a^2}{\Sigma}\sigma\right]\sigma
d\phi^2
-\frac{4rMa}{\Sigma}\sigma  dt d\phi,
\eea
In the Boyer-Lindquist (BL)  coordinates
\( \{t,r,\theta ,\phi \}\)\footnote{We adopt the
geometrical  units $c=1=G$ and  the $(-,+,+,+)$ signature, Latin indices run in $\{0,1,2,3\}$.  The radius $r$ has unit of
mass $[M]$, and the angular momentum  units of $[M]^2$, the velocities  $[u^t]=[u^r]=1$
and $[u^{\phi}]=[u^{\theta}]=[M]^{-1}$ with $[u^{\phi}/u^{t}]=[M]^{-1}$ and
$[u_{\phi}/u_{t}]=[M]$. For the seek of convenience, we always consider the
dimensionless  energy and effective potential $[V_{eff}]=1$ and an angular momentum per
unit of mass $[L]/[M]=[M]$.},
where
\bea\label{Eq:delta}
\Delta\equiv a^2+r^2-2 rM\quad\mbox{and}\quad \Sigma\equiv a^2 (1-\sigma)+r^2,\quad  \sigma\equiv \sin^2\theta.
\eea
 Parameter  $a=J/M\geq0$ is the metric spin, where  total angular momentum is   $J$  and  the  gravitational mass parameter is $M$.
 The Kerr naked singularities (\textbf{NSs}) have  $a>M$.
  A Kerr black hole (\textbf{BH}) is defined  by the condition   $a\in[0,M]$  with Killing horizons horizons $r_-\leq r_+$ where  $r_{\pm}\equiv M\pm\sqrt{M^2-a^2}$).
  The extreme Kerr \textbf{BH}  has dimensionless spin $a/M=1$ and the non-rotating   case $a=0$ is the   Schwarzschild \textbf{BH} solution.

The  outer  and inner stationary  limits $r_{\epsilon}^\pm$ (ergosurfaces), are  given by
\bea\label{Eq:sigma-erg}
r_{\epsilon}^{\pm}\equiv M\pm\sqrt{M^2- a^2 (1-\sigma)}
\eea
respectively. We also introduce the function $\sigma_{erg}: r=r_{\epsilon}^\pm$:
\bea
 \sigma_{erg}\equiv \frac{(r-2M) r}{a^2}+1
\eea
The region $[r_\epsilon^-,r_\epsilon^+]$ is the ergoregion.  There is   $r_\epsilon^-=0$  and
  $r_{\epsilon}^+=2M$  in the equatorial plane $\theta=\pi/2$ ($\sigma=1$),
and  $r_+<r_{\epsilon}^+$ on   $\theta\neq0$.
More precisely  there is
\bea\nonumber &&
g_{tt}>0\quad \mbox{for}:a\geq M\quad  \left(\sigma=\sigma_\epsilon^+, r=r_\epsilon^-\right);  \left(
\sigma\in]\sigma_{\epsilon}^+,1[,  r\in[r_\epsilon^-, r_\epsilon^+]\right);  \left(\sigma=1, r\in ]0,2M]\right);
\\
&&\label{Eq:sigmaepsilonplus}
\mbox{where}\quad
\sigma_\epsilon^+\equiv\frac{a^2-M^2}{a^2}.
\eea
\textbf{NS}  poles  $\sigma=0$ are associated to $g_{tt}<0$ for $a\geq0$  and $r\geq0$.
%
%
In the following, where appropriate,  to easy the reading of   complex expressions, we will  use  the   units with $M=1$ (where $r\rightarrow r/M$  and $a\rightarrow a/M$).

The  equatorial  circular geodesics are confined on the equatorial  plane as a consequence of the metric tensor symmetry under reflection through  the plane $\theta=\pi/2$.
We consider  the  following  constant of geodesic  motions
\bea&&\label{Eq:EmLdef}
\Em=-(g_{t\phi} \dot{\phi}+g_{tt} \dot{t}),\quad \La=g_{\phi\phi} \dot{\phi}+g_{t\phi} \dot{t},\\&&
\nonumber\mbox{where}\quad  g_{ab}u^a u^b=\kappa \mu^2,\quad u^t= \frac{g_{\phi\phi} \Em+ g_{t\phi} \La}{g_{t\phi}^2-g_{\phi\phi} g_{tt}},\quad u^\phi= -\frac{g_{t\phi} \Em+ g_{tt} \La}{g_{t\phi}^2-g_{\phi\phi} g_{tt}},
\eea
with   $u^a\equiv\{ \dot{t},\dot{r},\dot{\theta},\dot{\phi}\}$ where
$\dot{q}$ indicates the derivative of any quantity $q$  with respect  the proper time (for  $\mu>0$) or  a properly defined  affine parameter for the light--like orbits (for $\mu=0$), and  $\kappa=(\pm,0)$ is a normalization constant ($\kappa=-1$ for  test particles).

 Quantity  $\xi_{\phi}=\partial_{\phi}$ is the rotational  Killing field
     and  $\xi_{t}=\partial_{t}$ is the Killing field
representing the stationarity of the  Kerr background.
 The constant $\La$ in Eq.\il(\ref{Eq:EmLdef}) can be interpreted       as the axial component of the angular momentum  of a test    particle following
timelike geodesics and $\Em$  represents the total energy of the test particle
 related to the  radial infinity, as measured  by  a static observer at infinity.
The specific  angular momentum is
 \bea&&\label{Eq:flo-adding}
\ell\equiv\frac{\La}{\Em}=-\frac{g_{\phi\phi}u^\phi  +g_{\phi t} u^t }{g_{tt} u^t +g_{\phi t} u^\phi} =-\frac{g_{t\phi}+g_{\phi\phi} \Omega }{g_{tt}+g_{t\phi} \Omega},\quad\Omega \equiv\frac{u^\phi}{u^{t}}= -\frac{g_{t\phi}+ g_{tt} \ell}{g_{\phi\phi}+ g_{t\phi} \ell}
\eea
where  $\Omega$ is the relativistic angular velocity.
 For a circularly orbiting test particle,    particle counter--rotation  (co-rotation) is \emph{defined} by $\La a<0$ ($\La a>0$).
 The  Kerr \textbf{NS} background    geodesic structure is constituted by the
radii $\{r_{\gamma}^{\pm},r_{mbo}^{\pm},r_{mso}^{\pm}\}$,  marginally circular (photon) orbit, marginally bounded orbit,  and marginally stable  orbit   respectively for counter-rotating $(+)$  and co-rotating $(-)$ motion--see Figs\il(\ref{Fig:PlotparolA2}). The spacetime geodesic structure regulates test particles circular motion and some key aspects of accretion disks physics--see Table\il(\ref{Table:tori-counter-rotating}).
 In Sec.\il(\ref{Sec:accretion-Tori}) we discuss further the notion of co-rotating and counter-rotating accretion tori in \textbf{NSs} spacetimes.
\subsection{Co-rotating and counter-rotating accretion tori in \textbf{NSs} spacetimes}\label{Sec:accretion-Tori}
 \begin{table}
 \resizebox{.71\textwidth}{!}{%
\begin{tabular}{ll}
  \hline
${r}_{[mbo]}^{\pm}:\;\ell^{\pm}(r_{mbo}^{\pm})=
 \ell^{\pm}({r}_{[mbo]}^{\pm})\equiv {\ell_{mbo}^{\pm}},
\quad
  r_{[\gamma]}^{\pm}: \ell^{\pm}(r_{\gamma}^{\pm})=
  \ell^{+}(r_{[\gamma]}^{+})\equiv \ell_{\gamma}^{+}$
 \\
$r_{\gamma}^{\pm}<r_{mbo}^{\pm}<r_{mso}^{\pm}<
 {r}_{[mbo]}^{\pm}\quad\mbox{and}\quad
 {r}_{[mbo]}^{+}< r_{[\gamma]}^{+}<r_\gamma^-\equiv 0$,\\
\hline
 $\ell^\pm\in \mathbf{L_1^\pm}\equiv]\ell^\pm_{mbo}, \ell^\pm_{mso}[$: quiescent  and cusped tori. $r_{\times}\in]r_{mbo}^\pm,r_{mso}^\pm]$, $r_{center}\in]r_{mso}^\pm,r_{[mbo]}^\pm]$;\\
\hline
 $\ell^\pm\in \mathbf{L_2^\pm}\equiv[ \ell^\pm_{\gamma},\ell^\pm_{\mbo}[$: quiescent  tori and proto-jets.  $r_{j}\in]r_{\gamma}^\pm,r_{mbo}^\pm]$   $r_{center}\in]r_{[mbo]}^\pm,r_{[\gamma]}^\pm]$;\\
\hline
 $\ell^\pm\in \mathbf{L_3^\pm}: \ell^\pm<\ell^\pm_{\gamma}$: quiescent  tori,   $r_{center}>r_{[\gamma]}^\pm$.
\\
\hline
  \end{tabular}}
  \caption{Co-rotating and counter-rotating tori in Kerr \textbf{NSs}. $\ell$ is the fluid specific angular momentum.  Radii $r_{mso}^-=\{\bar{r}_{mso}^-,\tilde{r}_{mso}^-\}$, are the marginally stable orbits related to the motion  $\ell^-\gtrless 0$, where $
r_{mso}^-=\bar {r} _ {mso}^- $ for spin  $a>a_1$ and $r_{mso}^-=
\tilde {r} _ {mso}^-$  for  $a\in]M, a_ 1]$  and $
\bar {r} _ {mso}^ -= \tilde {r} _ {mso}^-$ on  $a_ 1$--Figs\il(\ref{Fig:PlotparolA2}).
Spin $a_1$ is defined in Table\il(\ref{Table:corotating-counter-rotatingspin}).
For simplicity of notation, when it is not necessary to specify,
we will use the abbreviated notation  $r_{mso}^-$ for $\{\bar{r}_{mso}^-,\tilde{r}_{mso}^-\}$.  We adopt the notation $q_{\bullet}\equiv q(r_{\bullet})$ for any quantity $q$ evaluated on a radius $r_{\bullet}$.   Radii  $\{r_{\gamma}^{\pm},r_{mbo}^{\pm}\}$,  are the marginally circular (photon) orbit and the marginally bounded orbit for $\ell^+$  and $\ell^-$ specific angular momentum. Relation between radii   $\{r_{mso}^\pm, r_{\gamma}^{\pm},r_{mbo}^{\pm}\}$  characterizes fluids  with  $\ell^+$ and $\ell^-$  at \textbf{NSs} spins $a>a_2$--see Table\il(\ref{Table:corotating-counter-rotatingspin})--where there is   $r_\gamma^-\equiv 0$. Radii $\{r_{[mbo]}^\pm,r_{[\gamma]}^\pm\}$ constitute  the Kerr geometry extended geodesic structure. There is $\ell_\gamma^-=a$ and $r_{[\gamma]}^-=a^2$--see also Figs\il(\ref{Fig:PlotparolA2}).
}\label{Table:tori-counter-rotating}
  \end{table}
We analyze two families of accretion  tori,  governed by the distribution of specific angular momentum on the equatorial plane,
\bea\label{Eq:pian-l-equa-l}
\ell^{\mp}\equiv\frac{a^3\mp r^{3/2} \Delta-a (4-3 r) r}{a^2-(r-2)^2 r},
\eea
\cite{abrafra,submitted,ringed,pugtot}.
Toroids are full  GRHD (barotropic) Polish doughnut  (PD) models,  composed by one -particle-species  fluids orbiting on the equatorial plane (the tori  symmetry  plane) of   the central attractor, and   parameterized with constant specific angular momentum $\ell^\pm=$constant.

The  fluid dynamics  is  governed by the Euler equation only, expressible through an effective potential function, $V_{eff}(r;\ell,a)$,  encoding the centrifugal and gravitational components of the force balance.
The pressure gradients are  regulated by  the gradients of an effective potential function for the fluid.
Toroidal configurations are characterized by a maximum of the pressure and density (torus center $r_{center}$) and, eventually, a vanishing  of pressure point, tori cusps $r_{\times}$ and proto-jets cusps $r_j$. Proto-jets  are open, cusped,  HD  toroidal configurations, with matter funnels along the attractor rotational axis.
The  torus  edges are solutions of $V_{eff}^2=K^2=$constant   and there is   $K=K_\times$ for the  cusped tori  edges.
There is $K_\times<1$ for tori cusps and $K_\times\geq1$ for proto-jets cusps--see Table\il(\ref{Table:tori-counter-rotating}), these values can be obtain from the function $K(r)\equiv V_{eff}(r;\ell(r),a)$.

Geometrically thick tori considered in this analysis are well known and widely used in literature--see for example \cite{abrafra,PLUS}.
In this section  we outline some fundamental aspects for the analysis of the turning points, constituting  the constraints used in the second part of this work, while in Sec.\il(\ref{Sec:fluid-effe-pot}) we summarize some in-depth aspects of these models.

We shall  investigate the  "\emph{disk--driven}" free--falling accretion flow composed by matter and photons,  and the case of  "\emph{proto-jets (or jets)  driven}" flows.

In this section we discuss further the notion of  co-rotation and counter-rotation in \textbf{NSs} spacetimes.

\begin{figure}
\centering
      \includegraphics[width=7.65cm]{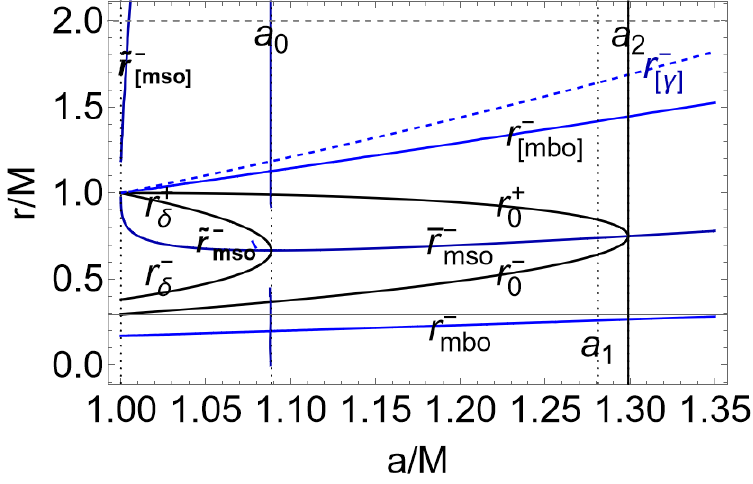}
\includegraphics[width=7.65cm]{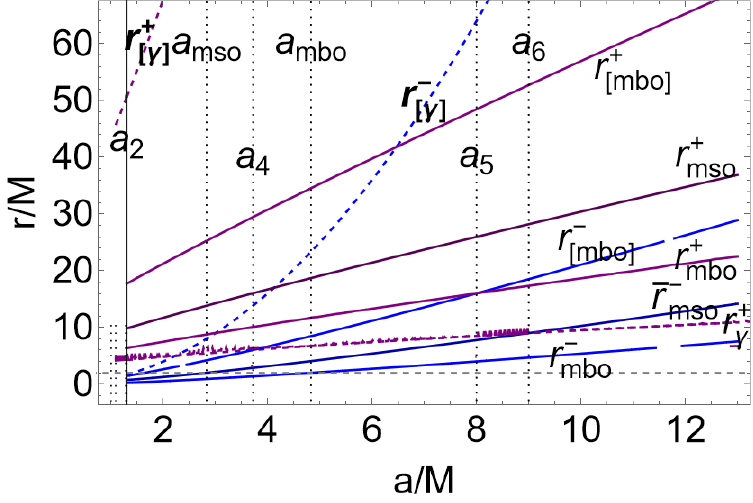}
    \includegraphics[width=7.65cm]{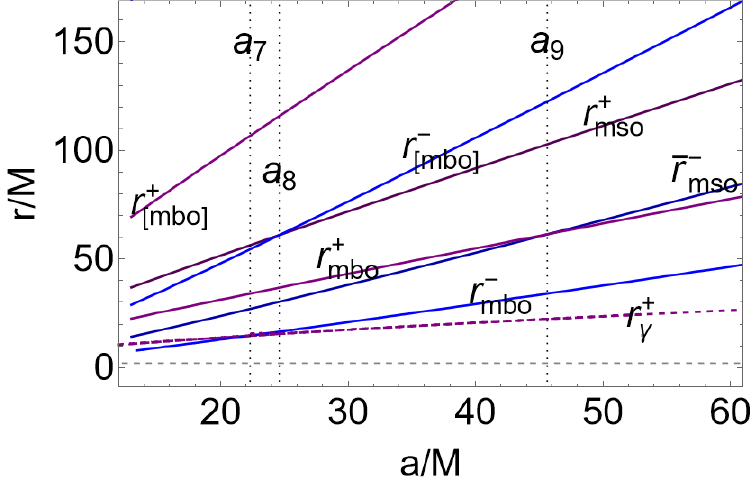}
    \includegraphics[width=7.65cm]{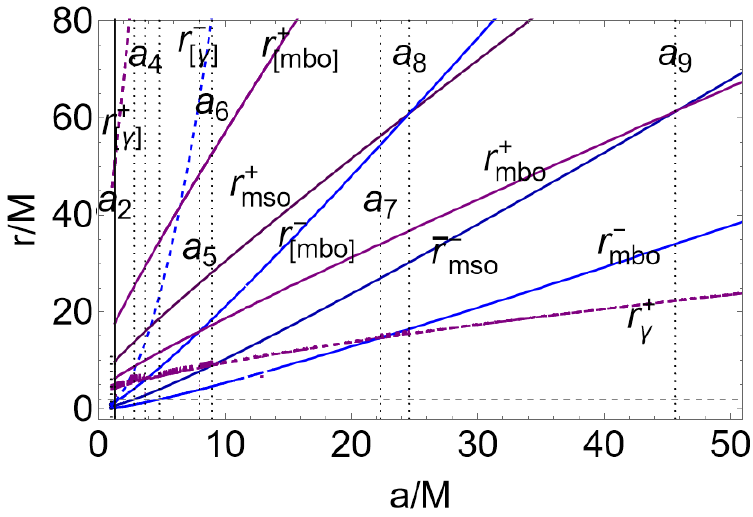}
    \caption{Analysis of the  extended geodetic structure of the Kerr \textbf{NS}, constraining   tori with fluid angular momentum $\ell=\ell^\pm$.  Spins $\{a_0,a_1,a_2,a_4,a_5,a_6,a_7,a_8,a_9\}$ are defined in Table\il(\ref{Table:corotating-counter-rotatingspin}).  Radius $r_{\gamma}^+$ is the counter--rotating last circular orbit relative to counter-rotating fluids with $\ell=\ell^+<0$, radius $r_\epsilon^+=2M$, is the outer ergosurface on the equatorial plane. $\{r_{mso}^\pm,r_{mbo}^{\pm}\}$ are the marginally stable orbit and marginally bounded orbits relative to fluids with  $\ell^\pm$ respectively.   Radii $\{r_{[mso]}^{\pm}, r_{[mbo]}^{\pm},r_{[\gamma]}^+\}$ are defined on Table\il(\ref{Table:tori-counter-rotating}). On radii  $r_0^\pm$  momenta are $\La(r_0^\pm)=0$,  and there is  $\Em(r_\delta^\pm)=0$. }\label{Fig:PlotparolA2}
\end{figure}

We introduce the   radii    $r_0^\pm<r_\epsilon^+=2M:\La=\ell=0$,
 and
 $r_\delta^\pm: \Em=0,\La>0$ where $r_{0}^-< r_\delta^-<r_\delta^+< r_0^+$, showed in  Figs\il(\ref{Fig:PlotparolA2}) (explicit forms are in  \cite{submitted}) and spins $a_j$ with $j\in\{0, 9\}$ and $\{a_{\mathbf{i}},a_{\mathbf{ii}},a_{
 \mathbf{iii}},a_{\mathbf{iv}}\}$, in Table\il(\ref{Table:corotating-counter-rotatingspin}),
\begin{table}
\begin{tabular}{l|l|l}
  \hline
  $a_0\equiv \frac{4}{3} \sqrt{\frac{2}{3}}M=1.08866M: r_\delta^+=r_\delta^-$,&$ a_1 \equiv1.28112M$,& $a_2\equiv \frac{3 \sqrt{3}}{4}=1.29904M: r_0^+=r_0^-$
  \\
$a_3\equiv \sqrt{3}M$&$a_4\equiv 3.7195M:  r_\gamma^+=r^-_{[mbo]}$,&
$a_5\equiv8M: r^-_{[mbo]}= r_{mbo}^+$\\
$a_6\equiv 9M: r_\gamma^+=\bar{r}_{mso}^-$&
$a_7\equiv 22.3137M: r_\gamma^+=r_{mbo}^-$&
$a_8\equiv 24.6082M: r^-_{[mbo]}= r_{mso}^+$\\
$a_9\equiv 45.6274M: r_{mbo}^+=\bar{r}_{mso}^-$
&
 $a_{\mathbf{i}}\equiv 4.12311M: r_{mso}^+={r}_{[\gamma]}^-$&
 $a_{\mathbf{ii}}\equiv 3M: r_{mbo}^+={r}_{[\gamma]}^-$\\
  $a_{\mathbf{iii}}\equiv 2.23607M: r_{\gamma}^+={r}_{[\gamma]}^-$
&$a_{\mathbf{iv}}\equiv 1.03886 M: r_{mso}^+=\tilde{r}_{[mso]}^-$,&
\\
  \hline
\end{tabular}
\caption{Notable spins in the Kerr NS spacetime. Radii $r$, defining the spin sets are in Table\il(\ref{Table:tori-counter-rotating}). Radii $r_0^\pm$ and $r_\delta^\pm$  are defined in Sec.\il(\ref{Sec:accretion-Tori})-- see also Figs\il(\ref{Fig:PlotparolA2}).}\label{Table:corotating-counter-rotatingspin}
\end{table}
defined by the geodesic properties of the spacetime, showed in  Figs\il(\ref{Fig:PlotparolA2}).
Fluid specific angular momentum $\ell=\ell^+<0$ defines counter-rotating fluids  while  $\ell=\ell^-\lesseqgtr0$  distinguishes co-rotating and counter-rotating tori  respectively and  we  summarize the situation  in Table\il(\ref{Table:corotating-counter-rotating}).
 \begin{table}
 \resizebox{1\textwidth}{!}{%
\begin{tabular}{l|l}
  \hline
  \textbf{Counter-rotating tori  $(\ell<0)$}&\textbf{Co-rotating tori  ($\ell>0$)}
  \\   $\ell=\ell^+<0$,  $\ell=\ell^-<0$ in the ergoregion for  $a\in[M,a_2]$& $\ell=\ell^->0$ $(\La>0,\Em>0)$,  for $(r>0,a>a_2)$
  \\
 \textbf{(I)}  $\ell=\ell^+<0$ $(\La<0,\Em>0)$ (out of  the ergoregion) & $(r\in]0,r_0^-[\cup]r_0^+,+\infty[,a\in]a_0,a_2[)$ $(\La>0,\Em>0)$
 \\ \textbf{(II)}    $\ell=\ell^-<0$,  $(\La<0,\Em>0)$, in
$]r_0^-,r_\delta^-[\cup]r_\delta^+,r_0^+[$, for  $a\in[M, a_0]$.& $(]r_\delta^-,r_\delta^+[,a\in]M, a_0[$ $(\La<0,\Em<0)$
\\
 and in $]r_0^-,r_0^+[$, for  $a\in]a_0,a_2]$&
$ r\in]0,r_0^-[\cup r>r_0^+,a\in]M, a_0[$ $(\La>0,\Em>0)$\\
\hline
  \end{tabular}}
  \caption{Co-rotating and counter-rotating tori in Kerr \textbf{NSs}, $\ell$ is the fluid specific angular momentum.  Radii $r_0^\pm$ and $r_\delta^\pm$  are defined in Sec.\il(\ref{Sec:accretion-Tori})-- see also Figs\il(\ref{Fig:PlotparolA2}), spins are defined in Table\il(\ref{Table:tori-counter-rotating}). $\Em$ and $\La$ are the test particles energy and angular momentum.}\label{Table:corotating-counter-rotating}
  \end{table}
However, $\ell^-<0$ with  $(\La<0,\Em>0)$  in the \textbf{NS} ergoregion,   and  with   co-rotating solutions  $\ell^->0$  with $(\La<0,\Em<0)$ with negative energy $\Em$   are all co-rotating with respect to the static observers at infinity as they correspond  to  the relativistic angular velocity with respect to static observers at infinity,    $\Omega>0$.

The Kerr geodesic structure regulates the accretion disk physics bounding the accreiting disk inner edge (tori cusps) and proto-jets cusps. The cusps are point of the minimum pressure and density in the  barotropic  toroids--see Sec.\il(\ref{Sec:fluid-effe-pot}). The \emph{extended geodesic structure}  includes      radii  $\{r_{[\gamma]}^{\pm},r_{[mbo]}^{\pm},r_{[mso]}^{\pm}\}$ defined Table\il(\ref{Table:tori-counter-rotating}) and   governing   the location of the   toroidal configurations centers (point of maximum density and pressure in the barotropic tori) \cite{letter,dsystem,multy,dragged}.

Table\il(\ref{Table:tori-counter-rotating})  shows  the situation  for fluids with  $\ell^+$, and  for fluids with $\ell=\ell^-$  orbiting in  \textbf{NS} spacetimes with  spins $a>a_2$.

 \textbf{NSs} with  spin $a\in]M,a_2]$ and fluids with $\ell^-$ are characterized by a more articulated structure and we summarized this situation  in Figs\il(\ref{Fig:PlotparolA2}).
  For tori   centers and  cusps located on  $r_{0}^\pm$   there is $\ell=\La=0$ (tori in these  limiting cases  are considered   in \cite{submitted}). On  $r_\delta^{\pm}$ ("center" and "cusp" respectively)  there is  $\Em =0$ ($\ell$ is not well defined).
For  $a\in[M, a_2]$, there can be    tori   with $\ell^-<0$ and ($\Em>0,\La<0$) in the ergoregion  and
  tori with  $\ell^->0$ and ($\Em<0,\La<0$). There can be  double tori system  with $\ell=\ell^+=\ell^-<0$ or $\ell=\ell^->0$, and this case is detailed  in Sec.\il(\ref{Sec:double-tori}).
\begin{figure}
 \includegraphics[width=5.75cm]{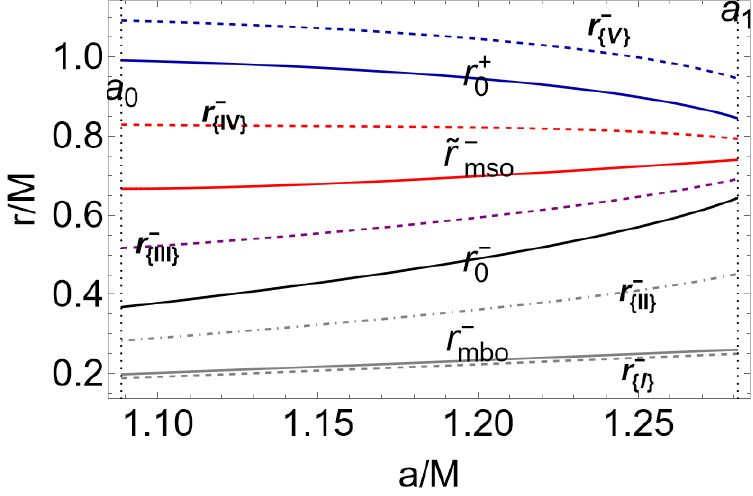}
   \includegraphics[width=5.75cm]{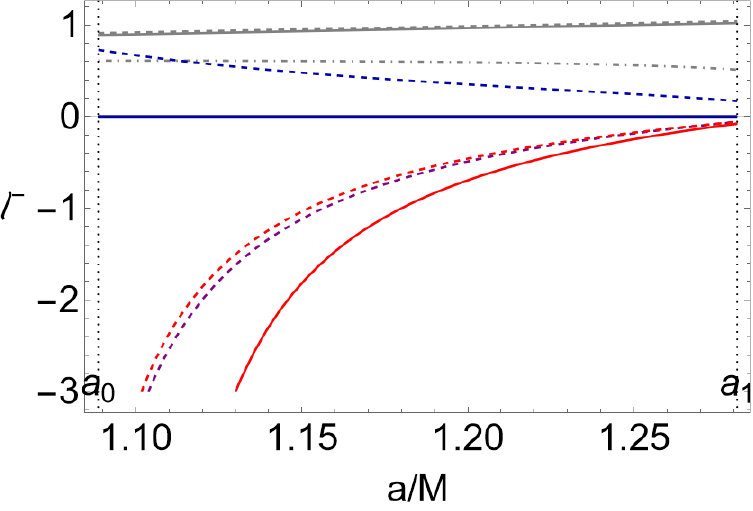}
   \includegraphics[width=5.75cm]{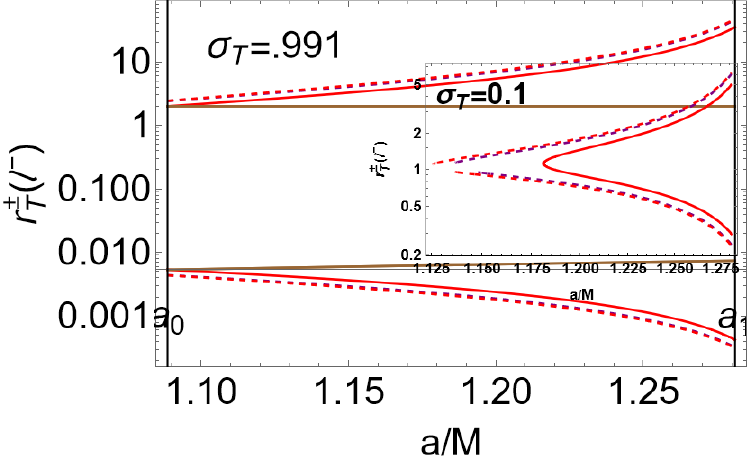}
  \includegraphics[width=5.75cm]{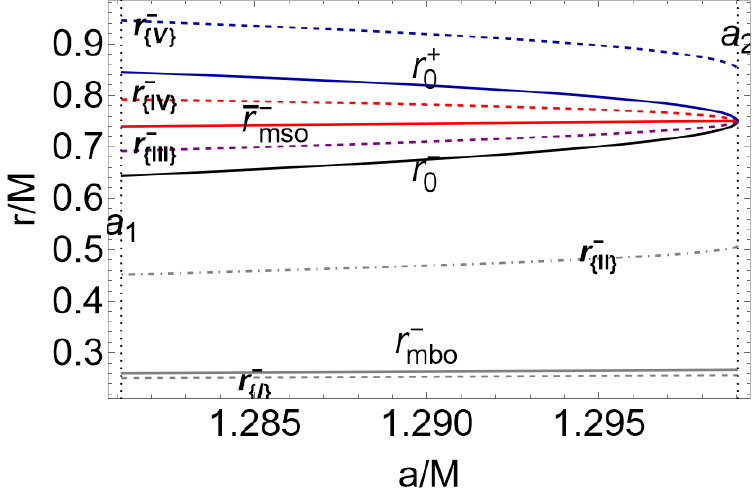}
   \includegraphics[width=5.75cm]{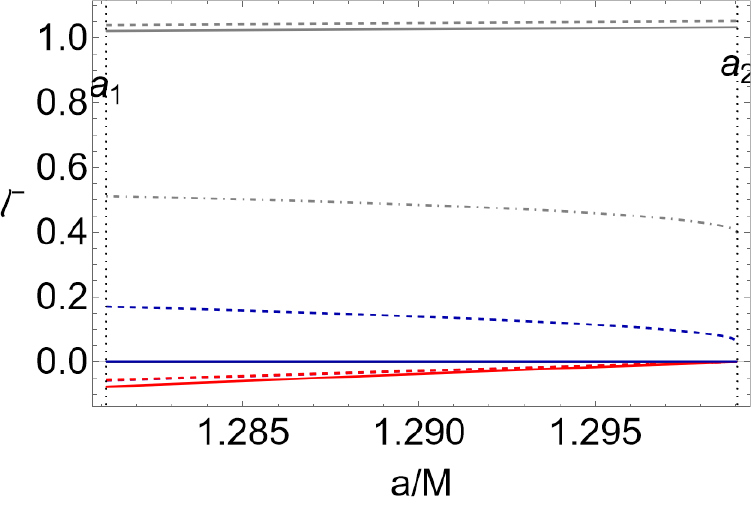}
   \includegraphics[width=5.75cm]{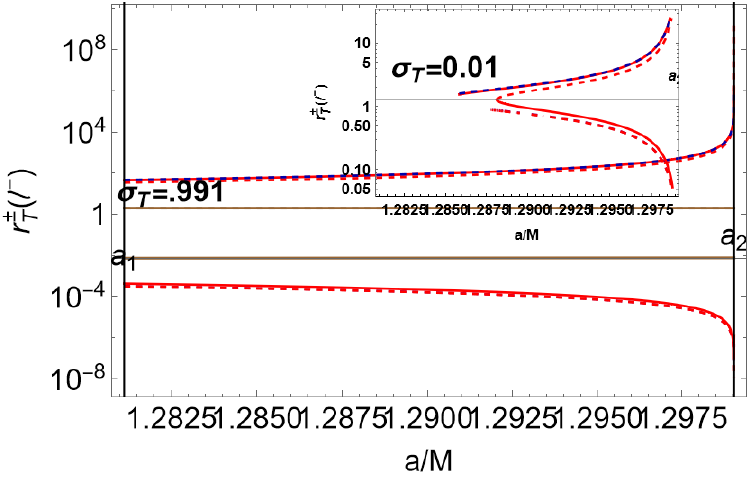}
 \caption{Analysis of the flow inversion point radius $r_\Ta^\pm$ with fluid specific angular momentum $\ell=\ell^-$.  (For $\ell^->0$  inversion points are only for spacelike particles with $(\Em<0,\La<0)$.) Spins $\{a_0,a_1,a_2\}$ are defined in Table\il(\ref{Table:corotating-counter-rotatingspin}). Radii $r_{mso}^-=\{\bar{r}_{mso}^-,\tilde{r}_{mso}^-\}$ are the marginally stable orbit, radius $r_{mbo}$ is the marginally bounded orbit.
 There is $\sigma\equiv \sin^2\theta$ ($\sigma=1$ is the equatorial plane). There is  $r_0^\pm: \La=0 $, where $\La$ is the test particle angular momentum and $r_\delta^\pm:\Em=0$ where $\Em$ is the test particle energy.
 Radii  $r/M$ (left panels) and fluid specific angular momentum $\ell^-$ (center panels) are plotted as functions of $a/M$, inversion point radius $r_\Ta^\pm$ is shown  for different planes signed on the panels (right panels).   See also Figs\il(\ref{Fig:Plotdacplot2r3}). There is $r_{\{I\}}^-\equiv r_{mbo}^--0.01,r_{\{II\}}^-\equiv  \left(r_{0}^--r_{mbo}^-\right)/2+r_{mbo}^-,r_{\{III\}}^-\equiv r_{mso}^-- \left(r_{mso}^--r_{0}^-\right)/2,r_{\{IV\}}^-\equiv r_{0}^+- \left(r_{0}^+-r_{mso}^-\right)/2,r_{\{V\}}^-\equiv r_{0}^++0.1$.}\label{Fig:Plotdacplot1}
\end{figure}
\begin{figure}
  \includegraphics[width=5.75cm]{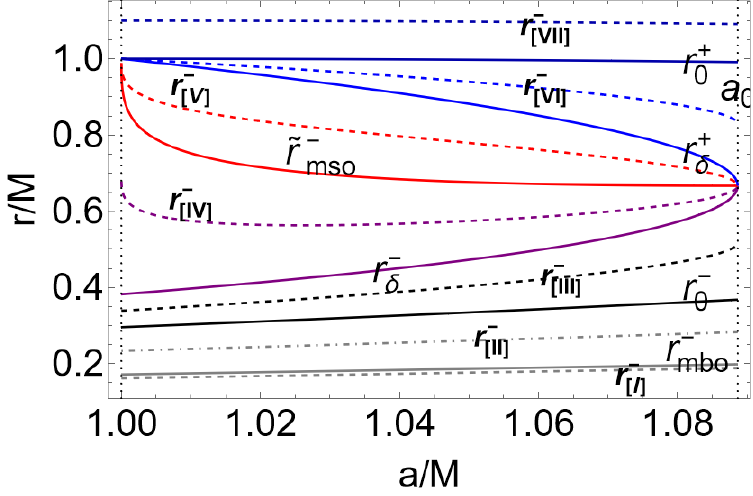}
   \includegraphics[width=5.75cm]{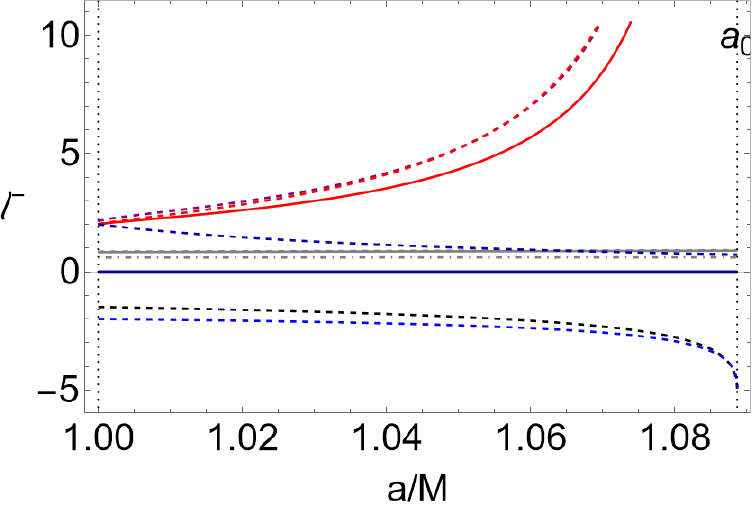}
   \includegraphics[width=5.75cm]{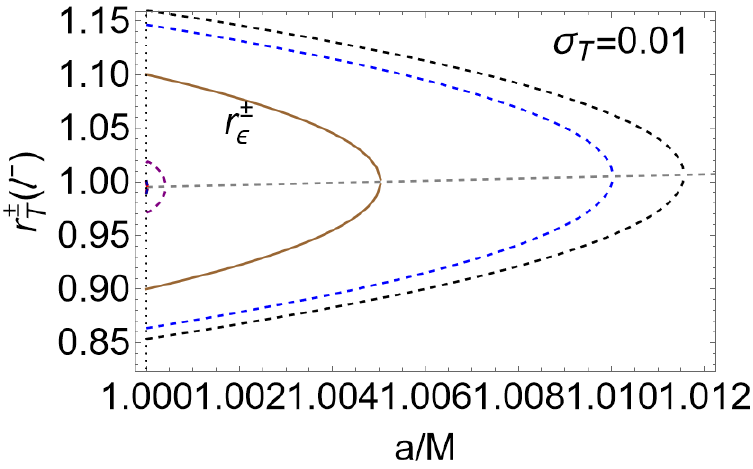}
   \includegraphics[width=5.75cm]{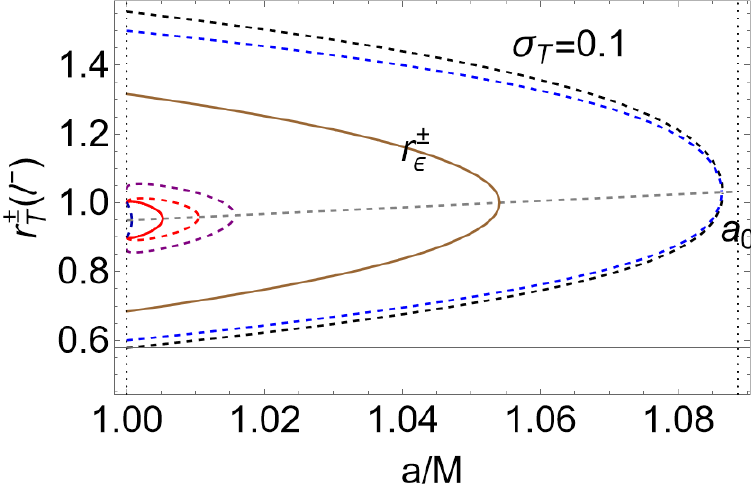}
   \includegraphics[width=5.75cm]{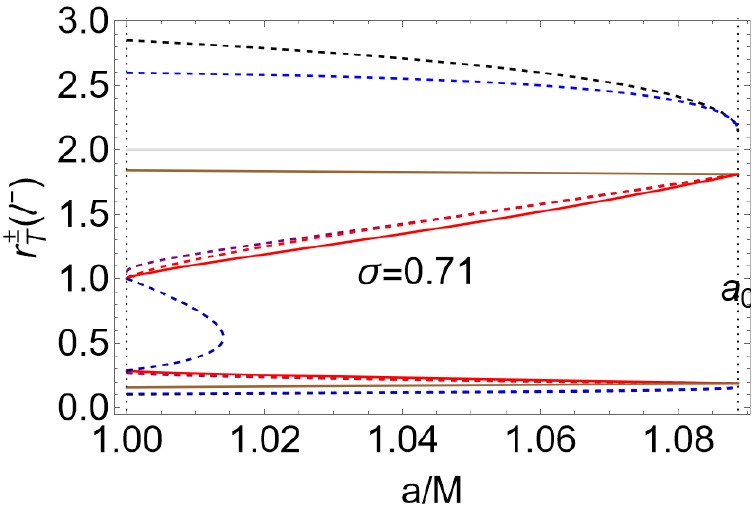}
   \includegraphics[width=5.75cm]{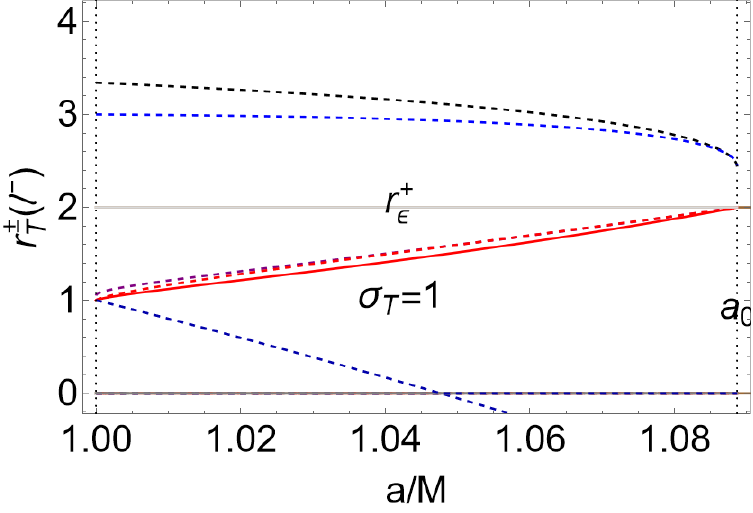}
 \caption{Analysis of the flow inversion point radius $r_\Ta^\pm$ for different planes $\sigma\equiv\sin^2\theta$ signed on the panels,  in the spin range is $a\in[M, a_0]$, with fluid specific angular momentum $\ell=\ell^-$. Radii $r_{mso}^-=\{\bar{r}_{mso}^-,\tilde{r}_{mso}^-\}$ are the marginally stable orbit, radius $r_{mbo}$ is the marginally bounded orbit.
  There is  $r_0^\pm: \La=0 $, where $\La$ is the test particle angular momentum and $r_\delta^\pm:\Em=0$ where $\Em$ is the test particle energy. Left panels show radii $r/M$ as functions  of the \textbf{NS} spin $a/M$. Center panels show the fluid specific angular momentum $\ell^-$, evaluated on the left panel radii, as function of $a/M$. Right panels: inversion radius  $r_\Ta^+>0$ on different planes, evaluated on the momenta and radii of the center and left panels.
 Spins $\{a_0,a_1,a_2\}$ are defined in Table\il(\ref{Table:corotating-counter-rotatingspin}). There is $r_{[I]}^-\equiv r_{mbo}^--0.01,r_{[\text{II}]}^-\equiv \left[r_{0}^--r_{mbo}^-\right]/2+r_{mbo}^-,r_{[\text{III}]}^-\equiv  \left[r_{\delta}^--r_{0}^-\right]/2+r_{0}^-,r_{[\text{IV}]}^-\equiv r_{mso}^-- \left[r_{mso}^--r_{\delta}^-\right]/2,r_{[V]}^-\equiv r_{\delta}^+- \left[r_{\delta}^+-r_{mso}^-\right]/2,r_{[\text{VI}]}^-\equiv r_0^+-\left[r_0^+-r_{\delta}^+\right]/2,r_{[\text{VI}]}^-\equiv r_0^++0.1$.  Brown curves are the outer and inner ergosurfaces $r_\epsilon^\pm$ respectively.  See also Figs\il(\ref{Fig:Plotdacplot2r3}). (For $\ell^->0$  inversion points are only for spacelike particles with $(\Em<0,\La<0)$.)}\label{Fig:Plotdacplot2r3}
\end{figure}
 \begin{figure}
\centering
    \includegraphics[width=5.75cm]{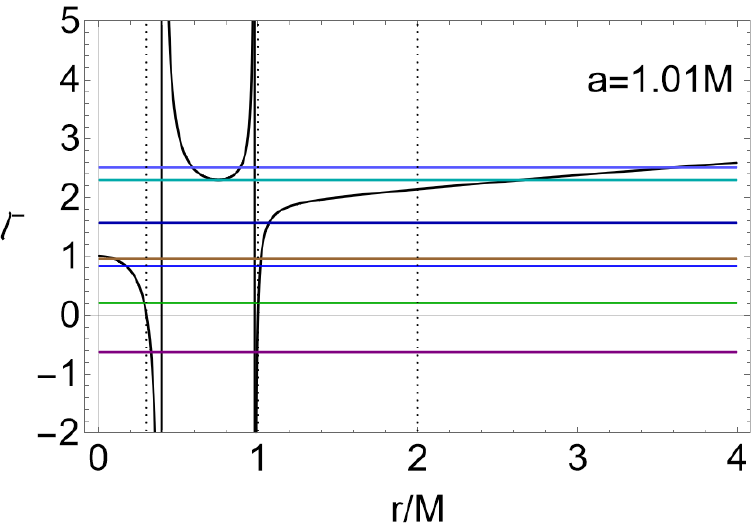}
    \includegraphics[width=5.75cm]{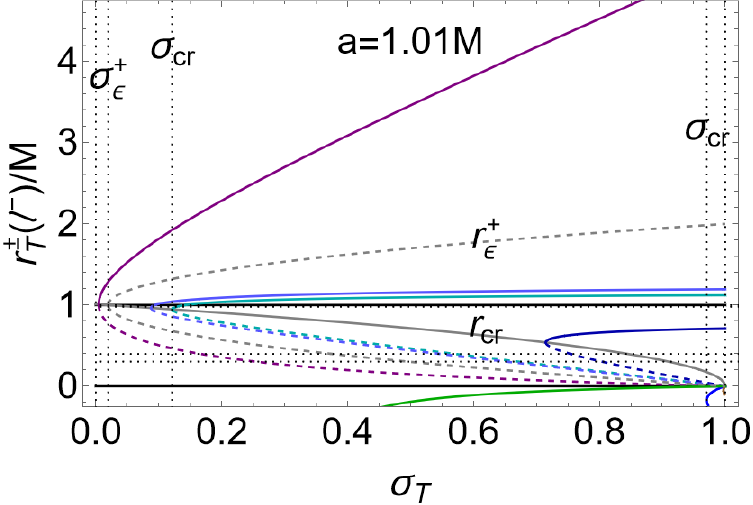}
        \includegraphics[width=5.75cm]{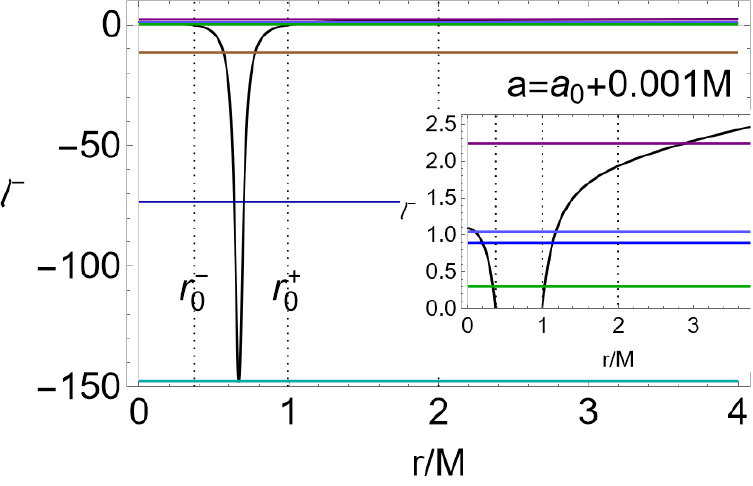}
         \includegraphics[width=5.75cm]{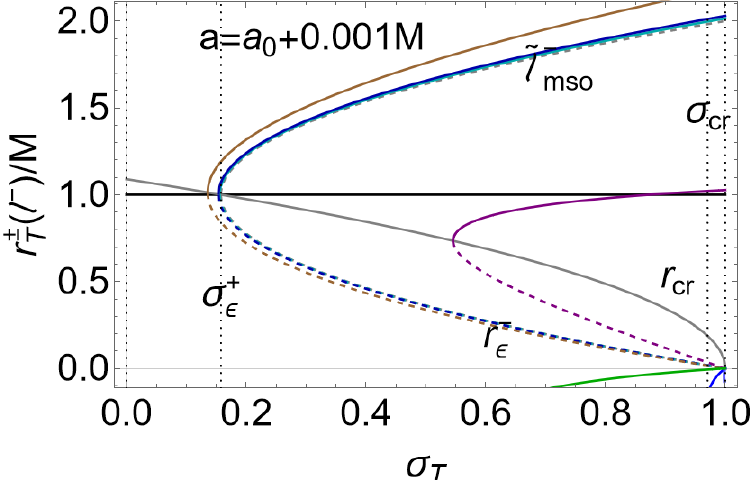}
          \includegraphics[width=5.75cm]{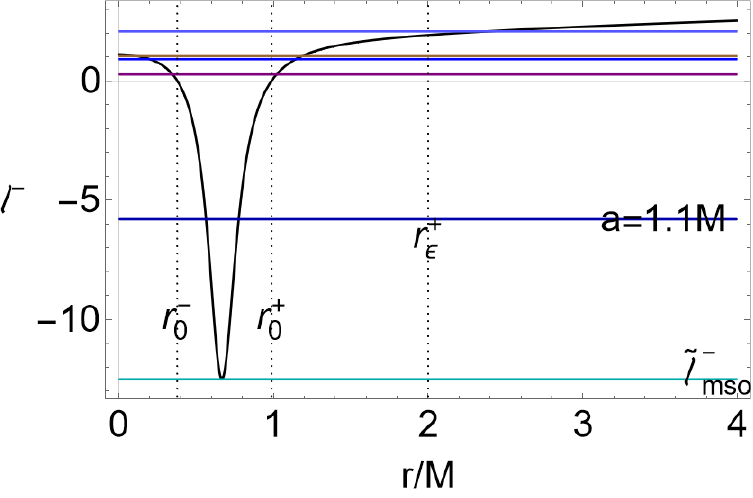}
           \includegraphics[width=5.75cm]{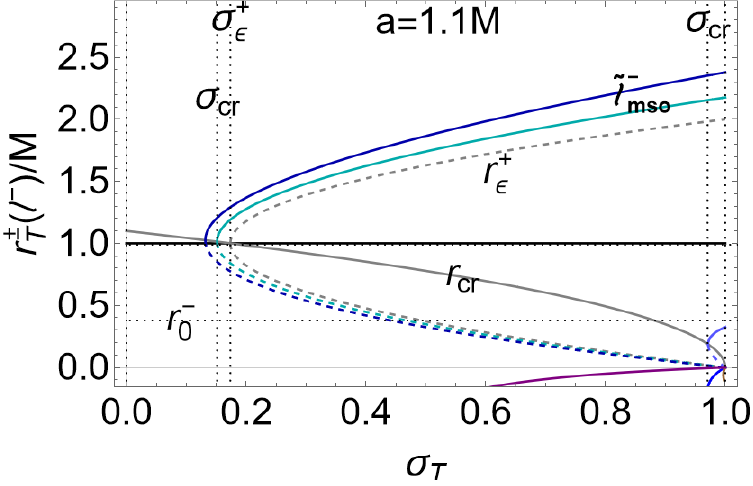}
                  \includegraphics[width=5.75cm]{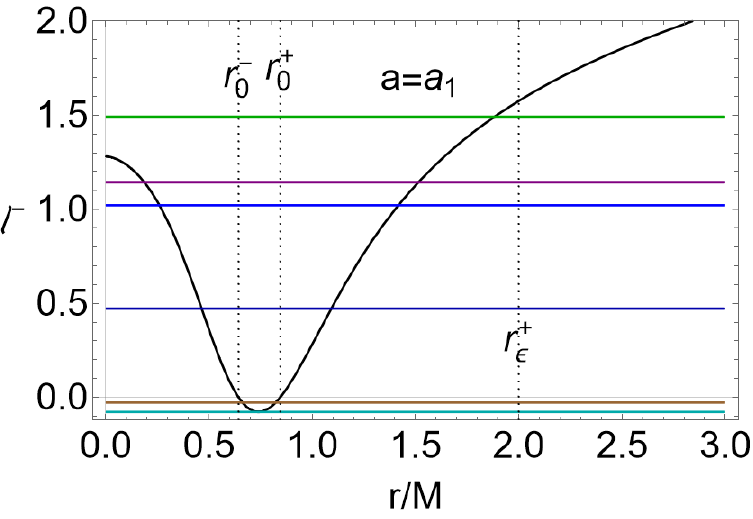}
       \includegraphics[width=5.75cm]{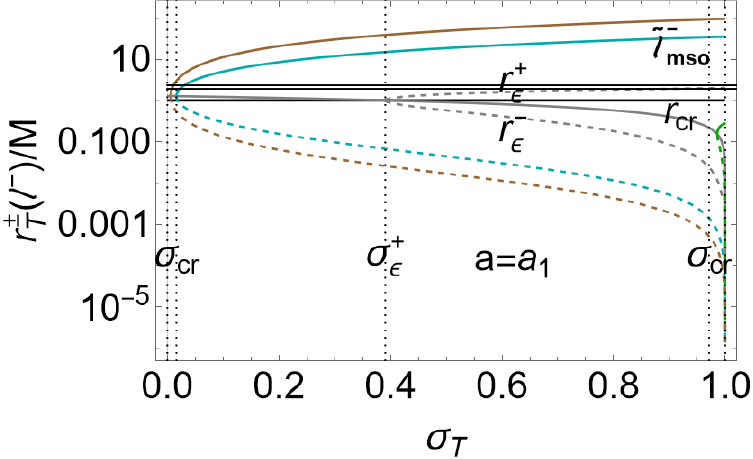}
                               \includegraphics[width=5.75cm]{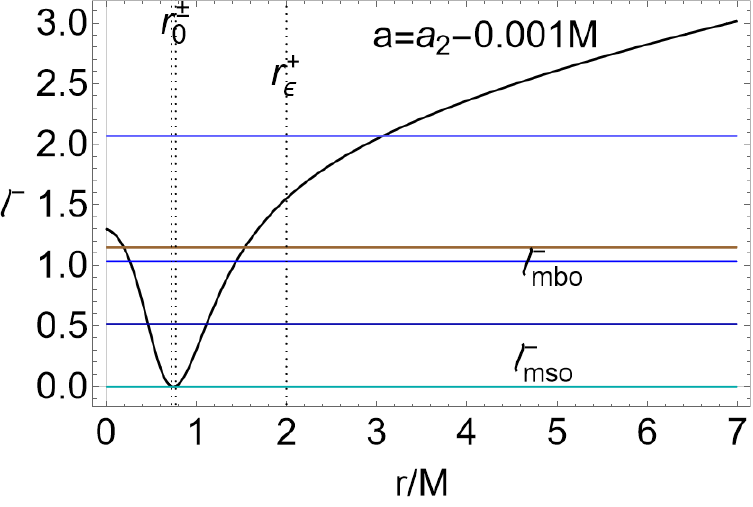}
                                 \includegraphics[width=5.75cm]{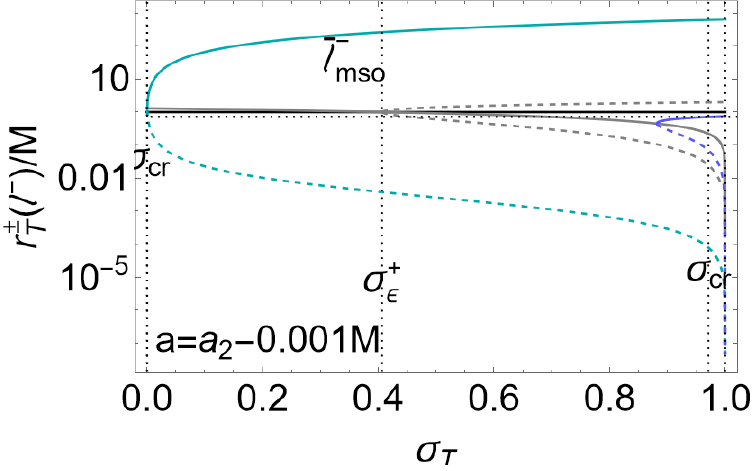}
      \includegraphics[width=5.75cm]{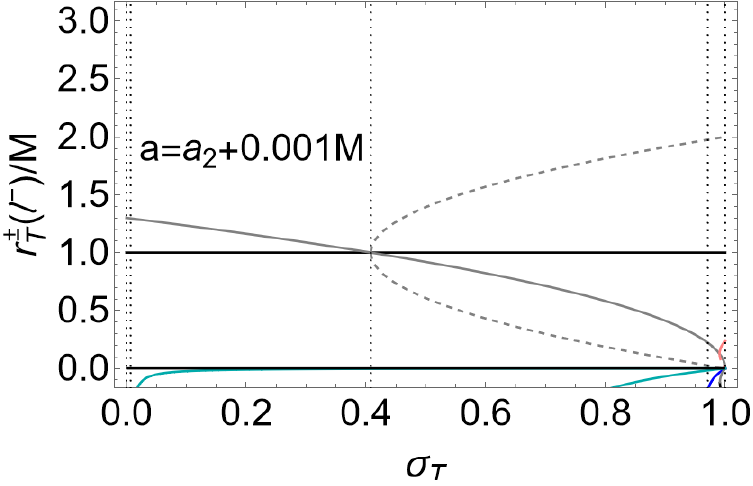}
      \caption{Fluid specific angular momentum $\ell^-$ for different spacetime spin $a/M$ signed on the panels, darker-cyan curve is $\ell_{mso}^-$, light-blue curve is $\ell_{mbo}$. Marginally circular orbit  radius is $r_{mbo}^-$  and marginally stable orbits $r_{mso}^-=\{\tilde{r}_{mso}^-,\bar{r}_{mso}^-\}$. Horizontal lines select $\ell$ values used for the evaluation of the inversion radii. Inversion points $r_\Ta^{\pm}$ of Eq.\il(\ref{Eq:lLE})  of the co-rotating  and counter-rotating  accretion with $\ell=\ell^-$ are  shown as functions of the  planes $\sigma\equiv \sin^2\theta$, where $\sigma=1$ is the equatorial plane, for different  \textbf{NS} spin-mass ratio $a/M$ signed on the panels.   Plain curves are $r_\Ta^+$ while $r_\Ta^-$ are dashed curves.   For $\ell^->0$  inversion points are only for space-like particles with $(\Em<0,\La<0)$. Spins $\{a_0,a_1,a_2\}$ are defined in Table\il(\ref{Table:corotating-counter-rotatingspin}). Radius $r_\epsilon^\pm$ are the outer and inner ergosurfaces.
  Plane curve $\sigma_\epsilon^+$ is defined in Eqs.\il(\ref{Eq:sigmaepsilonplus}) and radius $r_{cr}$ is in Eq.\il(\ref{Eq:sigmatur}). Limiting plane $\sigma_{cr}$ is function $\sigma_{\sigma a}$ of Eqs\il(\ref{Eq:sas}) on the selected fluid specific angular momentum. There are no inversion points for co-rotating fluids in the geometries  with $a>a_2$.}\label{Fig:Plotvinthalloa5p5}
\end{figure}
 \section{Flow inversion points from orbiting  tori   with  $\ell=\ell^\pm$}\label{Sec:contain-inner-torus}
In this section we analyze the accretion dirven and proto-jets driven flow inversion points.
In Sec.\il(\ref{Sec:FIP}) the  flow inversion points  are defined, while  in Sec.\il(\ref{Sec:fromaccretionflows}) we discuss the flow inversion points from the orbiting accretion tori and proto-jets configurations.
\subsection{Flow inversion points}\label{Sec:FIP}
 The flow inversion points  are points of vanishing  axial velocity of the flow motion as related to distant static observers, therefore defined by the condition   $u^{\phi}=0$ (or equivalently $\Omega=0$) on the flow particles and  photons velocities (relativistic angular velocity).

From the definition of  $\ell$  and  $\Em$, Eqs\il(\ref{Eq:EmLdef}) and  Eq.\il(\ref{Eq:flo-adding}), and the  inversion point definition   we obtain:
\bea\label{Eq:lLE}
&&\ell=-\left.\frac{g_{t\phi}}{g_{tt}}\right|_\Ta=-\frac{2 a r_{\Ta} \sigma_{\Ta} }{\Sigma_\Ta -2 r_\Ta},
\quad \mbox{where}
\\&&\nonumber
r_\Ta^\mp=\mp\sqrt{a^2 \left(\sigma_\Ta +\frac{\sigma_\Ta ^2}{\ell ^2}-1\right)-\frac{2 a \sigma_\Ta }{\ell }+1}-\frac{a \sigma_\Ta }{\ell }+1,
\eea
and
\bea
&&
\label{Eq:sigmatur}
\sigma_\Ta\equiv\frac{\ell \Delta_\Ta}{a (a \ell -2 r_\Ta)},\quad  r_\Ta^-= r_\Ta^+=r_{cr}\equiv a\sqrt{1-\sigma_\Ta},\\&&\nonumber
\Em_\Ta \equiv - g_ {tt}(\Ta) \dot {t}_\Ta,\quad \La_\Ta \equiv g_ {t\phi}(\Ta)\dot{t}_\Ta,
\eea
where  notation $q_\Ta$ or $q(\Ta)$ is for  any quantity $q$ considered at the inversion point, and $q_0=q(0)$ for  any  quantity $q$ evaluated at the initial point of the (free-falling) flow trajectories.

Therefore, $\Em_\Ta$ and  $\La_\Ta$ are the energy   and momentum of the flow particles at the inversion point. We stress that  quantities in Eq\il(\ref{Eq:lLE})  and Eq.\il(\ref{Eq:sigmatur})  are independent from  the normalization condition,  being a consequence of the definition of constant  $\{\ell,\mathcal{E},\laa\}$.
Radius $r_\Ta$ identifies a surface,\emph{ inversion  sphere}, which is  a general property of the (photon and timelike) orbits in the Kerr  \textbf{NS} spacetime, surrounding the central singularity,
depending only on $\ell$ where $\Omega=0$. Furthermore, radius  $r_{cr}$,
of Eq.\il(\ref{Eq:sigmatur})  is   a background property,
 independent  explicitly from  $\ell$.
It si worth noting that there are no timelike and photon-like inversion points with  $\ell>0$.

 Contrary  to the \textbf{BH} case, in the \textbf{NS}  there can be  two inversion points radii $r_\Ta^\pm$.
The inversion radii on the poles are well defined  in the \textbf{BH} case only ($\lim_{\sigma_\Ta\rightarrow0}r_{\Ta}^{\pm}=r_{\pm}$).
Note that a very large $\ell$ in magnitude is  typical of proto-jets emission or  quiescent   toroids orbiting  far from the central attractor for counter-rotating tori (with $\ell=\ell^+<0$) or some co-rotating tori with $\ell=\ell^->0$) in a class of slow rotating \textbf{NSs}, in this case
the inversion point approaches    the ergosurface
i.e.
$\lim_{\ell\rightarrow \pm \infty} \sigma_\Ta=\sigma_{erg}$ and $\lim_{\ell\rightarrow (\pm \infty)} r_{\Ta}^{\pm}=r^\pm_{\epsilon}$, we address this issue with more details in  Secs\il(\ref{Sec:ergo-inversion-in-out}).

Increasing  the   spin $a/M>0$,  the inversion point approaches  the equatorial plane $(\sigma_\Ta\leq 1)$. However, for a very faster spinning \textbf{NS}  ($a\rightarrow+\infty$) the limit of  $r_\Ta=r_\Ta^{\pm}$ for    is not well defined.
\subsection{Accretion flows turning points}\label{Sec:fromaccretionflows}
In Sec.\il(\ref{Sec:FIP}) we discussed  the existence of the inversion points  considering the condition of    constant  $\ell=\ell^\pm$, independently by the causal conditions on matter at the turning point.

In this section we   explore the  inversion spheres  properties for flows from the  accretion disks  orbiting   \textbf{NSs}, depending on  the  central attractor   spin-mass ratios $a/M$ and   fluid specific angular momenta $\ell^\pm$.

\begin{figure}
\centering
    \includegraphics[width=6.75cm]{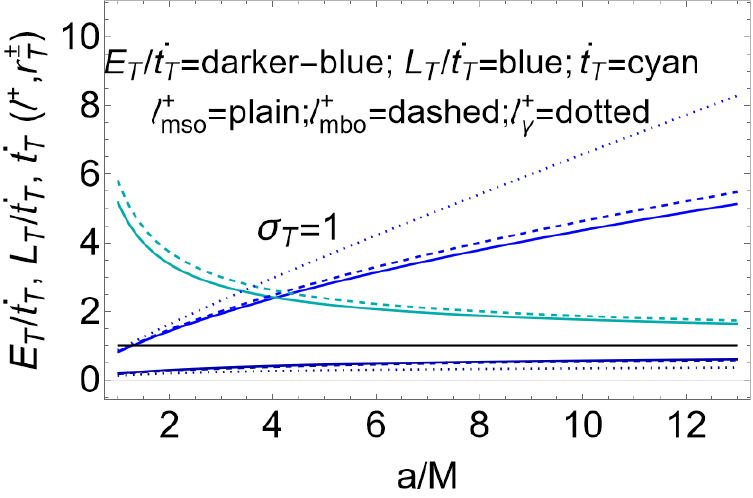}
          \includegraphics[width=6.75cm]{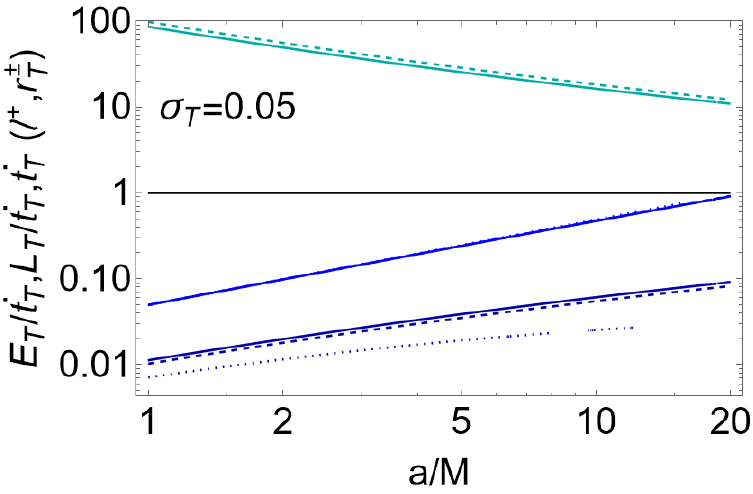}
  \caption{Analysis of the counter-rotating flows inversion points $r_\Ta^{\pm}$ of
Eq.\il(\ref{Eq:lLE}) with fluid specific  angular momentum $\ell^+<0$  of the plane $\sigma\equiv \sin^2\theta$, for different \textbf{NS} spin-mass ratios signed on the panels, where $\sigma=1$ is the equatorial plane.  Notation $(mso)$ is for quantities evaluated on the marginally stable orbit, $(mbo)$  refers  to the marginally bounded orbit, $(\gamma)$ indicates the marginally circular orbit.  Particles energy ratio $\Em_\Ta/\dot{t}$ and  angular momentum $\La_\Ta/\dot{t}$  at the inversion points $r_\Ta^\pm$
of Eqs\il(\ref{Eq:lLE}) (where $u^t=\dot{t}$) are shown for two different inversion planes $\sigma_\Ta$.  (The most general solutions $r_\Ta^\pm$ and $\sigma_\Ta$ are shown not considering    energy and momentum constraints of Sec.\il(\ref{Sec:fromaccretionflows}).) }\label{Fig:Plotvinthallene}
\end{figure}

We  focus on the necessary conditions for the existence of the inversion points of the counter-rotating and co-rotating flows, considering first the condition $\ell=$constant and then $\{\ell,\Em,\La\}$ constant, where a more general
necessary condition for the occurrence of the inversion point is $\La<0$.
We assume   constance of  $(\Em_\Ta, \La_\Ta)$  (implied by  $\ell=$constant) evaluated
at the inversion point with $\dot{t}$>0.
We then consider the  normalization condition at the  inversion point, $g_{\alpha\beta}u^\alpha u^\beta=\kappa$ (with  $u^\phi=0$) distinguishing particles and photons in the flow.
 The matter flow is then related to the orbiting structures,  considering   toroids with $\ell=\ell^+<0$ $(\La<0,\Em>0)$,  centered at $r>r_\gamma^+$-- Figs\il(\ref{Fig:Plotvinthallene}) and tori with momentum $\ell=\ell^-<0$ $(\La<0,\Em>0)$, in
$]r_0^-,r_\delta^-[\cup]r_\delta^+,r_0^+[$, for spacetimes $a\in[M, a_0]$, and in the orbital region $]r_0^-,r_0^+[$, for \textbf{NSs} with spin $a\in]a_0,a_2]$.

Inversion radius $r_\Ta$ and plane $\sigma_\Ta$ of Eqs\il(\ref{Eq:rtur-econ-bi-nign}) are not independent variables, and they can be found solving the equations of motion or using   further assumptions at any other points of the fluid trajectory. However, quantities  $(\sigma_\Ta (r_\Ta),r_{\Ta}(\sigma_{\Ta}))$ depend on the  constant of motion  $\ell$ only,
describing    both matter and photons, and they   are independent from the initial particles velocity  ($\{\dot{\sigma}_{\Ta},\dot{r}\}$, therefore their dependence on the tori models and accretion process is limited to the dependence on the fluid specific angular momentum $\ell$, and the results considered here are adaptable to a variety of different general relativistic accretion models.
We  note that the inversion sphere  describes also  outgoing particles,  with $\dot{r}_\Ta>0$, or particles with an axial velocity $\dot{\theta}\neq0$ along the \textbf{BH} rotational axis.

For flows from the orbiting tori  or proto-jets there is a maximum and a minimum boundary  $r_\Ta(a;\ell,\sigma_\Ta)$, associated to a maximum and minimum value of $\ell$. This region, as well as  its boundaries, will  be   called the \emph{inversion corona}.
  The extremes of $\ell$ parameters are determined by the tori models parametrized with $\ell$.
The orbiting structures  constrain  the range  of values for  $\ell$,   defining  the inversion corona for proto-jets or accretion driven flows, as background geometry properties, depending only on the spacetime spin.
The \emph{accretion driven  inversion corona} has boundaries defined by $r_\Ta^\pm$ (or $\sigma_\Ta$), evaluated on $\ell_{mso}^\pm$ and $\ell_{mbo}^\pm$, while the \emph{proto-jets driven coronas} has (in general) boundaries defined by $r_\Ta^\pm$ (or $\sigma_\Ta$), evaluated on $\ell_{mbo}^\pm$ and $\ell_{\gamma}^\pm$ .
For  \emph{accretion driven}, and  \emph{proto-jets driven} inversion points,  on the equatorial plane, there is $r_\Ta<r_\times$ or $r_\Ta<r_J$.
 Test particle energy and angular momentum at the inversion points are in Figs\il(\ref{Fig:Plotvinthallene}).

The structure of this section is as follows:
\medskip

In Sec.\il(\ref{Sec:double-tori}), we address the special case of inversion points from double tori systems orbiting \textbf{NSs} at equal fluid specific angular momentum.
Tori and inversion points at $\ell=\pm
a$ are explored in Sec.\il(\ref{Sec:l=pma}).
Inversion points and torus outer edge are  the focus of Sec.\il(\ref{Sec:outer-edge}).
The inversion point location in relation to the  ergoregion in Sec.\il(\ref{Sec:ergo-inversion-in-out}).
In Sec.\il(\ref{Sec:double-inversion-points})  the presence of  double inversion points is discussed.
Inversion points from proto-jets and the inversion points verticality are focused in Sec.\il(\ref{Sec:inversionpoint-proto-jets}).
Some notes on the inversion coronas thickness and slow counter-rotating  inversion spheres
are in Sec.\il(\ref{Sec:slow-Invers}).
Further notes on flows inversion points from orbiting tori are in Sec.\il(\ref{Sec:deta-tori-integr}).
Relative location of the  flows inversion points for different initial data and  the existence of more inversion points for flow  particle trajectory are investigated Sec.\il(\ref{Sec:relative-location}).
\subsubsection{Double tori system}\label{Sec:double-tori}
Slowly spinning  \textbf{NSs}  are characterized by the presence of  double orbiting tori  and also   double  {cusped} tori,  distinguishing  \textbf{NSs} from \textbf{BHs},  and  having  equal  fluid specific angular momentum and therefore one common  inversion radius  $r_\Ta^\pm(\sigma_\Ta)$.     Such tori have been detailed in \cite{submitted}, for convenience  we report below their main properties.
     The formation\footnote{This double system might eventually be   formed by the same original orbiting matter.   However it should be noted  that the toroidal components are in general characterized by a  very high difference of the energy at  the tori centers \cite{submitted}.}  of this  system in the geometries  with $a\in]M, a_2[$ is due to the presence of fluids with $\ell^-<0$,  and  with $\ell^->0$ and  $\Em<0$.
Double tori can be both co-rotating or both counter-rotating.

\begin{figure}
\centering
\includegraphics[width=6cm]{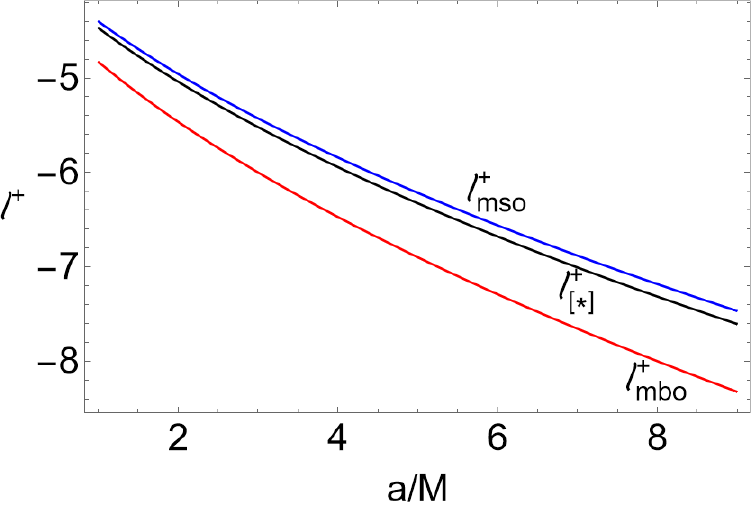}
 \includegraphics[width=6cm]{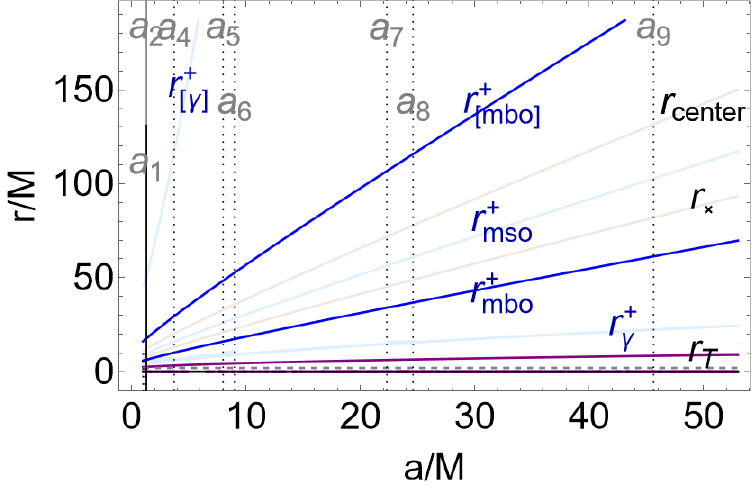}
 \includegraphics[width=6cm]{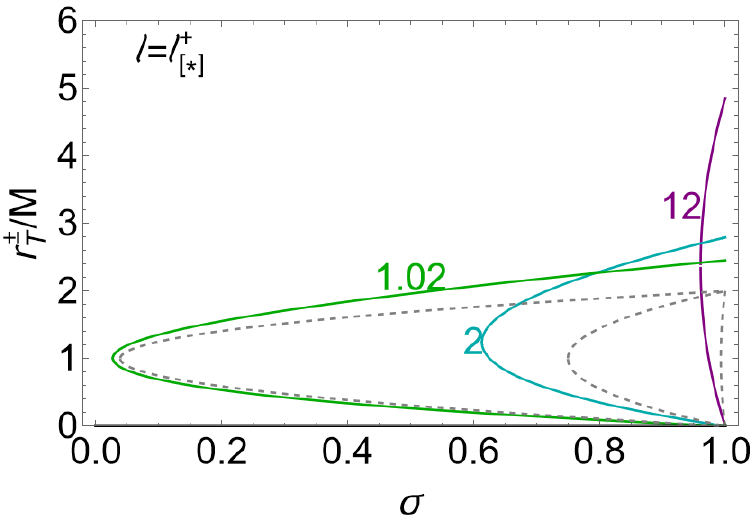}
\includegraphics[width=6cm]{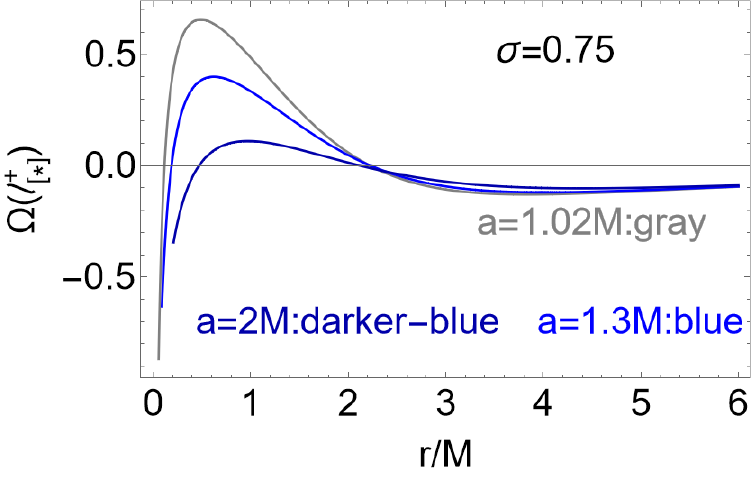}
 \caption{Inversion points for counter-rotating tori with fluid specific angular  momentum $\ell=\ell^+<0$. The analysis for  tori with fluid specific angular momentum $\ell^-$ is in Figs\il(\ref{Fig:Plotconfsud}). Spins $\{a_4,a_5,a_6,a_7,a_8,a_9\}$ are defined in Table\il(\ref{Table:corotating-counter-rotatingspin}). There is $\ell_ {[*]}^+\equiv \ell^-(r_ {[*]}^+)$, and   $r_ {[*]}^+\equiv r_{mso}^+-(r_ {mso}^+-r_ {mbo}^+)/2$. Upper left panel:  fluid angular momentum $\ell_{[*]}^+$ and $\ell_{mso}^+$, $\ell_{mbo}^+$ (on the marginally stable and marginally bounded orbit respectively) as functions of the \textbf{NS} spin-mass ratio. Upper right panel: geodesic structure of the Kerr \textbf{NS}  and  the inversion radius $r_\Ta$ for the torus model with center radius  $r_{center}$ and cusp $r_{\times}$, where $\ell=\ell_{[*]}^+$. Bottom-left panel: inversion point radius $r_\Ta^\pm$  as function of the plane $\sigma\equiv\sin^2\theta\in[0,1]$, for different \textbf{NS} spin-mass ratio signed on the curves for fluid specific angular momentum $\ell^+_{[*]}$. Dashed curves are the   ergosurfaces $r_{\epsilon}^\pm$. Bottom right panel: analysis of the flows inversion points in terms of the relativistic angular velocities $\Omega$  for particles with   fluid specific angular momentum $\ell=\ell_{[*]}^+$ for different  spins. There is $\Omega(r_	\Ta^{\pm})=0$. Note the presence of two inversion points.}.
 \label{Fig:PlotparolA0tun1}
\end{figure}

\begin{figure}
\centering
\includegraphics[width=5.75cm]{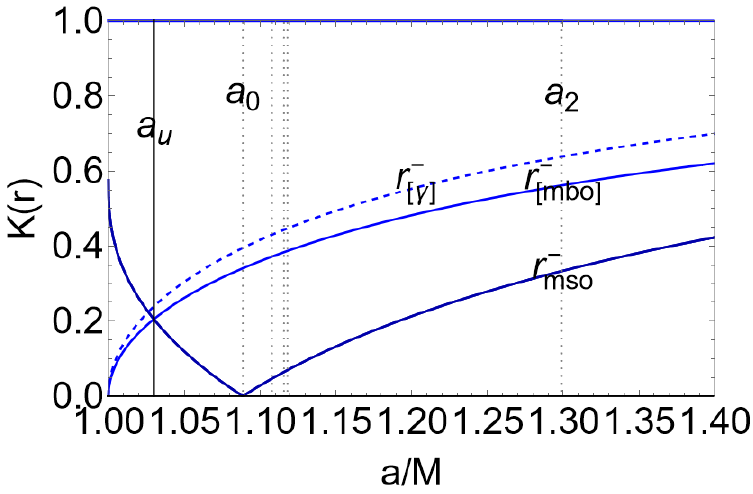}
                              \includegraphics[width=5.75cm]{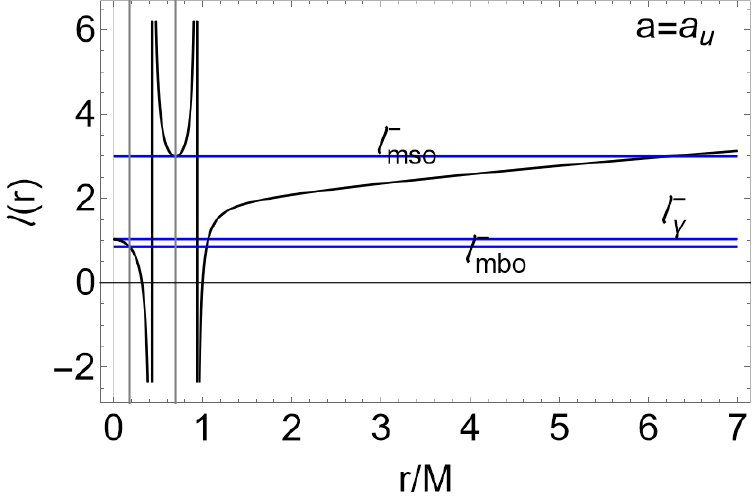}
    \includegraphics[width=5.75cm]{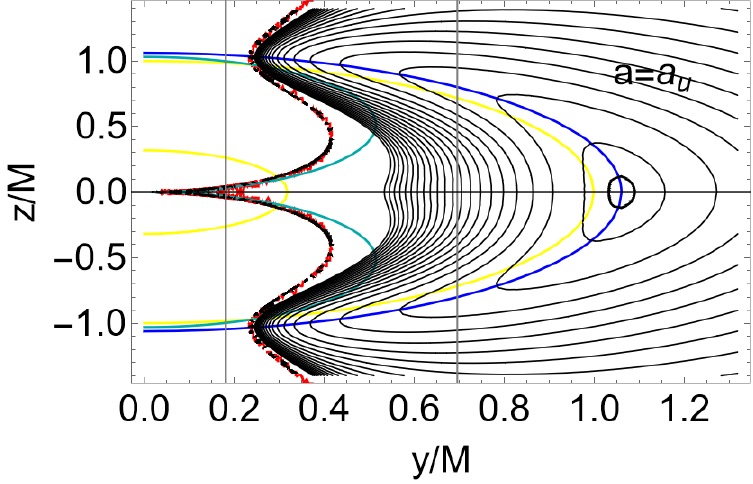}
   \caption{ Analysis of the double co-rotating toroids with $\ell=\ell^-$, possible for $\ell>\ell_{mso}^->0$ in $a\in]M,a_0[$.
   For $\ell^->0$  inversion points are only for space-like particles with $(\Em<0,\La<0)$ in the conditions of Sec.\il(\ref{Sec:co-space-like}).    Here the  most general solutions $r_\Ta^\pm$ are shown not considering the further     constraints of Sec.\il(\ref{Sec:fromaccretionflows}).  Spins $\{a_0,a_1,a_2\}$ are defined in Table\il(\ref{Table:corotating-counter-rotatingspin})  and  $a_u=1.03M$.
 Notation $(mso)$ is for marginally stable orbit, $(mbo)$ is for marginally bounded orbit, $(\gamma)$  is for marginally circular orbit. Radius $r_{\epsilon}^+=2M$ is the outer ergosurface on the equatorial plane.  Left panel: function $K(r)\equiv V_{eff}(\ell(r))$ ($V_{eff}$ is the fluid effective potential) as function of the spin $a/M$, evaluated for the radii of the extended geodetical structure. Center panel: fluid specific angular momentum $\ell(r)$ in the spacetime $a_u: K(r_{mso}^-)=K(r^-_{[mbo]})$. Double co-rotating tori are possible for $\ell>\ell_{mso}^-$. Right panel: toroidal configurations in the geometry $a_u$ with specific angular momentum $\ell=1.025$.  The proto-jets with cusp very close to the central singularity. There is $r=\sqrt{z^2+y^2}$ and $\theta=\arcsin y/\sqrt{z^2+y^2}$. Cyan curve is radius $r_{cr}=r_\Ta^+=r_\Ta^-$, defined in Eq.\il(\ref{Eq:sigmatur}). Blue curve contains pressure maxima (the tori centers, dependent on $\ell$ only) and tori geometric maxima (dependent on $K$ and $\ell$), solutions of $\partial_r V_{eff}^2=0$ in the plane $(y,z)$,  in this case it is $r=a^2$.}\label{Fig:Plotptim1}
\end{figure}

 More precisely, there are
  \begin{description}
  \item[\textbf{(+,+)}] Counter--rotating double  toroids having
 $\ell^-=\ell^+<0$ orbiting
\textbf{NS} with  $a\in]a_0,a_2[$----Figs\il(\ref{Fig:PlotparolA0tun1}).
The inner torus of the double system, centered in the ergoregion where $\Em>0$ and $\La<0$,  has  $\ell^-\in \mathbf{L_1^-}$ and it  can be cusped. The topology of outer counter-rotating torus depends on $\ell^+=\ell^-$
 and the \textbf{NS} spin.
 \item[\textbf{($-,-$)}] Co-rotating double  toroids with
 $\ell^->0$, orbiting \textbf{NSs} with   $a\in]M, a_0[$, where $\Em<0$ and $\La<0$ and $\ell>\ell_{mso}^->0$. The inner torus of the couple can be cusped. As there is $\ell_{mso}^->\ell_{\gamma}^->0$, see Figs\il(\ref{Fig:Plotgusrleona41},\ref{Fig:Plotgusrleona42},\ref{Fig:Plotptim1}), the outer torus of the pair  has $\ell\in\mathbf{L_3^-}$, and it is quiescent.
There are only \emph{spacelike} inversion points for $\Em<0$ and $\La<0$.
  \end{description}

More in details:

\medskip

\textbf{(+,+): Double counter--rotating  tori system with
 $\ell^-=\ell^+<0$}

It is convenient to introduce the following  spins.
 \bea&&\label{Eq:ba-pan-ba}
  a_\iota^0\equiv 1.1078:\ell^+_{\gamma}=\ell^-_{mso}<0,\quad
 a_\iota^- \equiv 1.1157:\ell^+_{mbo}=\ell^-_{mso}<0,\quad a_\iota^+ \equiv 1.118M:\ell^+_{mso}=\ell^-_{mso}<0.
\eea
%
%
(Note, there is  $\ell_{mbo}^->0$ for any  spin $a>M$).
Double tori with  equal $\ell=\ell^+=\ell^-<0$, are in the geometries with  $a < a_\iota^+$, for a selected range of   specific angular momentum $\ell<0$--see Figs\il(\ref{Fig:Plotmoocalpit15}), Figs\il(\ref{Fig:PlotgluT}) and Figs\il(\ref{Fig:PlotbluecurveherefNS1p}).

Double cusped tori, can form   in   \textbf{NSs}  spacetimes  with  $a<a_\iota^+$ with  $\ell^+\in\mathbf{L_1^+}=]\ell_{mbo}^+,\ell_{mso}^+[$ for cusped tori, or with, $\ell^+\in\mathbf{L_2^+}= ]\ell_\gamma^+,\ell_{mbo}^+[$ for tori and proto-jets, and quiescent tori in $\ell^+\in\mathbf{L_3^+}:\ell^+<\ell_\gamma^+$.

The situation is as follows:

\begin{description}
\item[--]
For $a\in]a_0,a_2[$, there is $\ell^-\in]\ell_{mso}^-,0[$, and the double tori formation is constrained by  $\ell_{mso}^-$, defining the limiting spins $\{a_\iota^\pm,a_0\}$.
\item[--]
For $a\in ]M, a_0[$, there is $\ell^-<0:\ell^-\in]-\infty,0[$ and double counter-rotating  tori are \emph{always} possible at $\ell^-<\ell_{mso}^+$.
(The inner tori are always in $\mathbf{L_1^-}$ and  entirely  contained in the ergoregion).
\end{description}
 In other words, for $a<a_0$, there is a very limited  range for the inner  tori  center and cusp location, according to a variation of the momenta $\ell^+$ in $\mathbf{L_1^+}, \mathbf{L_2^+}$ or $\mathbf{L_3^+}$.
For  $a<a_0$ there is  $\ell_{mso}^->0$, and co-rotating (cusped and quiescent) tori can be  with $\ell>\ell_{mso}^-$. In these  \textbf{NS} geometries,   there still can be (cusped and quiescent)
tori with  $\ell^-<0$, having cusps in $r_\times\in]r_{0}^-, r_{\delta}^-[$ and center in $r_{center}\in ] r_{\delta}^+, r_0^+
[$.

More in details the situation is as follows:
\begin{description}
\item[--]
For  $a\in]a_{\iota}^-,a_{\iota}^+[$, there is $\ell_{mso}^-\in]\ell_{mbo}^+,\ell_{mso}^+[=\mathbf{L_1^+}$ (cusped tori).
\item[--]
For $a\in ]a_\iota^0,a_\iota^-[$, there is $\ell_{mso}^-\in]\ell_{\gamma}^+,\ell_{mbo}^+[=\mathbf{L_2^+}$ (quiescent tori or proto-jets).
\item[--]
For $a\in ]a_0,a_\iota^0[$ there is $\ell_{mso}^-<\ell_\gamma^+$ and, in the range $\ell^+<\ell_\gamma^+$ ($\mathbf{L_3^+}$) (quiescent tori).
\end{description}
Note, for
$a\in [a_\iota^-,a_\iota^+]$ (there is  $a_\iota^->a_0$) there is $\ell_{mso}^-\in [\ell_{mbo}^+,\ell_{mso}^+]$, and therefore in these geometries   a double cusped tori system with   $\ell\in]\ell_{mbo}^+,\ell_{mso}^-[$  is possible.
For  $\ell^-=\ell^+<\ell_{\gamma}^+$, the outer disk is quiescent. For    $\ell^-=\ell^+\in]\ell_{\gamma}^+,\ell_{mbo}^+[$, the outer disk is quiescent, or there  is a  proto-jet. The inner torus  of the couple is quiescent or cusped.

In  the examples of
 Figs\il(\ref{Fig:Plotmoocalpit1})
  we can see also  the location of the inversion radius with respect to the ergosurfaces (further discussion  on this aspect is in Sec.\il(\ref{Sec:ergo-inversion-in-out})).

\medskip

\textbf{($-,-$) Double  co-rotating   tori with $\ell^->0$}

There are \emph{no} (time-like and photon-like) inversion points for $\ell=\ell^->0$, however for completeness we also  discuss  co-rotating  tori.

Double co-rotating tori  can orbits \textbf{NSs} with spins   $a\in]M, a_0[$, ($\Em<0,\La<0$). The inner torus of the pair,  having  $\ell^-\in]\ell_{mso}^-,+\infty[$, is always in $\mathbf{L_1^-}$. However,  the outer torus  is quiescent in $\mathbf{L_3^-}$, since $\ell_{mso}^->\ell_{\gamma}^-$.

 There is  $0<\ell_{mbo}^-<\ell_\gamma^-=a<\ell_{mso}^-$. Quiescent tori with  $\ell^-\in \mathbf{L_3^-}$ have center in  $r_{center}>r_{[\gamma]}^-$ (inner edge  is in $r>r_0^+$).  As there are no solutions of  $\ell(r)=\ell_\gamma^-=a$  in  $]r_{\delta}^-, r_{\delta}^+[$, therefore there are no double tori.

As noted in \cite{submitted}, there is
$K(r_{[\gamma]}^-)=\sqrt{1-{1}/{r_{[\gamma]}^-}}<1$, where $(r_\gamma^-=0, r_{[\gamma]}^-=a^2)$
regulate  the   co-rotating proto-jets formation   and define  the range  $\mathbf{L_3^+}\equiv \ell^->\ell_\gamma^-=a$,  where only quiescent tori are formed, where
$r_{[\gamma]}^-$  is related to the light surfaces  with frequency $\omega=1/\ell=1/a$.   These limiting configurations have also  a role in the case of tori with $\ell=0$, providing a case of toroidal  GRHD  (open)  cusped configurations with "axial cusp"-- \cite{submitted}. The limiting case of   $\ell=a$  is considered in  Sec.\il(\ref{Sec:l=pma}).

Since
$r_{\gamma}^-=0$, proto-jet cusps can be very close to the central  singularity--see Figs\il(\ref{Fig:Plotptim1}) and Sec.(\ref{Sec:inversion-proto-jet}).

More precisely: co-rotating proto-jets have  cusps in $r_j\in ]r_{\gamma}^-,r_{mbo}^-[$, center in $]r_{[mbo]}^-, r_{[\gamma]}^-[$ and momentum $\ell\in \mathbf{L_2^-}\equiv]\ell_{mbo}^-,\ell_{\gamma}^-[$.
The case of proto-jets emission is considered in details in  and Sec.(\ref{Sec:inversion-proto-jet}).

\medskip

We summarize the situation as follows:

\begin{description}
\item[--]\textbf{Tori couples orbiting in \textbf{NSs} with spin   $a \in]M, a_0[ $.}

There is
\bea\label{Eq:bar-rmso-m}
r_{[mso]}^->r_{[\gamma]}^->r_{[mbo]}^->r_{mso}^->r_{mbo}^->r_{\gamma}^-=0.
\eea
Radius  $r_{[mso]}^-\equiv \tilde{r}_{[mso]}^-:  \ell_{mso}^-\equiv\ell^-(r_{mso}^-)=\ell(r)$, see Figs\il(\ref{Fig:PlotparolA2}), regulates the  outer co-rotating torus center  location and constitutes a  main difference with the \textbf{BH} geometry.

The outer torus is  in $\mathbf{L_3^-}$, while the inner  torus is in  $\mathbf{L_1^-}\equiv \ell^->\ell_{mso}^->\ell_{\gamma}^->\ell_{mbo}^-$. The outer torus cannot be cusped, as
the center is located in $r>r_{[mso]}^->r_{[\gamma]}^->r_{[mbo]}^-$, and $\ell_{mso}^->\ell_{mbo}^-$-- Eqs\il(\ref{Eq:bar-rmso-m}).

The inner co-rotating tori, possible in the geometries with  $a\in]M, a_0[$, where $\Em<0$ and $\La<0$ and $\ell>\ell_{mso}^->0$,   have  for very large $\ell^-$,  inner edge  stretching  on $r_\delta^-$  (the center approaches $r_{\delta}^+$). The outer edge is studied in Sec.\il(\ref{Sec:outer-edge}).
 \begin{figure}
\centering
       \includegraphics[width=5.6cm]{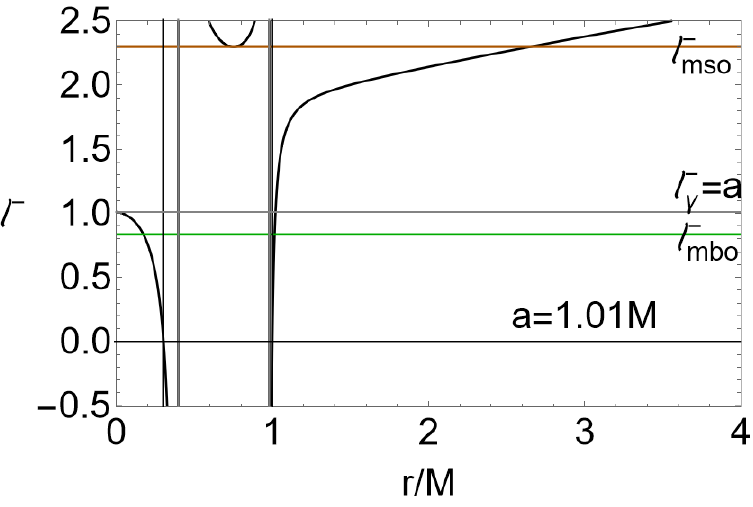}
     \includegraphics[width=5.6cm]{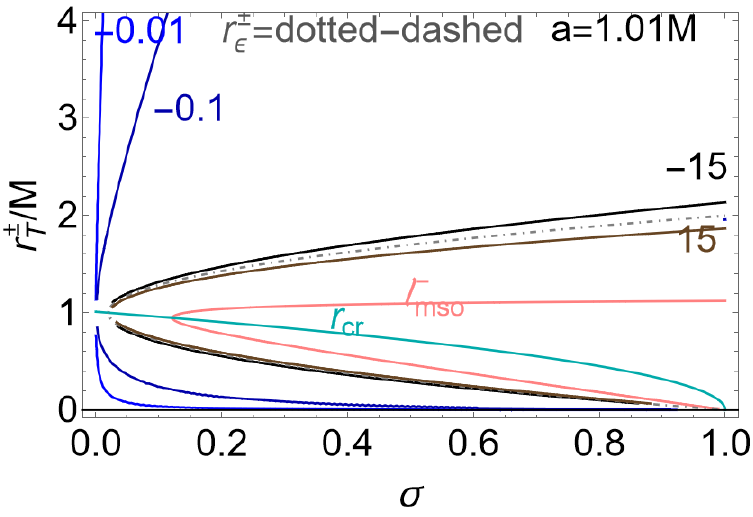}
           \includegraphics[width=5.6cm]{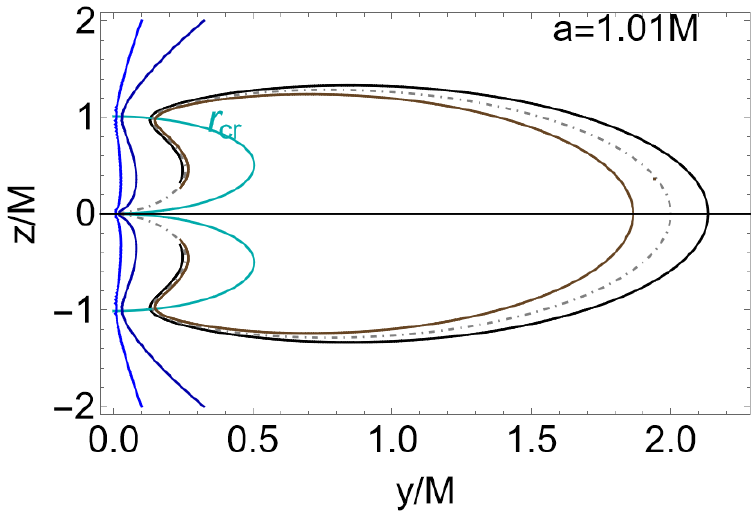}
      \caption{Double co-rotating tori system with fluid specific angular momentum $\ell=\ell^-$. Gray vertical lines are $r_\delta^\pm$, black vertical lines $r_0^\pm$. The specific angular momentum $\ell=\ell^-(r)$, the associated  effective potentials $V_{eff}$  for \textbf{NS} spin-mass ratio $a=1.01M$ and $\ell^-$ according to the colour notations of the $\ell^--r/M$ panels. The inversion points $r_\Ta^\pm$ are  also plotted as functions of the plane $\sigma=\sin^2\theta\in [0,1]$ (where $\sigma=1$ is the equatorial plane) for different $\ell$ signed on the curves.  Here the most general solutions $r_\Ta^\pm$ are shown not considering    constraints of Sec.\il(\ref{Sec:fromaccretionflows}).  For $\ell^->0$  inversion points are only for space-like particles with $(\Em<0,\La<0)$ in the conditions of Sec.\il(\ref{Sec:co-space-like}). Radius $r_\epsilon^\pm$ is the outer and inner ergosurfaces, and $r_{cr}$ is defined in Eq.\il(\ref{Eq:sigmatur}).   Notation $(mbo)$ is for marginally bounded orbit, $(mso)$ for marginally stable orbit. Momentum $\ell_\gamma^-=a$ refers on the limiting value of the specific angular momentum $\ell=\ell^->0$ for the occurrence of proto-jets (similarly to the $\ell^+$ case). There is $r=\sqrt{z^2+y^2}$ and $\sigma=y^2/(z^2+y^2)$.  The inversion  sphere  always closes on the equatorial plane, for small in magnitude $\ell$ closes at large $r$.}\label{Fig:Plotgusrleona41}
\end{figure}
 \begin{figure}
\centering
                      \includegraphics[width=5.6cm]{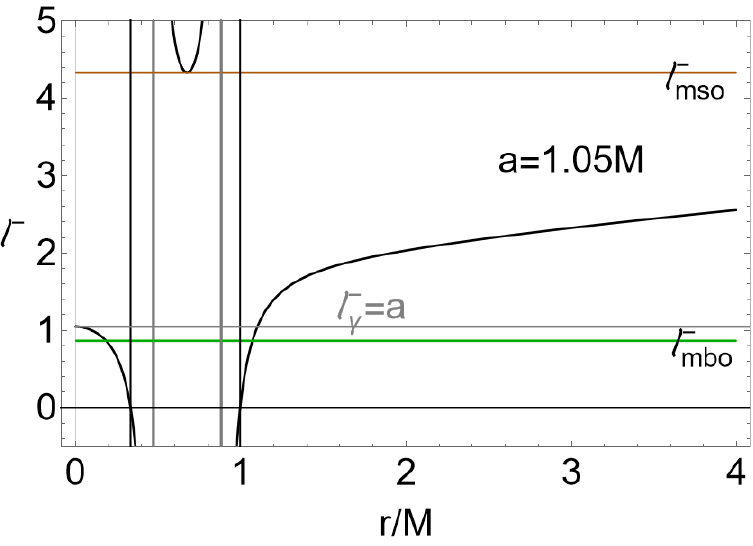}
                            \includegraphics[width=5.6cm]{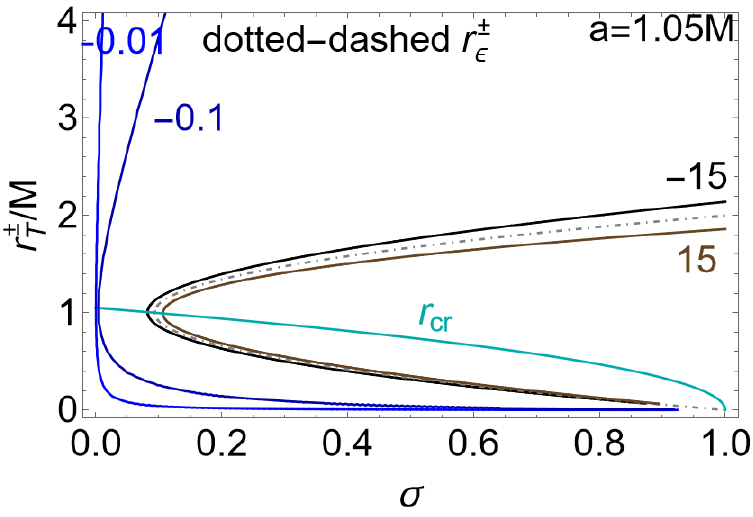}
                                \includegraphics[width=5.6cm]{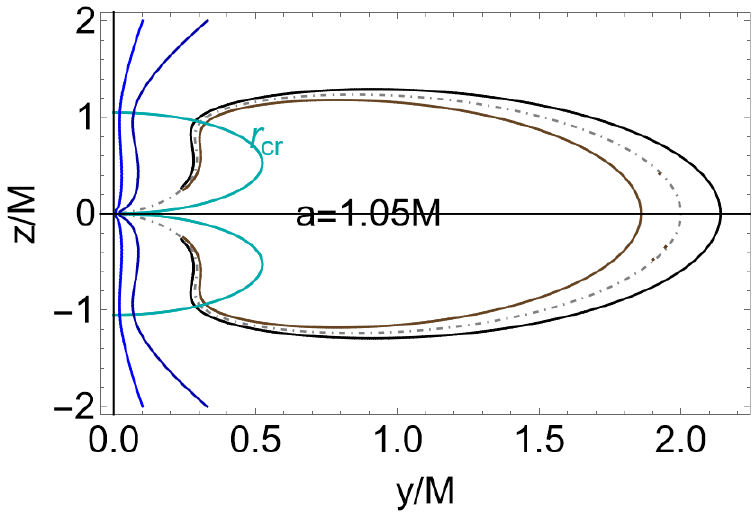}
      \caption{Double co-rotating tori system with fluid specific angular momentum $\ell=\ell^-$ for $a=1.05M$. See Figs\il(\ref{Fig:Plotgusrleona41}).
      }\label{Fig:Plotgusrleona42}
\end{figure}
 \begin{figure}
\centering
         \includegraphics[width=5.7cm]{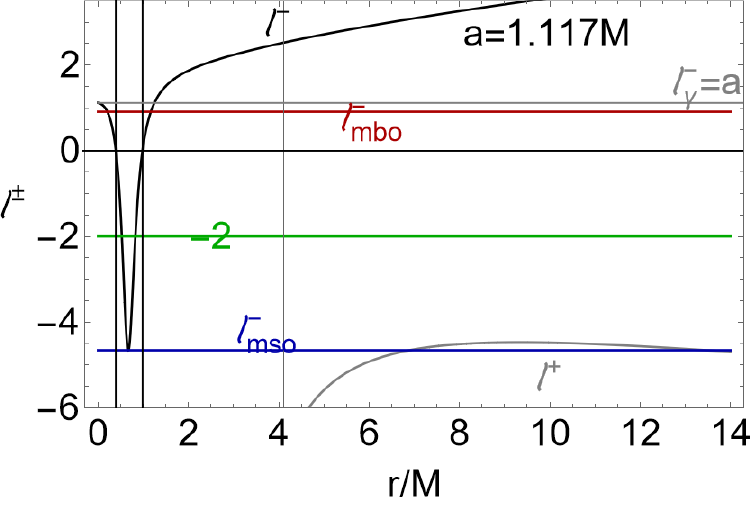}
                         \includegraphics[width=5.7cm]{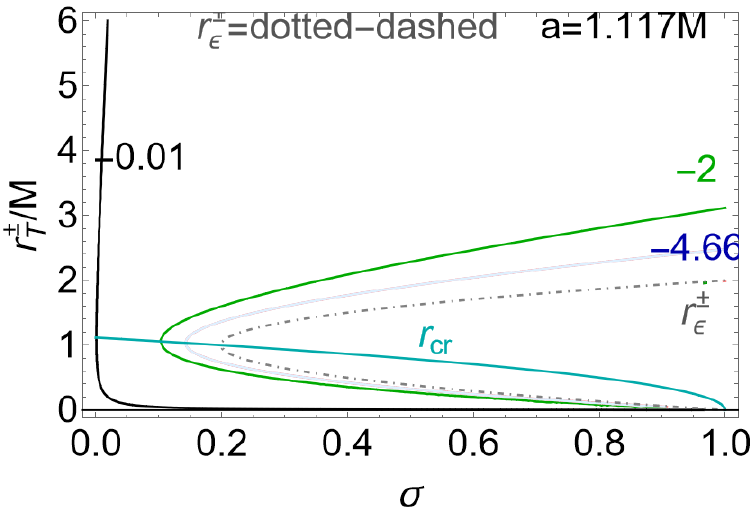}
                           \includegraphics[width=5.7cm]{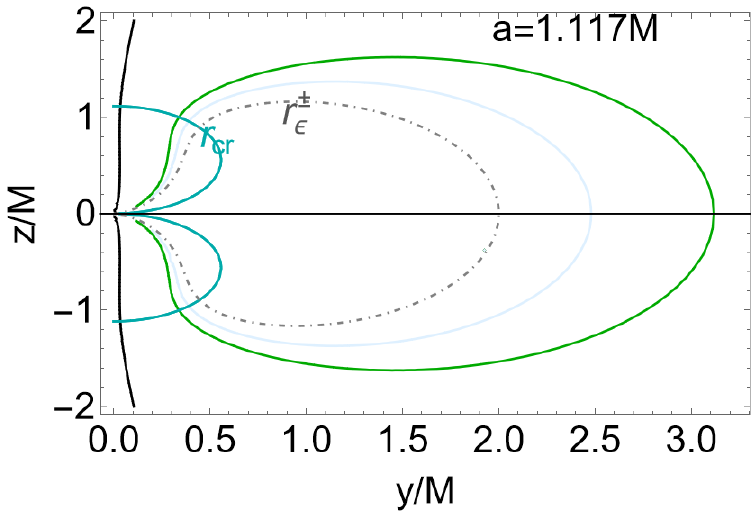}
      \caption{Double co-rotating tori system with fluid specific angular momentum $\ell=\ell^-$ for $a=1.117M$. See Figs\il(\ref{Fig:Plotgusrleona41}).
      }\label{Fig:Plotgusrleona43}
\end{figure}
From Figs\il(\ref{Fig:Plotgusrleona41})  and (\ref{Fig:Plotgusrleona42})  it is clear that for $a\in]M,a_0[$  there are   double co-rotating tori  only for     $\ell>\ell_{mso}^->\ell_{\gamma}^+=a>\ell_{mbo}^-$. The inner torus is cusped and is bounded by the light--surface with  $\omega=1/\ell$. The outer torus of the couple is in  $\mathbf{L_3^-}$  and therefore it is always quiescent and with center in $r>r_{[mso]}^-$.

For  $\ell\in ]\ell_{mbo}^-,\ell_{\gamma}^-[=\mathbf{L_3^-}$ there are proto-jets or quiescent tori with cusp in  $[0,r_{mbo}^-]$ and  center  in $[r_{[mbo]}^-,r_{[\gamma]}^-]$.
There are no double systems.
\\
\item[--]\textbf{Tori in the \textbf{NSs} spacetimes with spin  $]M,a_2]$.}

 In these spacetimes, the   range   $\ell\in [0,\ell_{mbo}^-]$  is  a   $\mathbf{L_1^-}$  range, where  cusped and quiescent co-rotating  tori are possible,  with cusp in $[r_{mbo}^-, r_0^-]$ and center in  $[r_0^+,r_{[mbo]}^-]$,  including  tori at  $\ell=0$ with cusp in $r_0^-$ and center in $r_0^+$. The  quiescent disk could have an inner edge in  the regions bounded by $(r_0^\pm, r_\delta^\pm)$, where test circular particle orbits have $\Em\gtrless0$ and $\La<0$.
\end{description}

\medskip

\textbf{Inversion points for double systems with $\ell^\pm<0$}

 Tori in a double   system with $\ell=\ell^\pm<0$ have same inversion spheres  (determined by $\ell$), while generally
proto-jets and accretion driven coronas are geometrically separated (there is  $\ell_{mso}^->\ell_\gamma^-$ at $a\in ]M,a_0[$)-- Sec.\il(\ref{Sec:inversionpoint-proto-jets}).
On the other hand, only  (particles and photons) flows with $\Em>0$ and $\La<0$ have inversion points.

In the interpretation of the inversion sphere as accretion   or proto-jets driven,  the tori surfaces interaction with the inversion corona is relevant. In Sec.\il(\ref{Sec:outer-edge}) this aspect is addressed in details.

For $\ell^-<0$, double counter-rotating tori are possible. Inner tori have cusp in  $]r_0^-, r_\delta^-[$ and center  in  $]r_\delta^+, r_{0}^+[$, therefore torus includes the region $]r_\delta^-,r_\delta^+[$.  In this region cusped tori can exist  with $\ell\in ]\ell_{mso}^+,+\infty[$.
(Fluid effective potential is not well defined on the light surfaces $r_s$ with frequencies $\omega=1/\ell$ where the potential diverges. These surfaces constrain the  torus edges).

\medskip

\textbf{Very large magnitude of fluid specific angular momentum}

In this section we consider the case of $\ell^\pm$ very large in magnitude, focusing particularly on slowly  spinning \textbf{NSs} with
 with $a\in ]M,a_0[$.
 In these spacetimes, there are  co-rotating and counter-rotating cusped tori ($\ell\in \mathbf{L_1^-}$) with   $\ell\to\pm \infty$  (center and cusps close to the radii  $r_\delta^\pm$ where  $\ell=\pm \infty$), with centers bounded by $]r_{mso}^+, r_{\delta}^+[$ (co-rotating fluids) and $]r_{\delta}^+, r_0^+[$ (counter-rotating fluids) and cusps in $]r_\delta^-,r_{mso}^-[$ (co-rotating fluids) and
 $]r_{0}^-, r_{\delta}^-[$ (counter-rotating fluids).

Tori with $\ell=\ell^-<0$ and   $\La<0$  are  only in the geometries  $]a_0,a_2[$-- Figs\il(\ref{Fig:Plotgusrleona41}),(\ref{Fig:Plotgusrleona42}) ,(\ref{Fig:Plotgusrleona43}).
In the limits $\ell\to\pm \infty$, there is
$r_\Ta^\pm=r_\epsilon^\pm$ and $\sigma_\Ta=\sigma_{erg}$. The inversion sphere approaches  the ergosurfaces from above, in the counter-rotating case, (and from below for the  co-rotating case for space-like particles, implying that the counter-rotating sphere is larger than the co-rotating inversion sphere).
The counter-rotating inversion spheres  decrease with the angular momentum magnitude, (while it  increases with $\ell^->0$), bounded by $r_\epsilon^\pm$.
The outer torus of a pair with  $\ell^+\to-\infty$  (and with $\ell^-\to+\infty$) are quiescent tori  in $\mathbf{L^+_3}$ ( $\mathbf{L_3^-})$, centered very far from the central singularity.

\begin{figure}
\centering
   \includegraphics[width=6cm]{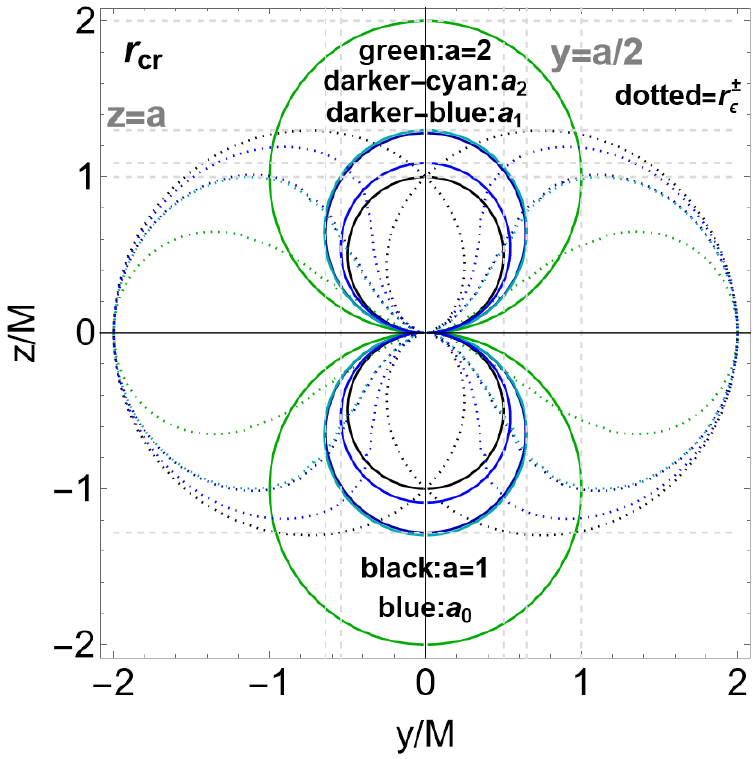}
     \caption{Radius $r_{cr}$ of Eq.\il(\ref{Eq:sigmatur}),  in the plane $(y,z)$ where $r=\sqrt{z^2+y^2}$ and $\sigma\equiv\sin^2\theta=y^2/(z^2+y^2)$. It is $y=\pm\sqrt{z (\mp a-z)}$  and $z=\pm\frac{1}{2} \left(\pm\sqrt{a^2-4 y^2}+a\right)$, for different spins signed on the panels. Dashed gray lines are $y=a/2$  and $z=a$. Radii $r_\epsilon^\pm$ are the outer and inner ergosurfaces. Notation $(mso, mbo,\gamma)$ refers the marginally stable, marginally bounded and last circular orbit respectively.}\label{Fig:PlotQCDHEP1b}
\end{figure}

The inversion spheres for  $\ell\to \pm \infty$ are similarly approaching from below and above the ergosurfaces respectively--Figs\il(\ref{Fig:Plotgusrleona41})  and (\ref{Fig:Plotgusrleona42}). (An interesting case is represented by  the solution   $u^\phi=0$ for very small  $\ell^-<0$  in magnitude. For the cusped tori with $\ell^-\in ]-\infty,0[$
(with cusp in $]r_0^-,r_{\delta}^-[$), the smaller in magnitude the fluid specific angular momentum is  the larger the sphere (see  Figs\il(\ref{Fig:PlotQCDHEP1b}) for the relative analysis of $r_{cr}$).)

Therefore:
in the spacetimes  with  $a\in]a_0,a_2[$
double cusped  counter-rotating configurations can exist for $\ell^-\in]\ell_{mso}^-,\ell_{mso}^+[$.

Inner cusped torus of the couple   has momentum  $]\ell_{mso}^-,\ell_{mbo}^-[$ ($\mathbf{L_1^-}$ range).
The  cusp  is located in $]r_0^-,r_{mso}^-[$ while the  center in $]r_{mso}^-,r_0^+[$  (there is $\ell^-<0$),   or  otherwise the cusp in
$]r_{mbo}^-,r_0^-[$ and center in $]r_0^+,r_{[mbo]}^-[$ (with  $\ell^->0$). We include   also tori at $\ell=0$, with critical points of pressure in $r_0^\pm$.

For $\ell\in ]\ell_{mbo}^-,\ell_{\gamma}^-[$, there are only proto-jets with cusps in $]0,r_{mbo}^-[$. For $\ell^->\ell_\gamma^-=a$, there are quiescent tori (this is an $\mathbf{L_3^-}$ range) with center in $r>r_{[\gamma]}^-=a^2$--see  Figs\il(\ref{Fig:Plotgusrleona43}).
Inversion points are all \emph{outside} the ergoregion.

For very small momenta magnitude $\ell\leq0$, there are counter-rotating tori with $\ell=\ell^-\neq\ell^+$  in $\mathbf{L_1^-}$. In this case the inversion coronas, for accretion driven flows, is  very  large and located out of  the ergoregion.

 However the inversion sphere  is always a closed surface (there is always a inversion point on the equatorial plane).
(Further constrains to the inversion points due to the conditions $(\Em, \La)$ constant on the flow particles are not considered.).

For $\ell=\ell^+$ the largest  inversion sphere is at $\ell_{mso}^+$ for  $a\lessapprox M$, discussed  in more details in Sec.\il(\ref{Sec:inversion-verti}).
The maximal extension on the equatorial plane is studied in \cite{submitted}.
\begin{figure}
\centering
\includegraphics[width=7.5cm]{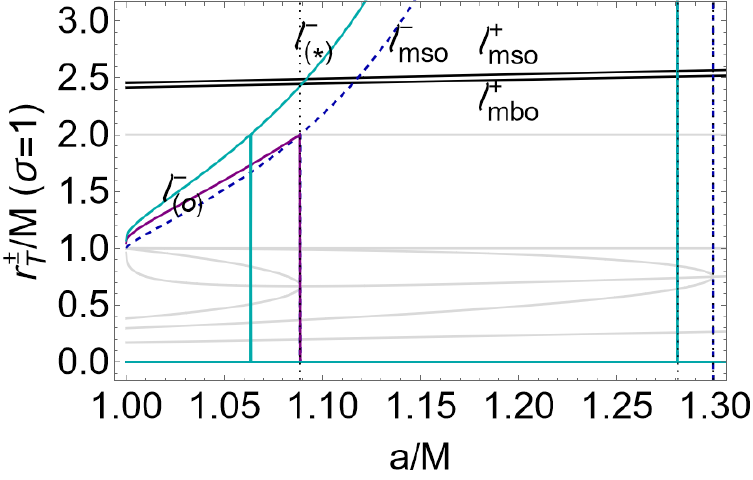}
       \includegraphics[width=7.5cm]{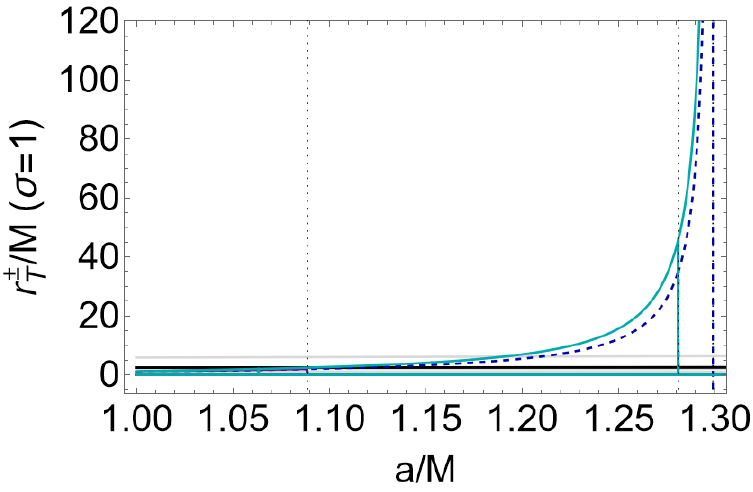}
     \includegraphics[width=7.5cm]{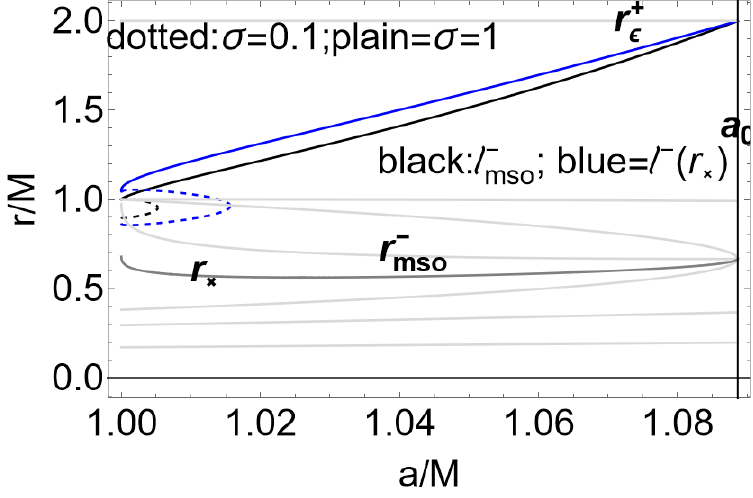}
    \includegraphics[width=7.5cm]{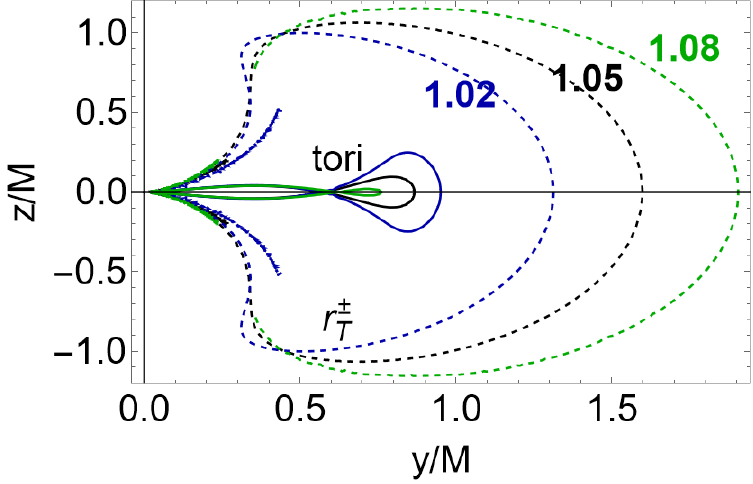}
  \caption{Upper line: Inversion radius $r_\Ta^{\pm}$ on the equatorial plane $\sigma=1$ as function of the \textbf{NS} spin-mass ratio $a/M$.  (The most general solutions $r_\Ta^\pm$, further    constraints are in Sec.\il(\ref{Sec:fromaccretionflows}).)  Gray curves are the Kerr \textbf{NS} geodesic structure. The inversion radius is evaluated on different flow  specific angular momenta $\ell$, signed on the curves. For $\ell^->0$  inversion points are only for space-like particles with $(\Em<0,\La<0)$ in the conditions of Sec.\il(\ref{Sec:co-space-like}).  Left panel is a close-up view of the left panel.  There is $\ell_{(o)}^-\equiv \ell^-(r_{(o)}^-)$ where
  $r_{(o)}^-\equiv r_ {mso}^--(r_ {mso}^--r_ {\delta}^-)/2$, while  $\ell_{(*)}^-\equiv \ell^-(r_{(*)}^-)$ where
  $r_{(*)}^-\equiv r_ {mso}^--(r_ {mso}^--r_ {0}^-)/2$. Radius $r_{mso}^-$ is the marginally stable orbit for fluids with specific angular momentum $\ell^-$, on $r_0^-$ there is $\La=\ell=0$ (test particle circular angular momentum and fluid momentum), on $r_\delta^-$ there is $\Em=0$ (energy of test particle circular orbits). $\ell_{mso}^+<0$ and $\ell_{mbo}^+<0$ are the specific angular momenta for fluid with $\ell=\ell^+<0$ on the marginally stable orbit and marginally bounded orbit respectively. Bottom left panel: light-gray curves are the geodesic structure for $a<a_0$, blue and black curves are the inversion  radius for particle flows with $\ell=\ell_{mso}^->0$ and $\ell^-(r_\times)>0$ signed on the panels. Radius $r_\epsilon^+=2M$ is the outer ergosurface on the equatorial plane. Bottom right panel: tori (plain curves) and inversion points (dashed curves) for selected values of the \textbf{NS} spin-mass ratios $a/M$ signed on the panels, according to the tori models of the left panel. There is $r=\sqrt{z^2+y^2}$ and $\sigma\equiv\sin^2\theta=y^2/(z^2+y^2)$.}\label{Fig:Plotgpass}
\end{figure}

In Figs\il(\ref{Fig:Plotgpass})   solutions  with  $\ell=\ell^->0$ are studied, where   $\Em<0$ and $\La<0$ at $a\in]M,a_0[$ (spacelike,  tachyonic particles inversion points), and     $\ell=\ell^-<0$ with  $\Em>0$, $\La<0$ at $a\in]a_0,a_2[$.
By increasing the spin, the inversion point sphere widens, extends outwards, and increases the vertical height. The associated  torus, on the other hand, shrinks, and approaches the central singularity, having an inner Roche lobe much larger than the outer Roche lobe.
The case of very low $\ell^-\in]\ell_{mso}^+,0[$ will be detailed in Sec.\il(\ref{Sec:inversion-verti}).
\subsubsection{Tori and inversion points at $\ell=\pm
a$}\label{Sec:l=pma}
Here we consider   inversion points with $\ell=\pm a$ and tori with $\ell=-a$.
There are \emph{no} double tori systems for $\ell=\pm a$. We can details the situation as follows

\begin{figure}
\centering
       \includegraphics[width=6.75cm]{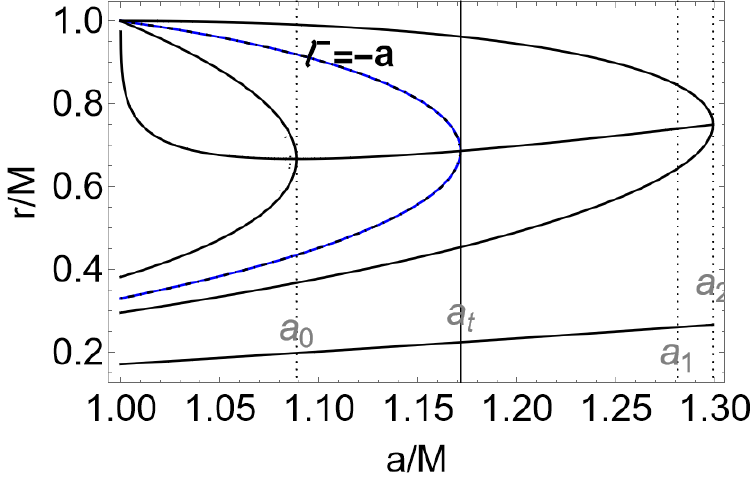}
          \includegraphics[width=6.75cm]{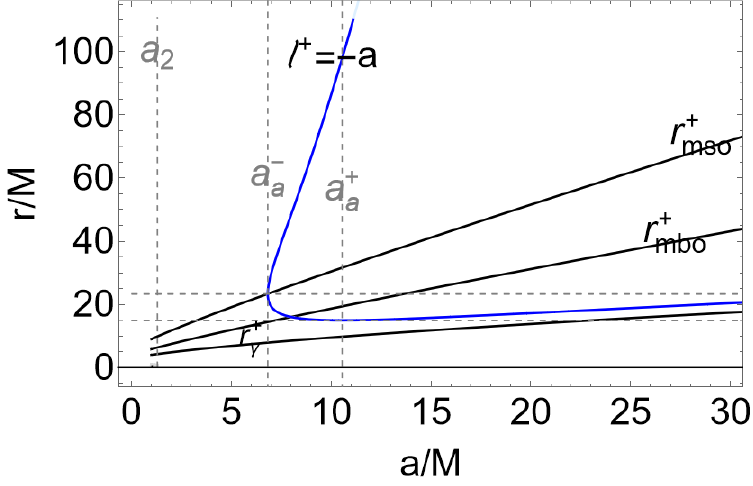}
    \includegraphics[width=6.75cm]{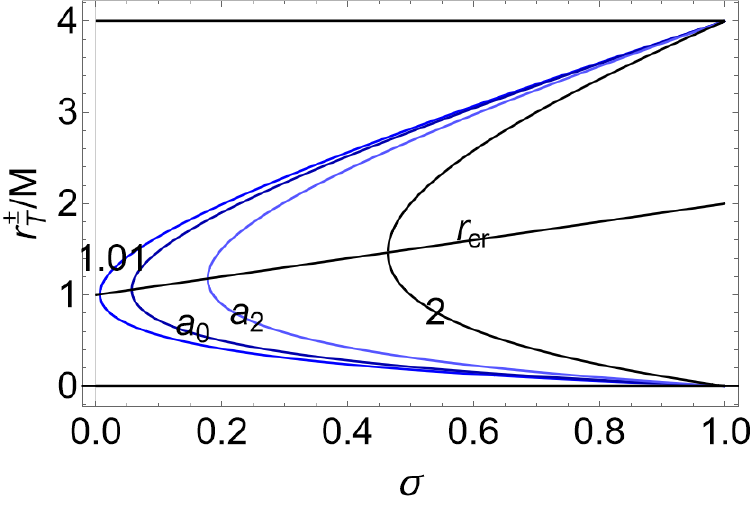}
          \includegraphics[width=6.75cm]{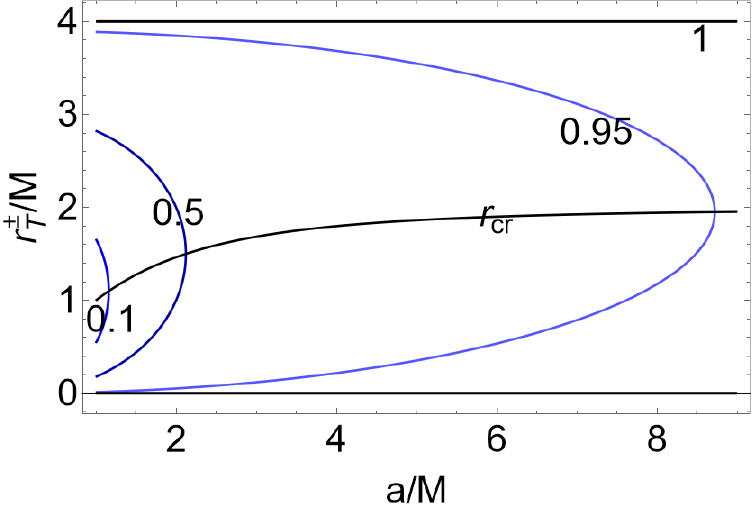}
  \caption{Analysis of the tori (upper line) and  flow inversion points (bottom line) with $\ell=-a$. There is $a_t\equiv2 \left(2-\sqrt{2}\right)M$, $a_a^+\equiv 2 \sqrt{2} \left(\sqrt{3}+2\right)M$ and $a_a^-\equiv 2 \left(\sqrt{2}+2\right)M$--see Eqs\il(\ref{Eq:rssisis}).
  There is $\sigma=\sin^2\theta\in [0,1]$ (equatorial plane is $\sigma=1$). Bottom line, inversion radius at $\ell=-a$ as function of $\sigma$ for different $a$ signed on the curves  (left) and as function of $a$ for different $\sigma$ (right).
 Upper line: critical points of pressure
 for tori with $\ell^-=-a$ (left) and $\ell^+=-a$ (right).
 Black curves are the geodesic structures. Spins $\{a_0,a_1,a_2\}$ are defined in Table\il(\ref{Table:corotating-counter-rotatingspin}).
}\label{Fig:Plotpurpodec}
\end{figure}

\begin{figure}
\centering
       \includegraphics[width=7.75cm]{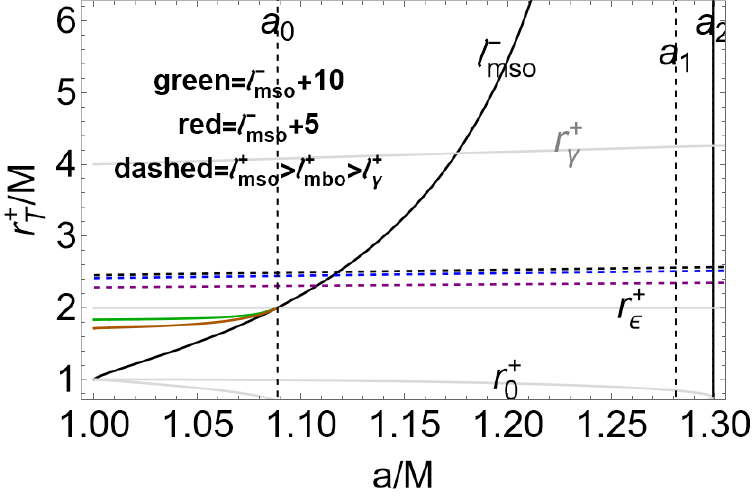}
   \caption{Analysis of the inversion radius $r_\Ta^+$ on the equatorial plane $\sigma=1$, for fluids with specific angular momentum $\ell=\ell^+<0$ and $\ell=\ell^-\gtreqless0$ signed on the panel, as functions of the \textbf{NS} spin-mass ratios $a/M$. Spins $\{a_0,a_1,a_2\}$ are defined in Table\il(\ref{Table:corotating-counter-rotatingspin}). Notation $(mso)$ is for marginally stable orbit, $(mbo)$ is for marginally bounded orbit, $(\gamma)$  is for marginally circular orbit.  The most general solutions  are shown not considering    constraints of Sec.\il(\ref{Sec:fromaccretionflows}).)
}\label{Fig:Plotptim}
\end{figure}

\begin{description}
\item[\textbf{Tori with  $\ell=a$.}]

 There are no double tori with $\ell=a$ as  $\ell_{mso}^->\ell_\gamma^-=a$ for $a\in ]M, a_0[$  -- Figs\il(\ref{Fig:Plotptim1}).

 More precisely for $\ell=\ell^-=\ell_\gamma^-=a$ there are (quiescent tori or)  proto-jets with cusp approaching  the central singularity (there is $r_\gamma^-=0$) and with  on center in $r_{[\gamma]}^-=a^2\in]r_{[mbo]}^-,r_{[mso]}^-[$.
In the \textbf{NS} with  $a<a_0$, tori with $\ell^->0$ have inner edge   in $]r_{\delta}^-,r_{mso}^-[$ and center  $]r_{mso}^-,r_\delta^+[$.

There are \emph{no} inversion points for $\ell=a$, since $\ell=\ell^-=a>0$ in $r=\{0,r_{[\gamma]}^-\}$  and, on $r_{[\gamma]}^-$  there is $\La>0$(spacelike inversion radius increases with  $\ell>0$, bounded from above by the  ergosurfaces--Figs\il(\ref{Fig:Plotgusrleona41})-- there is  $r_\Ta>r_{out}$--Figs\il(\ref{Fig:Plotptim}).
\\
\item[\textbf{Tori with $\ell=-a$}]
We distinguish the two cases $\ell=\ell^\pm=-a$ respectively.
\begin{description}
\item[--]
Tori with $\ell=\ell^+=-a$ can form   in the geometries with $a>a_a^-=2 \left(\sqrt{2}+2\right)M\approx 6.828M$ as cusped tori, quiescent tori or proto--jets according to the \textbf{NS} spins--Figs\il(\ref{Fig:Plotpurpodec}).
\\
\item[--]
Tori with $\ell=\ell^-=-a$ can form only in the spacetimes with $a\in]M,a_t[$, where $a_t\equiv2 \left(2-\sqrt{2}\right)M\approx1.17157M$,  there are quiescent or cusped tori,  with cups bounded by $r_0^-$ and $r_\delta^-$ or $r_{mso}^-$, and centers bounded by
$r_0^+$ and $r_\delta^+$ or $r_{mso}^+$, according to the \textbf{NS} spin--Figs\il(\ref{Fig:Plotpurpodec}).
\end{description}
\end{description}

\textbf{Inversion points with $\ell=-a$}

There are inversion points for $\ell=-a$      for
\bea&&\label{Eq:rssisis}
r_\Ta^\pm\in]0,4M]: \quad (\sigma=1, r_\Ta^+=4M), (\sigma_\Ta^S, r_\Ta=r_S^-),  (\sigma\in
]\sigma_\Ta^S, 1[,r_\Ta= r_S^\pm),\quad\mbox{where}
\\\nonumber
&&\sigma_\Ta^S\equiv \frac{1}{2} \left[a\sqrt{a^2+8}-(2+a^2)\right],
\quad
r_S^\pm\equiv\sigma +1\pm\sqrt{a^2 (\sigma -1)+(\sigma +1)^2},\quad r_{cr}=\frac{a}{2} \left[\sqrt{a^2+8}-a\right]
\eea
 increasing with the spin and the plane--Figs\il(\ref{Fig:Plotpurpodec}) (there is
$r_{cr}(\sigma)=1 + \sigma$).

 The analysis of the  inversion point location  with respect to the outer edge is done in Sec.\il(\ref{Sec:outer-edge})
\subsubsection{Inversion points and torus outer edge}\label{Sec:outer-edge}
Here we explore    the inversion spheres and the location of the flow inversion points on the equatorial plane  with respect  to the tori outer edge.

There can be, in a fixed spacetime  region,   more orbiting configurations with different  $\ell$--see for example\cite{ringed,pugtot,proto-jets,ella-correlation,colliding,long}. In this case, all proto-jets and cusped configurations have equal inversion coronas respectively,  depending only on the \textbf{NS} spin $a/M$.

The inversion corona is in general a narrow region surrounding the central singularity. As the conora thickness is generally small, inversion points location  in the corona    vary little   with   $\ell$ in the range of values defining the corona .
The inversion radius can be on the equatorial plane  at $r_\Ta<r_\times$ (or $r_J$), and this is always the case  for tori--driven and  proto-jets driven configurations with $\ell=\ell^+<0$ in Figs\il(\ref{Fig:Plotmoocalpit15})
and (\ref{Fig:Plotmoocalpit15mis},\ref{Fig:Plotmoocalpit1}), or otherwise the corona can   also "embed" the inner torus of a couple with  with $\ell=\ell^-$, for example in Figs\il(\ref{Fig:Plotmoocalpit15},\ref{Fig:Plotmoocalpit15mis},\ref{Fig:Plotmoutri},\ref{Fig:Plotgpass})).
Inversion corona  is interpreted as  accretion or proto-jets driven  if, on the equatorial plane,  $r_\Ta<r_\times$ or $r_\Ta<r_J$. On planes different from the equatorial, the situation depends on the particles initial conditions on the orbiting structure, on the jet collimation  angle,  on the torus morphological characteristics, as its height on the equatorial plane.

We can use the concept of  "excretion" inversion points if   $r_\Ta>r_{\times}$ on the equatorial plane which is the case for tori with  $\ell^-<0$ (at $a\in]M,a_2[$).
Fluids with $\ell=\ell^+$ have \emph{always} accretion or proto-jets driven inversion points on the equatorial plane i.e. $2M=r_\epsilon^+<r_\Ta<r_\gamma^+$.
(However, we stress that there is no "external cusp" or "extraction process" associated with the cusp,  as for example emerges in the models of thick discs analyzed in contexts in which  there is a toroidal magnetic field, in the presence of a cosmological constant or electric charge\cite{2014PhRvD..90d4029K,
1999GReGr..31...53S,1983BAICz..34..129S,1983BAICz..34..129S,2000CQGra..17.4541S,2005PhRvD..71b4037S}. Here the term excretion  distinguishes the case where an accretion driven or proto-jets driven interpretation  is possible from the case where  the inversion sphere or a portion of the inversion sphere embeds the torus or is external with  respect to the central singularity.).

\begin{figure}
\centering
       \includegraphics[width=5.75cm]{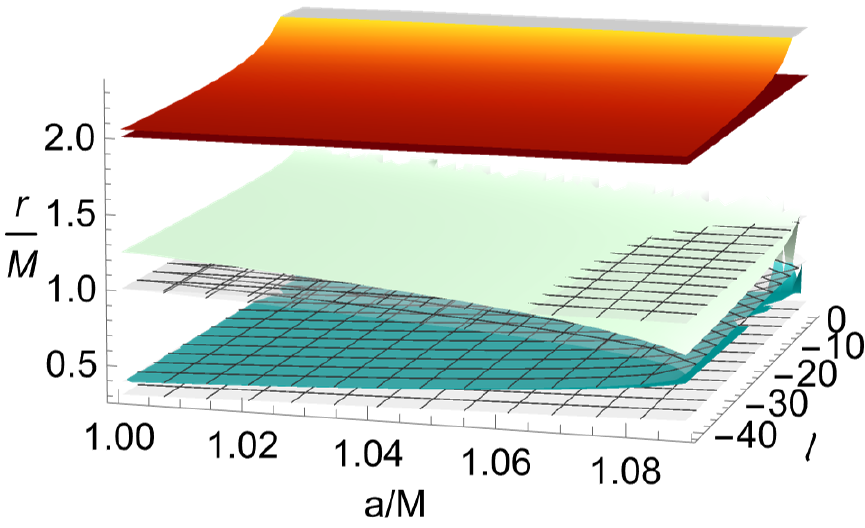}
           \includegraphics[width=5.75cm]{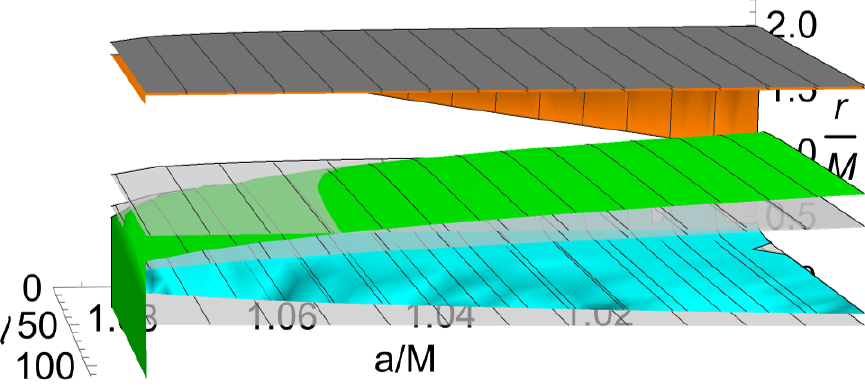}
          \includegraphics[width=5.75cm]{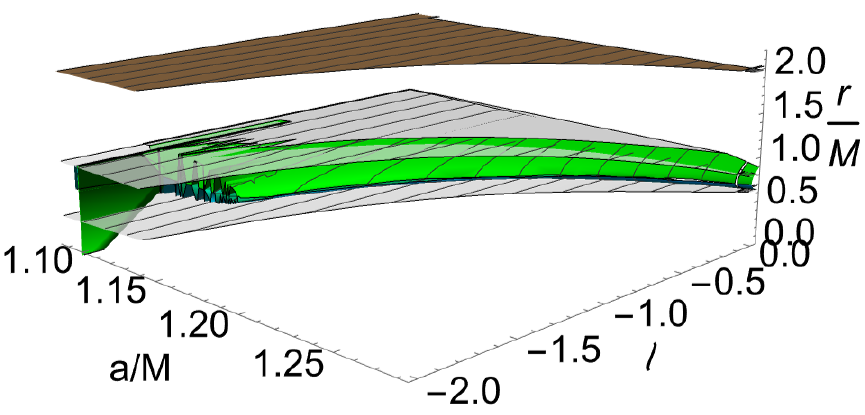}
  \caption{Analysis of the outer edge $r_{out}$ of the tori orbiting \textbf{NSs} with spins $a\in] M,a_2]$ with centers and cusps  bounded by $r_0^\pm$ and $r_{\delta}^\pm$.  Panels show the  cusped torus inner and outer edges and the inversion point $r_\Ta^+$ on the equatorial plane. There is  $r_{\times}<r_{out}<r_\Ta^+$.   There is no solution $r_{out}=r_{\epsilon}^+=2M$ on the equatorial plane. Radius $r_\epsilon^+$  is the outer ergosurface (gray-plane). Light-gray surfaces are $r_0^\pm$ and $r_\delta^\pm$.  Spins $\{a_0,a_1,a_2\}$ are defined inTable\il(\ref{Table:corotating-counter-rotatingspin}). Left panel: counter-rotating tori. Note the  inversion radius is always out of the ergoregion, the outer edge is inside the ergoregion crossing, for some $(a,\ell)$, the plane $r_0^+$.
  Center panel: co-rotating  tori. The inversion radius is always inside the ergoregion. Right panel: counter-rotating tori. The inversion radius is located out of the ergoregion.
}\label{Fig:Plotintrchge}
\end{figure}

A further aspect to be analysed is the location of the  inversion point, on the  equatorial plane, in relation with  the   torus  outer edge. The cusped tori outer edge (maximum outer torus  extension on the equatorial plane) is provided in \cite{submitted} with $K=V_{eff}(r_{\times},\ell)$ (where the cusp is $r_{\times p}$ for $a\in]a_0,a_1[$, and $r_\times$ for $a\in]a_1,a_2[$.  (In Figs\il(\ref{Fig:Plotintrchge}) we prove that, for $\ell^-\lessgtr0$, there  is $r_{outer}<r_\Ta$, the tori  outer edge is always inside the ergoregion.).
 In  \textbf{NSs} with $a\leq a_2$, the co-rotating and counter-rotating cusped tori outer edge  never crosses   the ergosurface  on the equatorial plane. (Tori are also confined by the light surface with $\omega=1/\ell$.) Counter-rotating tori  orbiting \textbf{NS} with $a<a_2$ can cross the radius $r_0^+$ for certain values of the specific angular momentum and  \textbf{NS} spin.
For co-rotating  tori in the \textbf{NSs} spacetimes with $a<a_0$, with centers and   cusps  bounded by $r_0^\pm$ and $r_{\delta}^\pm$, the outer edge can cross $r_0^+$ or $r_\delta^-$.
\subsubsection{Inversion points inside and outside the ergoregion}\label{Sec:ergo-inversion-in-out}
In this section we explore  inversion points location with respect to the geometry ergoregion. For $a\in]M,a_2]$, counter-rotating tori in the ergoregion are possible.  For counter-rotating fluids, inversion points are only out of the ergoregion --Eqs\il(\ref{Eq:mon-semb}).
(Contrary to the counter-rotating case,   inversion points with  $\ell >
   0$, possible only for spacelike  tachyonic--particles with $\La<0$ and $\Em<0$,  in the geometries with $a\in ]0,a_0]$,  exist \emph{only} in the ergoregion at $r\in ] r_\epsilon^-, r_\epsilon^+[$--see also
   Eqs\il(\ref{Eq:cap-sen-cam})).

\medskip

\textbf{Tori in the ergoregion}

We consider tori  with momentum $\ell=\ell^-\lesseqgtr0$ in the \textbf{NSs} ergoregion.
There are, for $a\in ]M,a_0[$, co-rotating cusped tori in the ergoregion.
There are also larger tori with cusp in the ergoregion
at $[r_0^-, r_{mbo}^-]$.
Proto-jets and co-rotating tori cusps are in the ergoregion, in   $]r_\gamma^-,r_{mbo}^-[$ and $[r_0^-, r_{mbo}^-]$
respectively.

\begin{figure}
\centering
 \includegraphics[width=5.765cm]{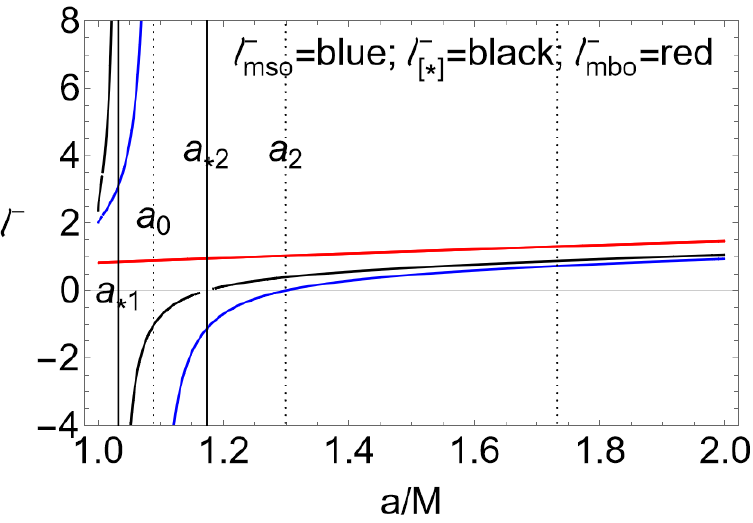}
 \includegraphics[width=5.765cm]{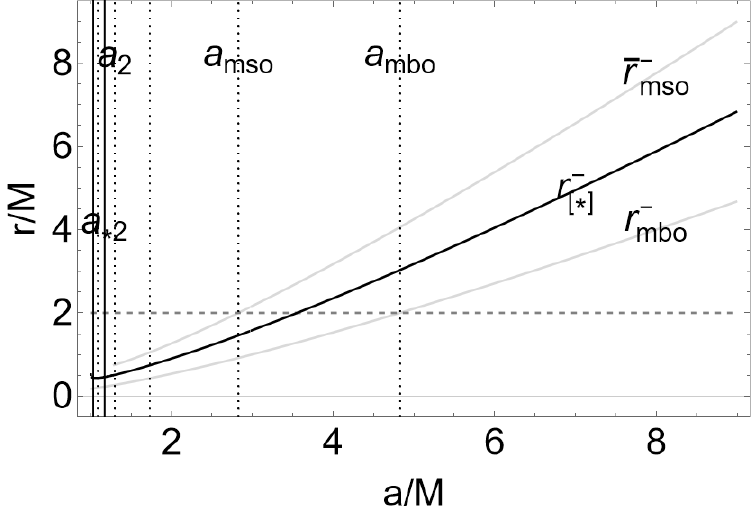}
\includegraphics[width=5.765cm]{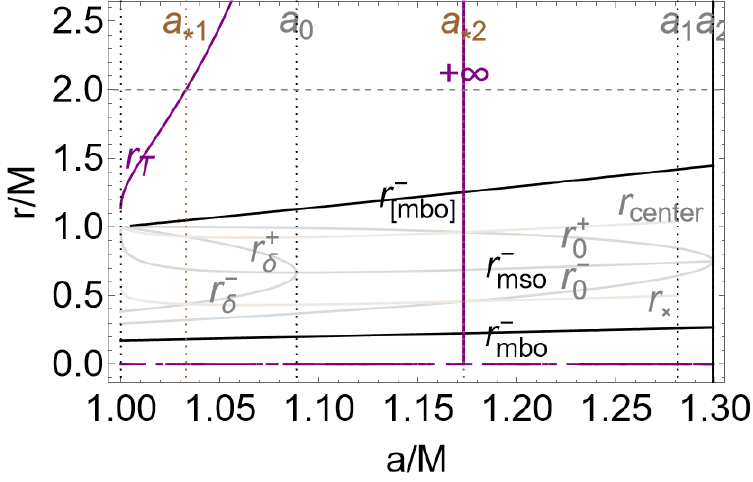}
\includegraphics[width=5.765cm]{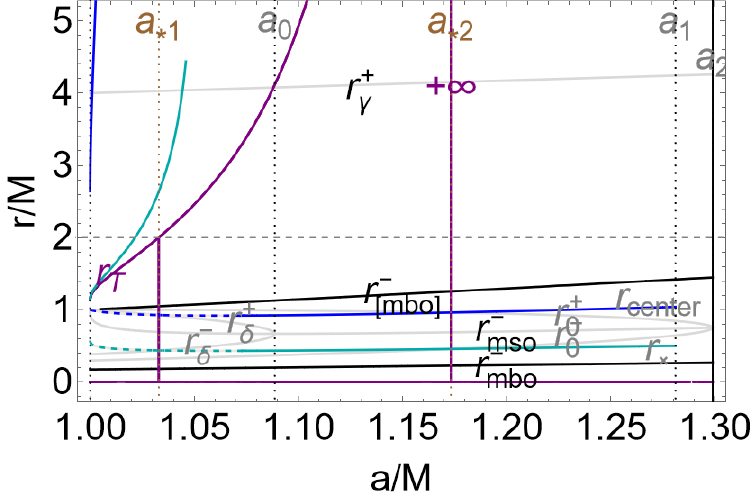}
\includegraphics[width=5.765cm]{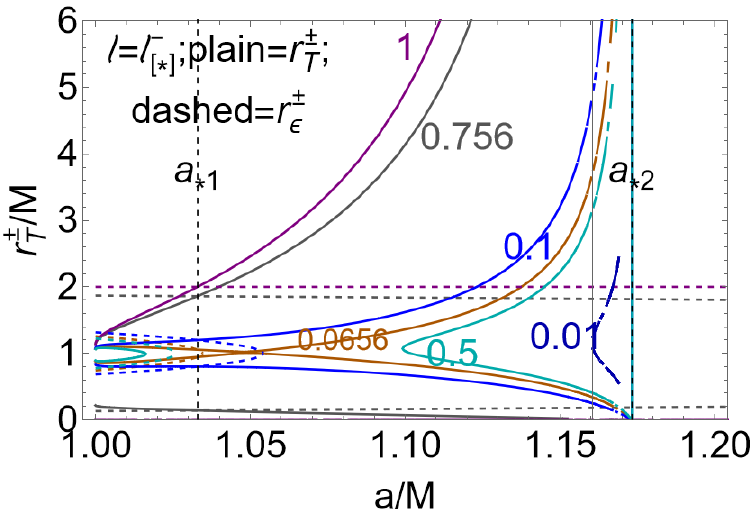}
\includegraphics[width=5.765cm]{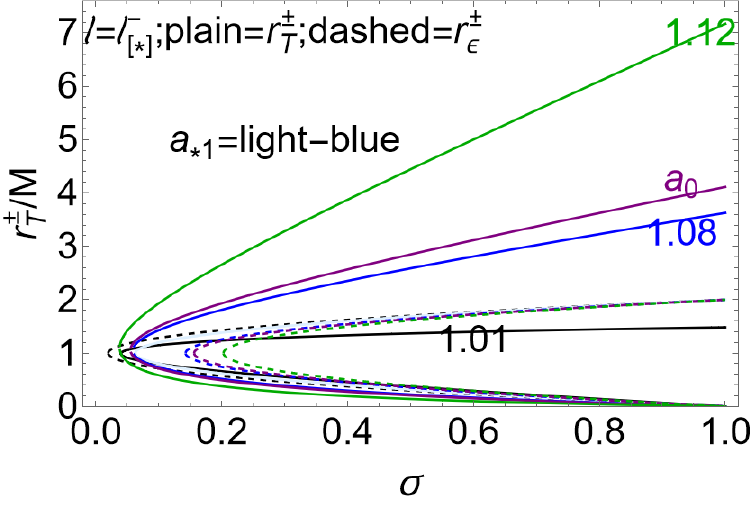}
\includegraphics[width=5.765cm]{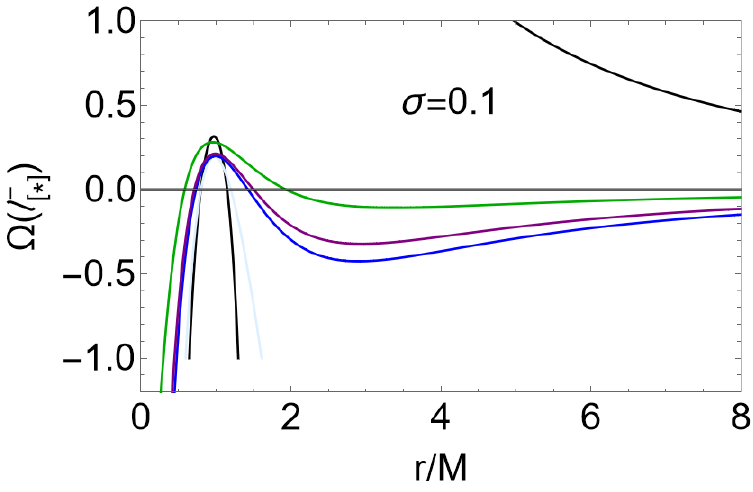}\includegraphics[width=5.765cm]{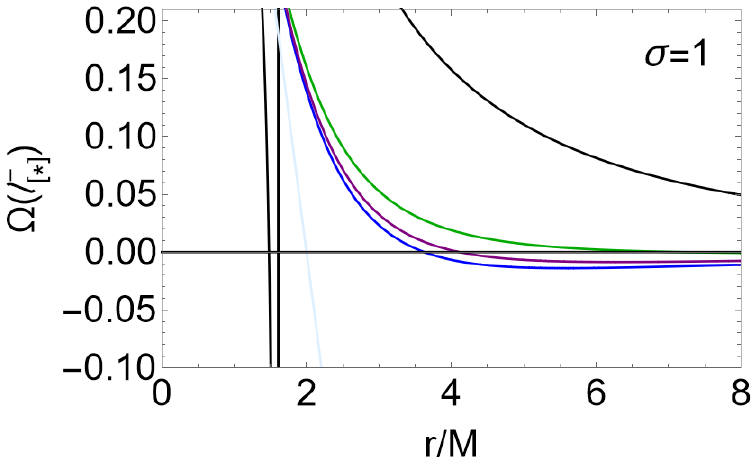}
\includegraphics[width=5.765cm]{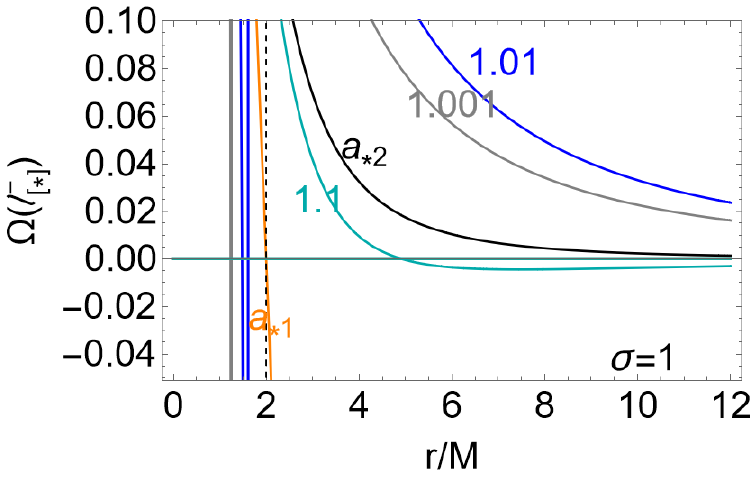}
 \caption{Inversion points for flows with fluid specific angular momentum $\ell^-$. For $\ell^->0$  inversion points are only for space-like particles with $(\Em<0,\La<0)$ in the conditions of Sec.\il(\ref{Sec:co-space-like}).   The most general solutions $r_\Ta^\pm$  are shown not considering    constraints of Sec.\il(\ref{Sec:fromaccretionflows}).  Spins $\{a_0,a_1,a_2\}$ are defined in Table\il(\ref{Table:corotating-counter-rotatingspin}). Spins $\{a_{*1},a_{*2}\} $ are defined in Eqs\il(\ref{Eq:spond-sasastar}), spins $\{a_{mso},a_{mbo}\}$ are defined in Eqs\il(\ref{Eq:amsoamno-avv}).  There is $\ell_ {[*]}^-\equiv \ell^-(r_ {[*]}^-), \quad r_ {[*]}^-\equiv r_{mso}^--(r_ {mso}^--r_ {mbo}^-)/2$. Upper left panel:  fluid angular momentum $\ell_{[*]}^-$ and $\ell_{mso}^-$, $\ell_{mbo}^-$ (on the marginally stable and marginally bounded orbit respectively) as functions of the \textbf{NS} spin-mass ratio. Upper Center panel: geodesic structure of the Kerr \textbf{NS}  and radius $r_{[*]}^-$ as function of $a/M$. Upper right panel  and center-line left panel show the inversion radius $r_\Ta$ (on the equatorial plane) for the torus model with center radius  $r_{center}$ and cusp $r_{\times}$, where $\ell=\ell_{[*]}^-$. Center-line right and center panels:: inversion point radius $r_\Ta^\pm$ for different planes $\sigma\equiv\sin^2\theta\in[0,1]$ ($\sigma=1$ is the equatorial plane), signed on the curves, as function of spin-mass ratio $a/M$ (center panel) and as function of the plane, for different \textbf{NS} spin-mass ratio signed on the curves (right panel) for fluid specific angular momentum $\ell^-_{[*]}$. Note the location of the inversion radius with respect to the  ergosurfaces $r_{\epsilon}^\pm$ (dashed curves). Bottom line panel: analysis of the flows inversion points in terms of the relativistic angular velocities $\Omega$ evaluated for particles with   fluid specific angular momentum $\ell=\ell_{[*]}^-$ for different plane sigma $\sigma$ and spins as signed on the planes (left and right panels) according to notation of the center-line right-panel. There is $\Omega(r_\Ta^{\pm})=0$.  Bottom right panel shows $\Omega(\ell_{[*]}^-)$ on the equatorial plane $(\sigma=1)$ for different spins $a/M$ signed on the curves. Note the presence of two inversion points for each curve at $\sigma<1$. The analysis for the counter-rotating tori with fluid specific angular momentum $\ell=\ell^+<0$ is in Figs\il(\ref{Fig:PlotparolA0tun1}). }
 \label{Fig:Plotconfsud}
\end{figure}

Let us introduce  the spins $\{a_{mso},a_{mbo}\}$ defined as:
\bea\label{Eq:amsoamno-avv}
a_{mso}\equiv 2.828M: \bar{r}_{mso}^-=2M\quad\mbox{and}\quad
a_{mbo}\equiv 4.828M: r_{mbo}^-=2M.
\eea
For $a\in ]a_2, a_{mso}[$,
 proto-jets and co-rotating  cusps are only in the ergoregion.
For  $a\in ]a_{mso},a_{mbo}[$, tori cusps are out of the ergoregion and  proto-jets cusps are inside the ergoregion.
 For  $a>a_{mbo}$, proto-jets and tori  cusps are out of the ergoregion--see Figs\il(\ref{Fig:Plotconfsud}) and Figs\il(\ref{Fig:PlotparolA2}).
From  Figs\il(\ref{Fig:Plotconfsud}) it is clear how, in the regions where  $\ell_{mso}^->0$ with $\La<0$,  there is $\ell_{mso}^->\ell_{mbo}^-$ at $a<a_0$.
 \begin{figure}
 \centering
       \includegraphics[width=5.75cm]{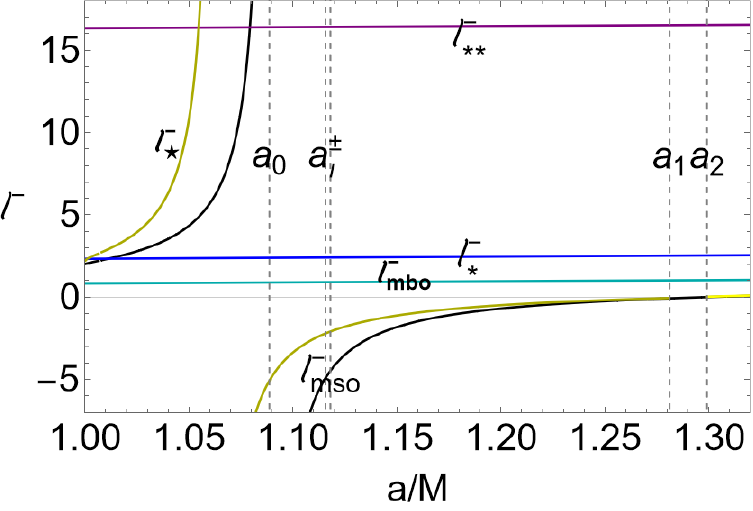}
     \includegraphics[width=5.75cm]{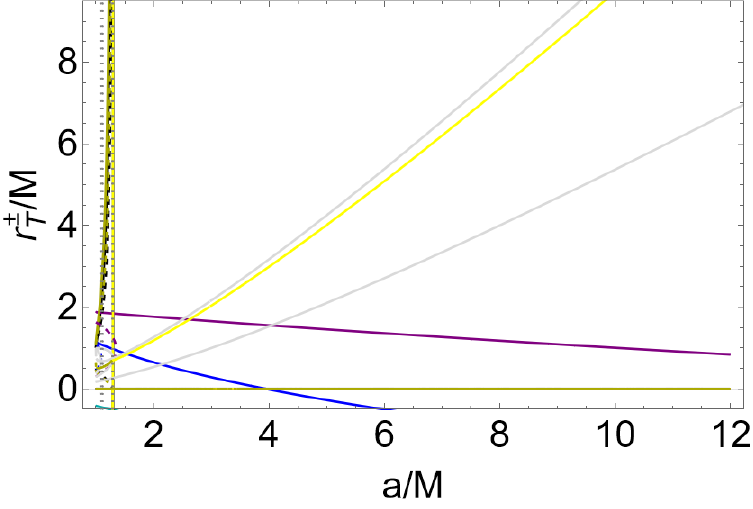}
      \includegraphics[width=5.75cm]{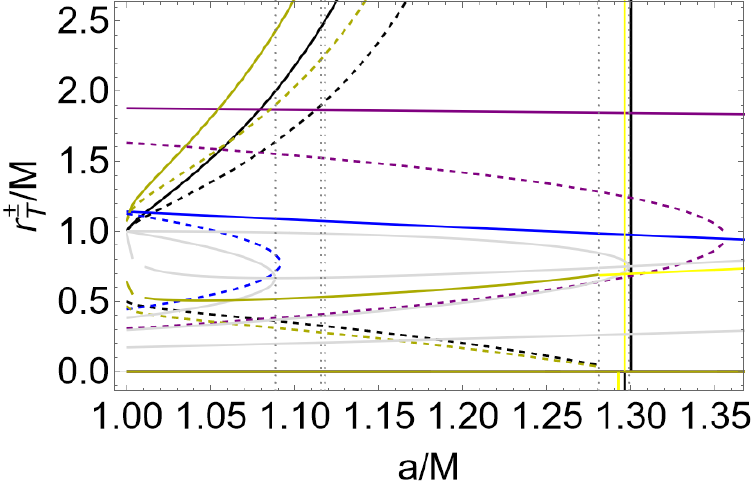}\\
       \includegraphics[width=5.75cm]{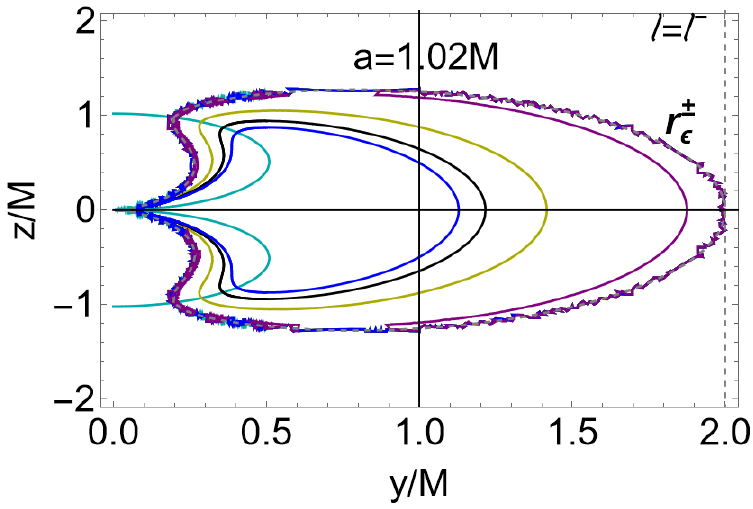}
        \includegraphics[width=5.75cm]{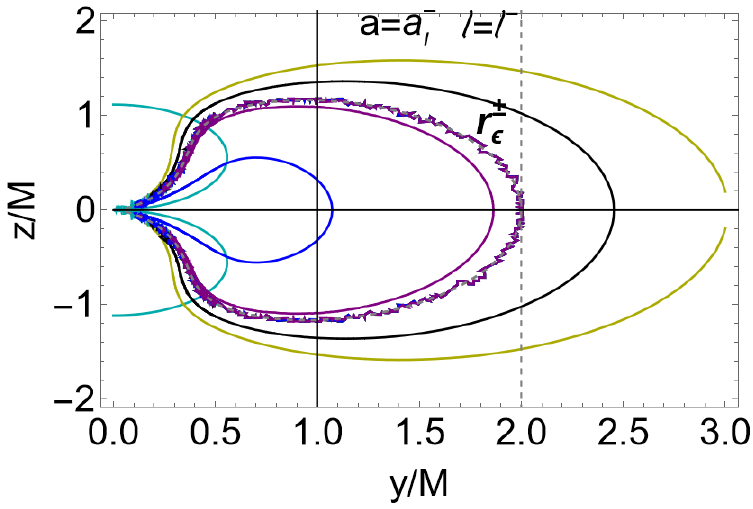}
           \includegraphics[width=5.75cm]{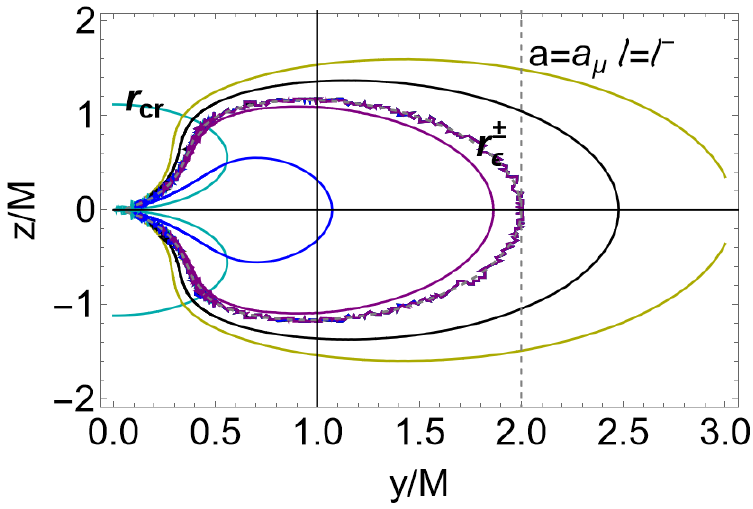}\\
     \includegraphics[width=4.25cm]{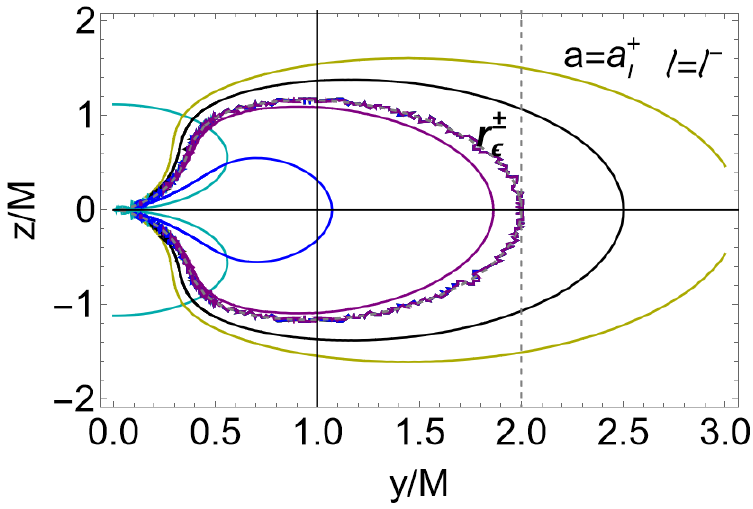}
          \includegraphics[width=4.25cm]{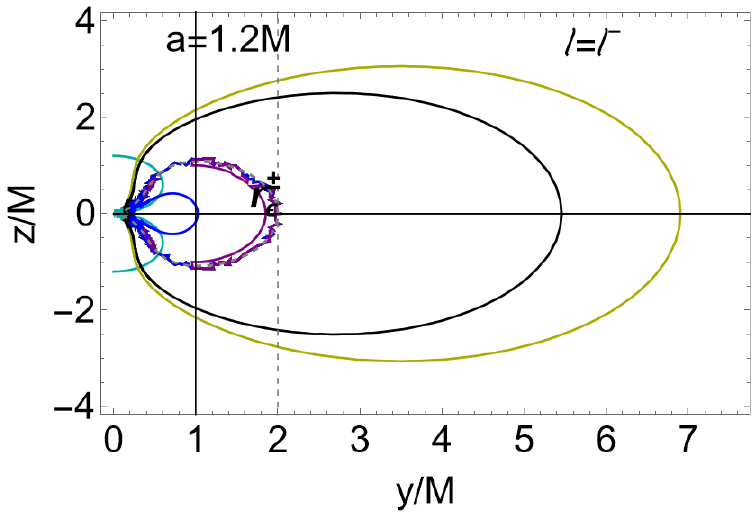}
      \includegraphics[width=4.25cm]{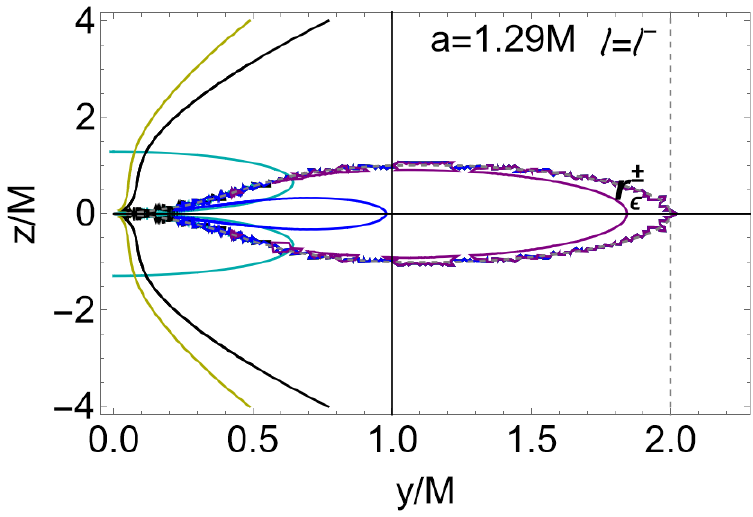}
    \includegraphics[width=4.25cm]{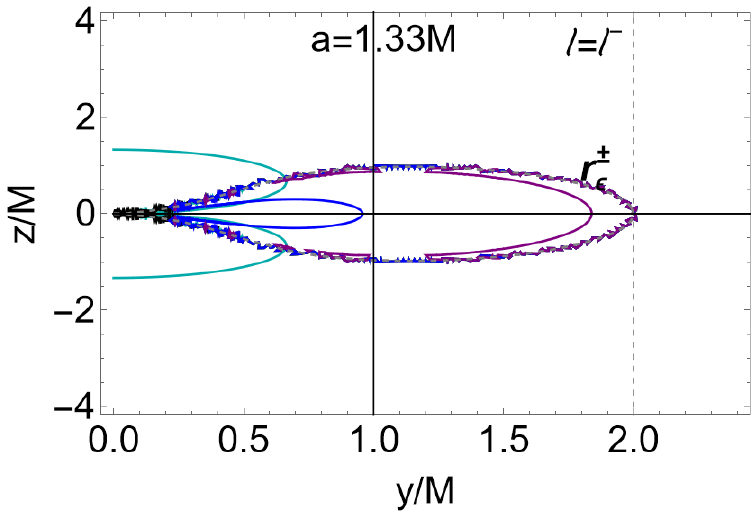}
  \caption{Analysis of  tori with fluids specific angular momentum $\ell=\ell^-$,    $\ell^-=\ell^+<0$ and flows inversion points. For $\ell^->0$  inversion points are only for space-like particles with $(\Em<0,\La<0)$ in the conditions of Sec.\il(\ref{Sec:co-space-like}).  Here  most general solutions $r_\Ta^\pm$ are shown, while further   constraints are in Sec.\il(\ref{Sec:fromaccretionflows}).  Spins $a_{\iota}^\pm$, are defined in Eqs\il(\ref{Eq:ba-pan-ba}). There is $a_{\mu}\equiv a_\iota^+ - (a_\iota^+ - a_\iota^-)/2$,  and  $(\ell_*^-,\ell_{**}^-)$ are $\ell_*^-\equiv \ell_{mbo}^-+1.5$, $
\ell_{**}^-\equiv \ell_{mbo}^-+15.5$, with  $\ell_\star^-\equiv \ell^-(r_\star^-)$,
where $r_\star^-\equiv\bar{r}_{mso}^--(\bar{r}_{mso}^--r_{mbo}^-)/9$, for  $a>a_1$,
and $r_\star^-\equiv\bar{r}_{mso}^--(\bar{r}_{mso}^--r_{0}^-)/2$ for  $a\in]M,a_1[$.  (\emph{mso} is for marginally stable orbit and \emph{mbo} is for marginally bounded orbit). See also  Figs\il(\ref{Fig:PlotgluT}), Figs\il(\ref{Fig:PlotbluecurveherefNS1p}).
Upper line left panel:  the fluid specific angular momentum $\ell^-$ as function of $a/M$.
 Radii $r_{\epsilon}^\pm$ are the outer and inner ergosurfaces.     Upper line right  panel is a close up view of the center panel:
 inversion radius $r_\Ta^{\pm}$ on  different planes $\sigma\equiv\sin^2\theta$, on the equatorial plane $\sigma=1$ (plain curves)  and  $\sigma=0.5$ (dashed curves)  and specific angular momenta  according to the colors notation of the left panel.  Remaining panels show  the inversion radius $r_\Ta^{\pm}$ on the plane $(y/M,z/M)$, where $\sigma=\sin^2\theta={y^2}/({z^2+y^2})$
 and $r=\sqrt{z^2+y^2}$, for different fluid specific angular momenta $\ell$, according to the notation signed on the upper line left panel. Tori models and accretion  flows inversion points are shown in Figs\il(\ref{Fig:Plotmoocalpit15mis}) and Figs\il(\ref{Fig:Plotmoocalpit15}). Note the inversion radius location  with respect to the ergosurface $r_\epsilon^\pm$.  Figs\il(\ref{Fig:Plotmoocalpit1}) show the analysis for fluids with momenta $\ell=\ell^+<0$. Radius $r_{cr}=r_\Ta^+=r_\Ta^-$ is defined in Eq.\il(\ref{Eq:sigmatur}). }\label{Fig:Plotmoutri}
\end{figure}
 \begin{figure}
 \centering
  \includegraphics[width=5.75cm]{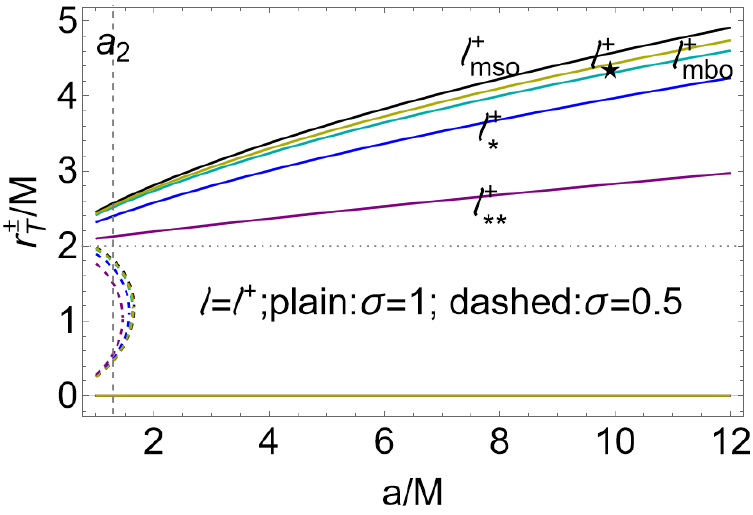}
    \includegraphics[width=5.75cm]{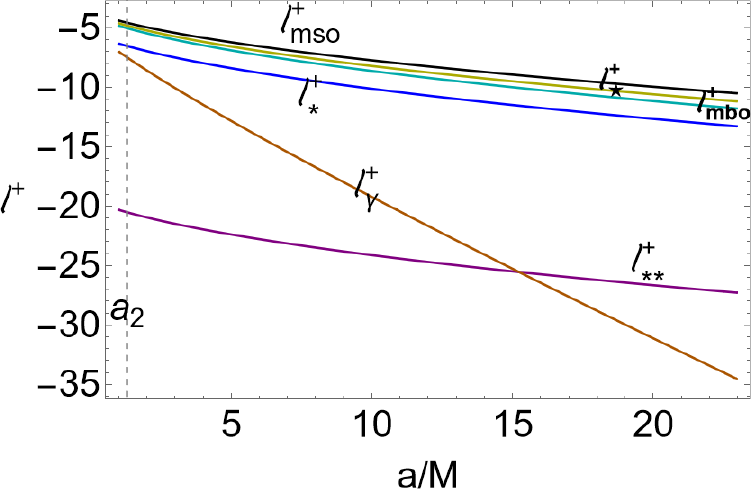}
 \includegraphics[width=5.75cm]{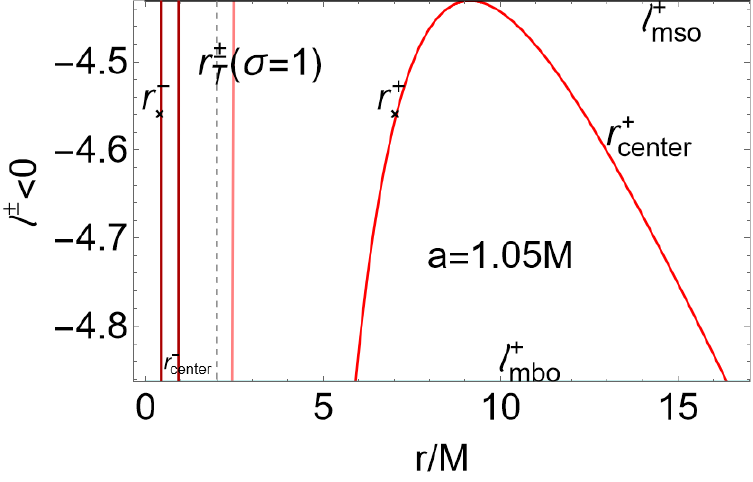}
 \\
   \includegraphics[width=5.75cm]{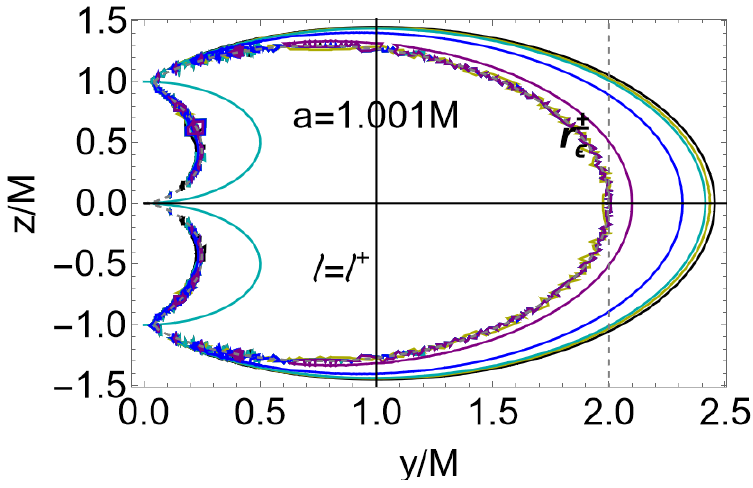}
        \includegraphics[width=5.75cm]{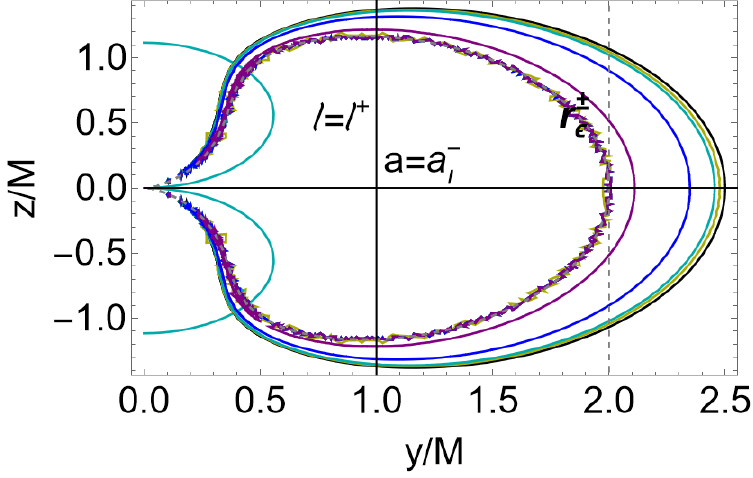}
          \includegraphics[width=5.75cm]{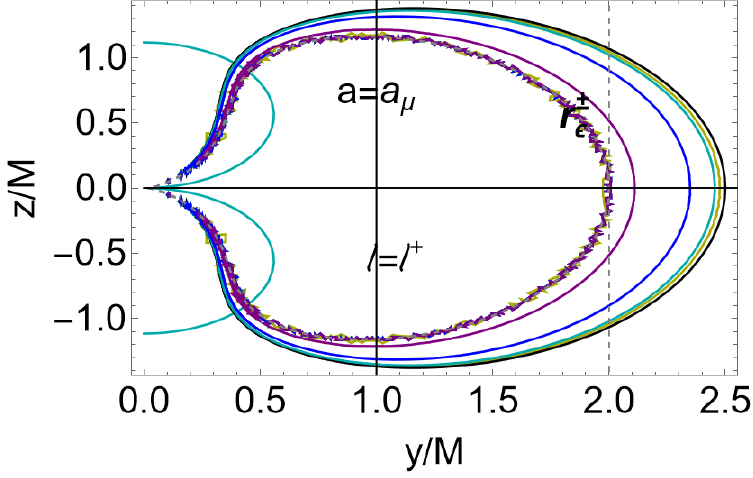}
     \includegraphics[width=5.75cm]{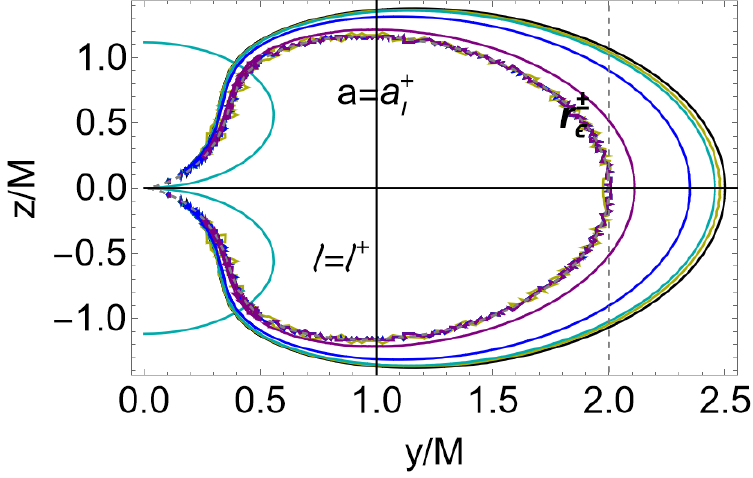}
     \includegraphics[width=5.75cm]{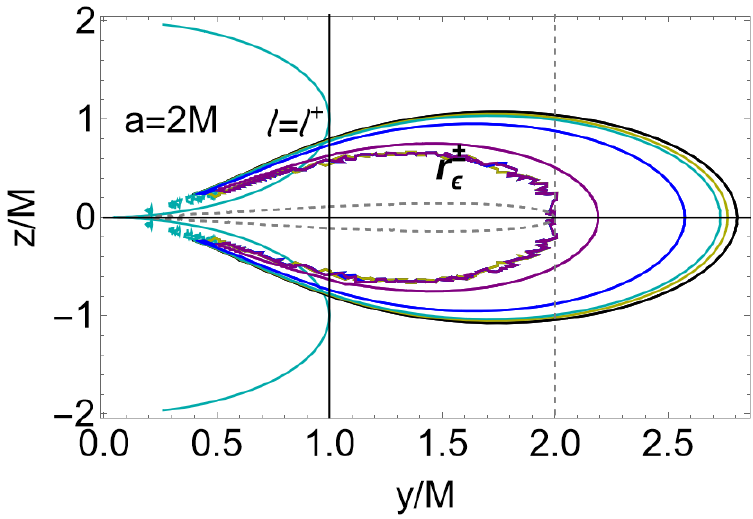}
       \includegraphics[width=5.75cm]{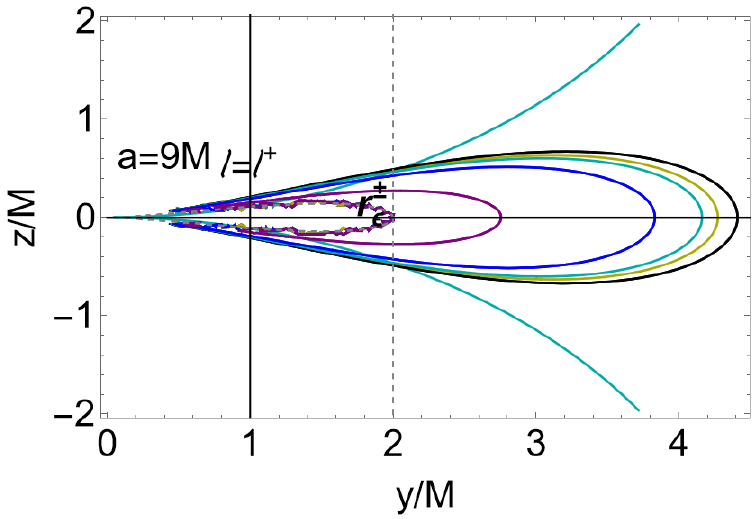}
  \caption{Tori with fluids specific angular momentum $\ell^+<0$ and $\ell^-=\ell^+<0$ and flows inversion points.  (The most general solutions $r_\Ta^\pm$  are shown, discussion on the  constraints are in Sec.\il(\ref{Sec:fromaccretionflows}).)  Spins $a_{\iota}^\pm$ are defined in Eqs\il(\ref{Eq:ba-pan-ba}). Radius $r_{cr}=r_\Ta^+=r_\Ta^-$ is defined in Eq.\il(\ref{Eq:sigmatur}). There is $a_{\mu}\equiv a_\iota^+ - ({a_\iota^+ - a_\iota^-})/{2}$,  and  the fluid specific angular momenta $(\ell_*^+,\ell_{**}^+)$ are $\ell_*^\pm\equiv \ell_{mbo}^+\mp1.5$, $
\ell_{**}^\pm\equiv \ell_{mbo}^+\mp15.5$. There is $\ell_\star^+\equiv \ell^+(r_\star^+)$, where
 $r_\star^+\equiv\bar{r}_{mso}^+-(\bar{r}_{mso}^+-r_{mbo}^+)/1.25$. Radii $r_{\epsilon}^\pm$ are the outer and inner ergosurfaces. Upper left panel:  inversion radius $r_\Ta^{\pm}$ on different planes $\sigma\equiv\sin^2\theta$ and specific angular momenta signed on the panel.  (mso is for marginally stable orbit and mbo is for marginally bounded orbit). Upper center panel: specific angular momenta as functions of $a/M$.  Upper right panel: specific angular momentum $\ell^{\pm}<0$ as function of $r/M$  for the \textbf{NS} spin  $a=1.05M$.  Remaining panels show  the inversion radius $r_\Ta^{\pm}$ on the plane $(y/M,z/M)$, where $\sigma=\sin^2\theta={y^2}/({z^2+y^2})$
 and $r=\sqrt{z^2+y^2}$, for different fluid specific angular momenta $\ell$, according to the notation signed on the upper line left panel. Tori models and accretion  flows inversion points are shown in Figs\il(\ref{Fig:Plotmoocalpit15mis}) and Figs\il(\ref{Fig:Plotmoocalpit15}). Note the inversion radius location  with respect to the ergosurfaces $r_\epsilon^\pm$. Figs\il(\ref{Fig:Plotmoutri}) show the analysis for fluids with momenta $\ell=\ell^-$. }\label{Fig:Plotmoocalpit1}
\end{figure}

\medskip

\textbf{Inversion points: inner and outer tori}

\begin{figure}
\includegraphics[width=6.75cm]{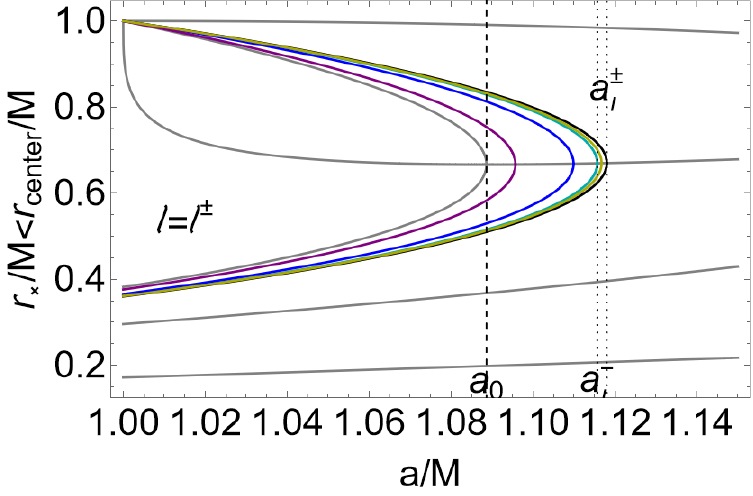}
\includegraphics[width=6.75cm]{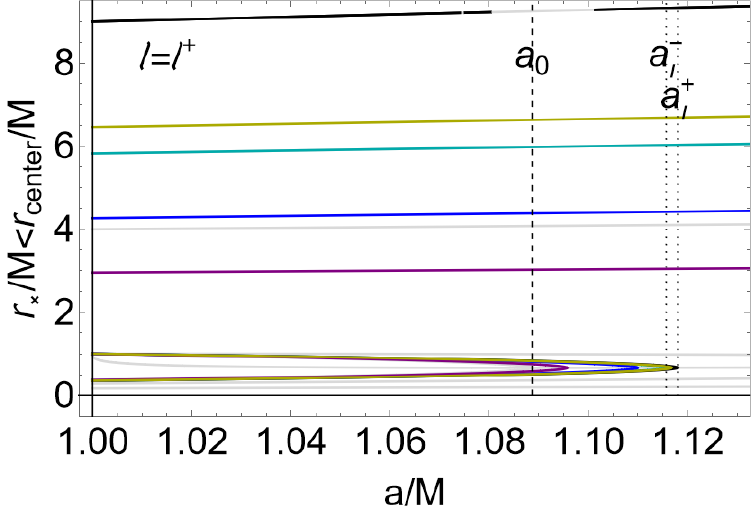}
 \includegraphics[width=6.75cm]{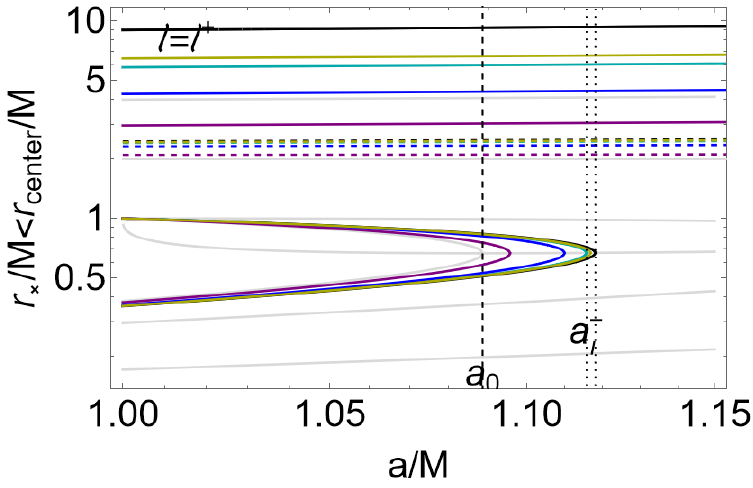}
\includegraphics[width=6.75cm]{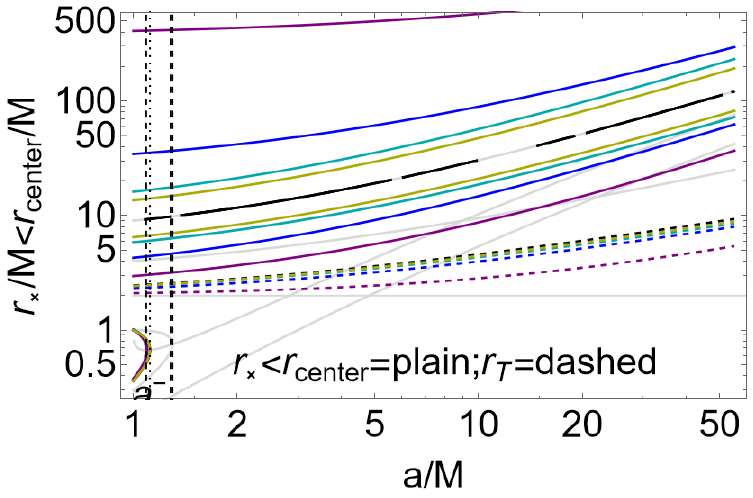}
\caption{Gray curves are the \textbf{NS} geodesic structure.  Spin $a_0$ is defined in Table\il(\ref{Table:corotating-counter-rotatingspin}) and spins $a_{\iota}^\pm$ in Eqs\il(\ref{Eq:ba-pan-ba}.) The center and inner edge of cusped tori $r_{center}>r_{\times}$ (plain curves) and the flow inversion point $r_\Ta$ (dashed curves), on the equatorial plane $\sigma=1$ are shown as functions of the \textbf{NS} spin--mass ratio for different fluid specific angular momentum according to the colors notation of  Figs\il(\ref{Fig:Plotmoocalpit1}).  (Here the  most general solutions $r_\Ta^\pm$  are shown not considering    explicitly the  constraints of Sec.\il(\ref{Sec:fromaccretionflows}).)  Tori configurations are shown in    Figs\il(\ref{Fig:Plotmoocalpit15}).  It is clear the presence of  double tori configurations.}\label{Fig:Plotmoocalpit15mis}
 \end{figure}
For  $\ell=\ell^+$  there is $r_\times>r_\gamma^+>r_\Ta^+$ on the equatorial plane, and $r_\Ta^+$ can be interpreted as   accretion driven inversion point on the equatorial plane--Figs\il(\ref{Fig:Plotmoocalpit15mis}). The distance cusps-inversion points increases with the \textbf{NS} spins, for tori  with cusp at  $r>r_\gamma^+$, and approaches the  central singularity  increasing   $\ell^+$ in magnitude.

In the double counter-rotating  tori  system with $\ell^-=\ell^+<0$ for  $a<a_\iota^+$, there is one inversion point on the  equatorial plane, and for the inner smaller torus of the double system  (having $\ell<0$ with $\La<0$ and $\Em>0$) it can be interpreted as "{excretion}" driven inversion point (in the sense of Sec.\il(\ref{Sec:double-tori})),    or for particle flows incoming from the outer region at $\ell$ constant, with general  $u^\phi$. On planes different from the equatorial  there are two inversion points, the inner one can be interpreted as accretion driven inversion point. Therefore, in the case of a double system, the inner region of the inversion corona, might be considered the more active part, in the sense that  particles  from different flows converge  with  equal  $\ell$  from the external tori (the flow crosses the outer inversion point)  and from the inner torus accretion flows  (particles can also have different $u^\phi$ \cite{submitted}).
 Consequently, the observation of the inversion point could provide indication of the orbiting structures.

We have inferred that   tori  with   $\ell^-<0$ have  no accretion driven  inversion point   on the equatorial plane.
It is simple to see, in the case $\ell^-<0$, as the inversion points are always out of the ergoregion, while the cusps are  always inside the ergoregion. For  $a\in]M,a_0[$  it has been proved that, while  $r_\times\in]r_\delta^-,\tilde{r}_{mso}^-[$, there is  $r_\Ta^+>r_0^+>r_\times$. Accretion driven interpretation is still possible   for planes different from the equatorial, where  there is a inversion point "internal" to the cusp, and therefore a particle coming from the disk towards the central singularity has a inversion radius. The inversion radius external to the torus (outer to  the outer edge) can play a role in the flux forming the torus  or  for   poorly collimated jets. These  counter-rotating tori are much smaller than the inversion  sphere, and this could imply that  the detection of particles (and photons) at the inversion sphere can be a tracer for the orbiting structures which are limited  at  $\ell\in]0,\tilde{\ell}_{mso}^-]$--(see the situation for $a<a_0$ and $r_\times\in ]r_\delta^-,\tilde{r}^-_{mso}[$).

As clear from Figs\il(\ref{Fig:Plotmoocalpit1}), the  inversion corona  from counter--rotating  accretion tori  is a narrow  region  with  maximum  thickness on the equatorial plane.

For  $a \gtrapprox 1$  the corona inner part differs from  for fast and slowly spinning central attractors.
In Figs\il(\ref{Fig:PlotbluecurveherefNS1p}) double inversion points re showed, and the possibility of in-falling or out-going particles crossing the inversion sphere.

\begin{figure}
\centering
 \includegraphics[width=5.75cm]{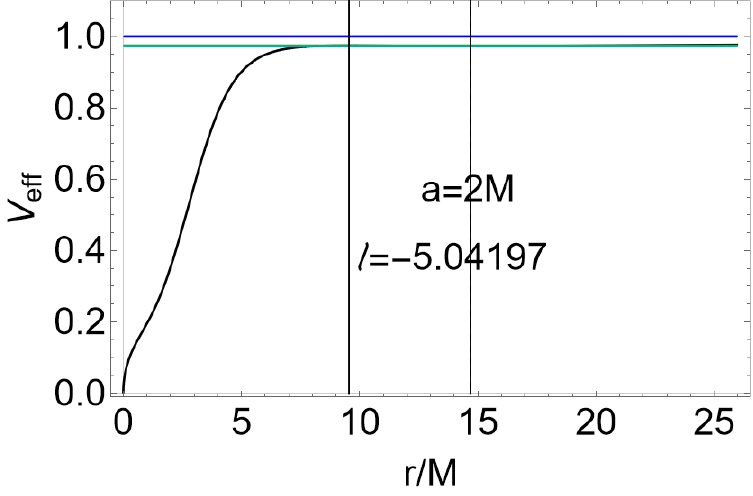}
                       \includegraphics[width=5.75cm]{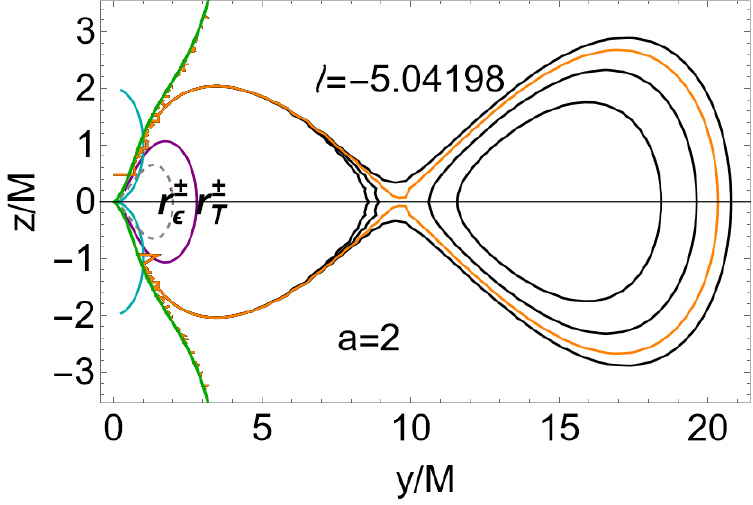}
                       \includegraphics[width=5.75cm]{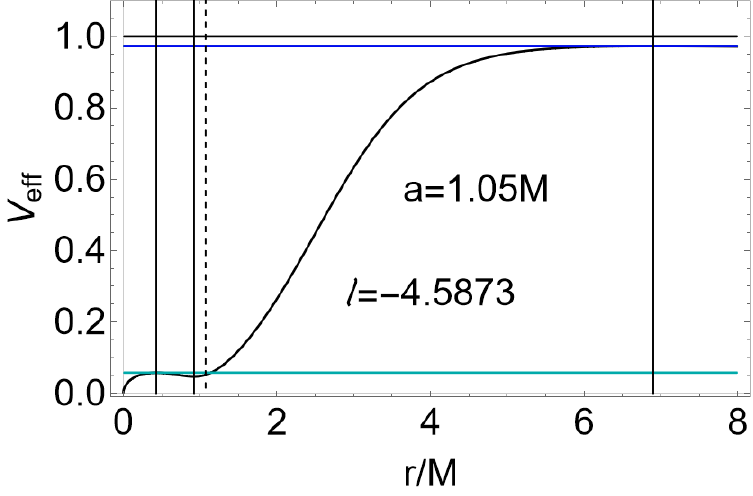}
                          \includegraphics[width=5.75cm]{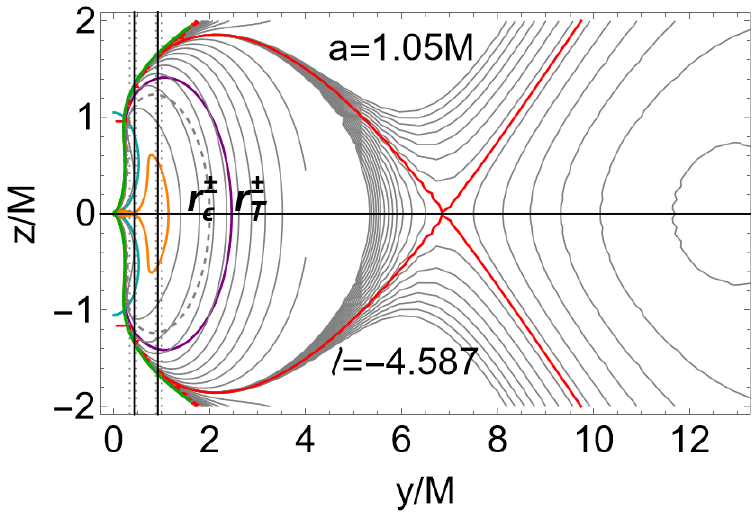}
                     \includegraphics[width=5.75cm]{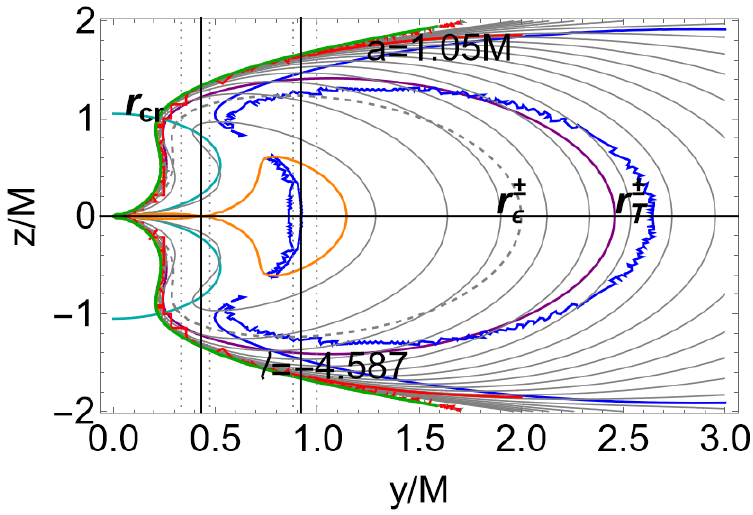}
           \includegraphics[width=5.75cm]{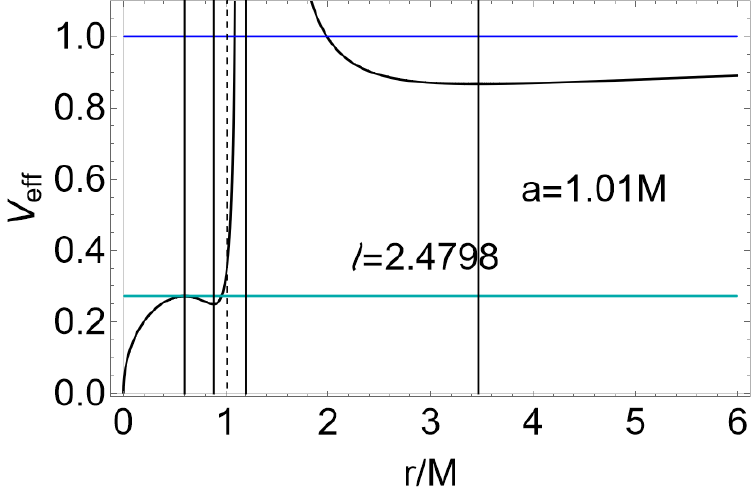}
                 \includegraphics[width=5.75cm]{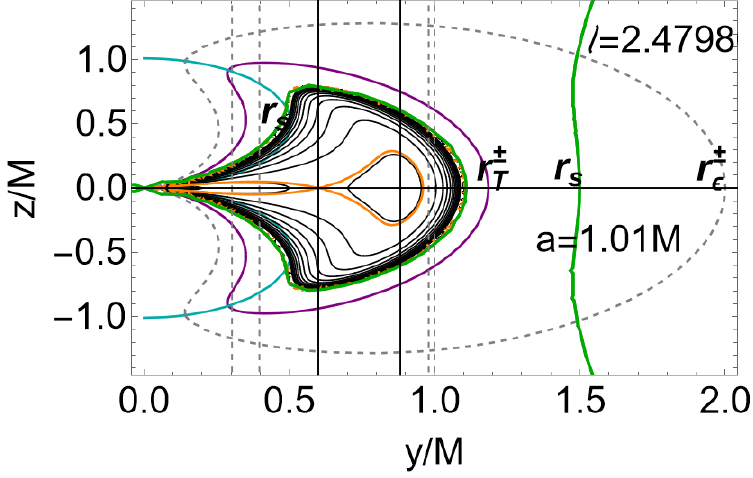}
             \includegraphics[width=5.75cm]{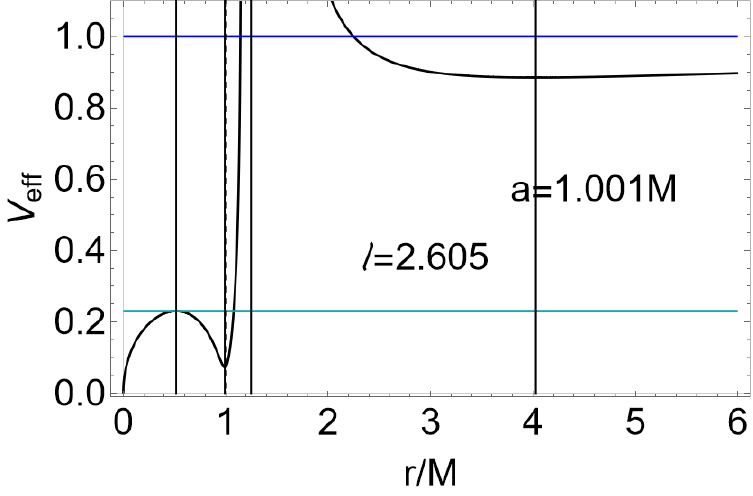}
                     \includegraphics[width=5.75cm]{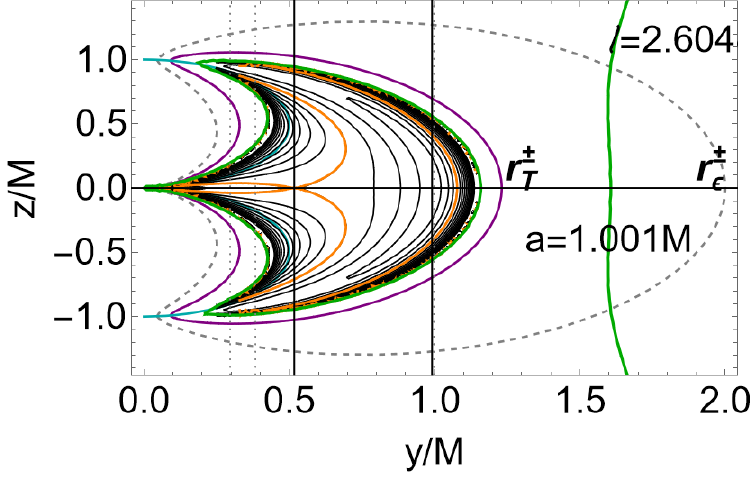}
  \caption{Inversion spheres in double tori systems.Tori effective potential for different spins and fluid specific  angular momentum as signed on the panels. The tori centers and cusps are also signed with vertical lines. It is clear the presence of an orbiting double tori systems.
 Tori configurations and the inversion points $r_{\Ta}^\pm$ are shown in the plane  $(y/M,z/M)$ where $\sigma=\sin^2\theta={y^2}/({z^2+y^2})$
 and $r=\sqrt{z^2+y^2}$ for different fluid specific angular momenta $\ell$.   (The most general solutions $r_\Ta^\pm$ and $\sigma_\Ta$ are shown, while discussion on the constrains is in Sec.\il(\ref{Sec:fromaccretionflows}).) Note the inversion radius location  with respect to the ergosurface $r_\epsilon^\pm$. For $\ell^->0$  inversion points are only for space-like particles with $(\Em<0,\La<0)$ in the conditions of Sec.\il(\ref{Sec:co-space-like}). The limiting boundary configuration for the inner  torus is the defined by the light surface $r_s$ with frequency $\omega=1/\ell$ (darker-green surface)--See also Figs\il(\ref{Fig:Plotmoocalpit15mis}) and Figs\il(\ref{Fig:Plotmoocalpit1}).}\label{Fig:Plotmoocalpit15}
\end{figure}

\begin{figure}
\centering
\includegraphics[width=5.75cm]{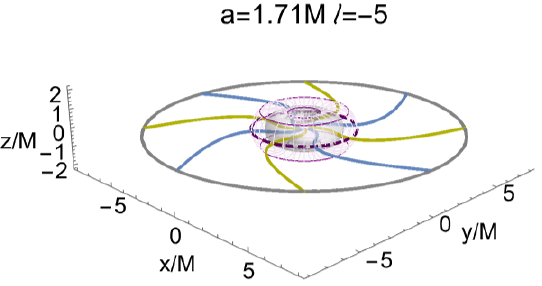}
       \includegraphics[width=4.cm]{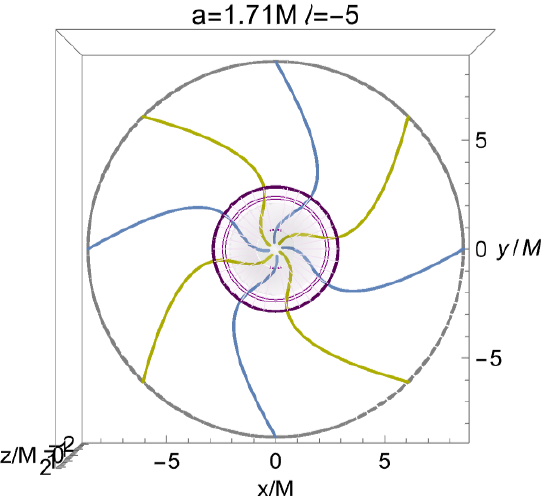}
    \includegraphics[width=5.75cm]{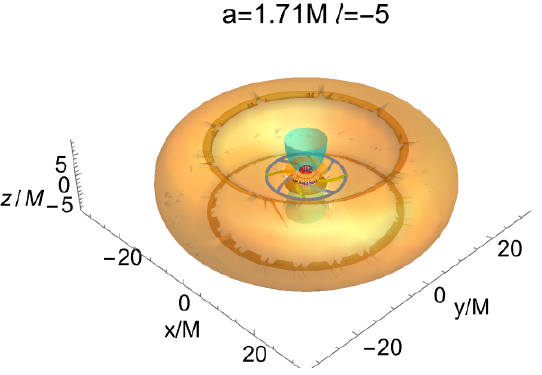}
    \includegraphics[width=5.75cm]{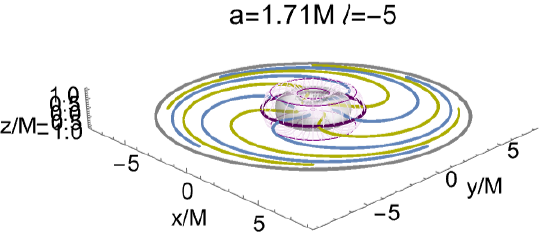}
       \includegraphics[width=4.cm]{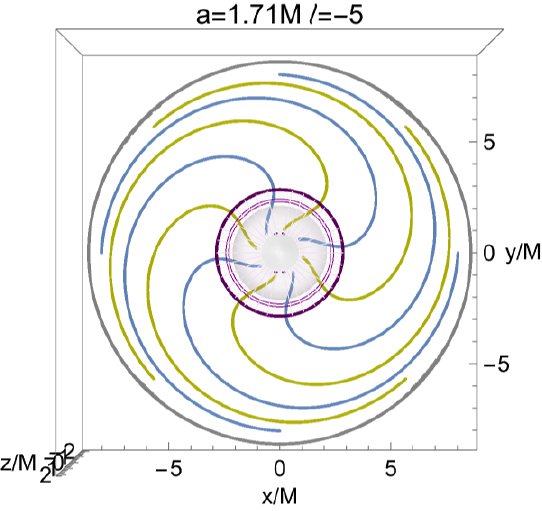}
    \includegraphics[width=5.75cm]{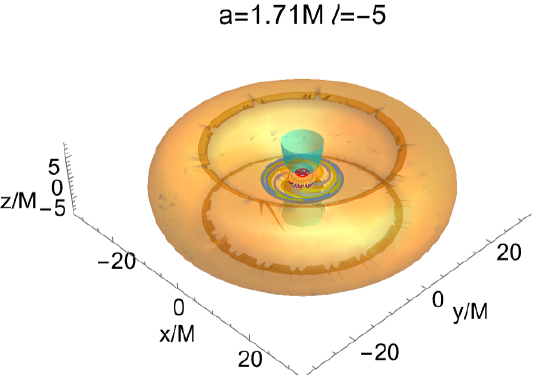}
    \includegraphics[width=5.75cm]{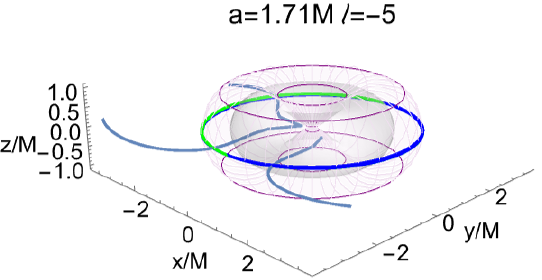}
       \includegraphics[width=4.cm]{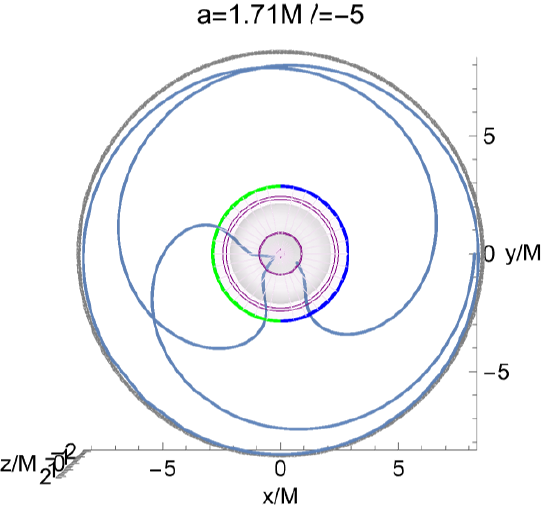}
    \includegraphics[width=5.75cm]{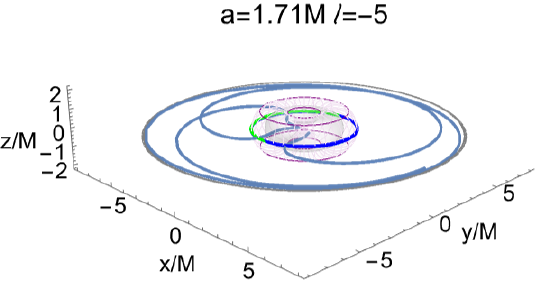}
  \caption{Photons  orbits (upper line) and test particle orbits (center line) with trajectories  inversion points  the \textbf{NS}  geometry  with spin $a=1.71M$. The  specific angular momentum is  $\ell=-5$.  The initial radius is  $r_0<r_\times$, where $r_\times$ is the torus inner edge. The inversion sphere $r_\Ta$ is light-purple surface. On the equatorial plane radius $r_\Ta^{\pm}$ is the purple curve. Gray surface is the geometry inner and outer ergosurfaces. Upper and center line: Center panel is the above view of the right panel.  Right panel shown the torus (orange surface) while the green surface is the light surface with rotational frequency $\omega=1/\ell$. Bottom line: Test particles analysis
where $\theta_0= \pi/2$  and  Carter constant is $Q=0.0035$.}\label{Fig:PlotbluecurveherefNS}
\end{figure}

As clear from Figs\il(\ref{Fig:PlotgluT},\ref{Fig:PlotbluecurveherefNS}),  the  outer torus can  also be  larger  than the inversion corona. The accretion driven coronas can also be rather small--Figs(\ref{Fig:Plotmoocalpit15}).
As these  tori are generally   geometrically thick tori  orbiting very strong attractors,  associated  to high accretion rates,   the  small region of the inversion corona  might  be  a very  active part of the accretion flows.
(The  inner disk of a double counter-rotating toroids with $\ell=\ell^+=\ell^-<0$, the inner Roche lobe
  can  be larger than the torus  outer Roche lobe and the disk is smaller of the "embedding"  inversion sphere--Figs\il(\ref{Fig:Plotmoocalpit1}) and Figs\il(\ref{Fig:Plotmoutri}).).

\medskip

\textbf{Example}

If  $\Em>0$,  the inversion radius is  outside  of the ergoregion (there is $\La<0$ and $\ell<0$). For $\ell>0$ there are no inversion point but spacelike inversion point for $(\Em<0,\La<0)$  in the ergoregion (for $s=-1$). Let us consider below  the more general solution $r_\Ta^\pm$.

In the example  considered in Figs\il(\ref{Fig:PlotparolA0tun1})--Figs\il(\ref{Fig:Plotconfsud}), we introduced the spin functions $\{a_{*1},a_{*2}\}$, and the fluid specific angular   momenta and radii   $(\ell_ {[*]}^\pm,r_ {[*]})$:
\bea&&\label{Eq:spond-sasastar}
a_{*2} \equiv 1.173M: \{r_{center}=r_0^+ ,r_\times=r_0^-\},\quad
a_{*1} \equiv 1.03297M: \{r_{center}=r_\delta^+, r_{\times}=r_\delta^-\},
\\\nonumber
&&
\ell_ {[*]}^\pm\equiv \ell^\pm(r_ {[*]}), \quad r_ {[*]}\equiv r_ {mso}^\pm-\frac{r_ {mso}^\pm-r_ {mbo}^\pm}{2}.
\eea
In   Figs\il(\ref{Fig:Plotconfsud}),  solutions $r_\Ta$ are only for  $a<a_{*2}$, as there is  $r_{center}^-<r_0^+$ and therefore $\La<0$. The inversion point is inside the ergoregion for $a<a_{*1}$,  where  $r_{center}<r_{\delta}^+$ and  $\La<0$ with $\Em<0$--see also Figs\il(\ref{Fig:Plotdacplot2r3},\ref{Fig:Plotvinthalloa5p5},\ref{Fig:Plotmoocalpit1},\ref{Fig:Plotmoocalpit15},\ref{Fig:PlotgluT}).

Inversion points $r_\Ta$  for fluids with specific angular momentum $\ell^-$ are  only in the geometries with spins  $a\in ]M, a_{*2}[$, where $\La<0$ and there is $\ell^-\gtreqless0$ (according to the sign of $\Em$). At $a=a_2$ there is the limiting condition $r_{\Ta}\to\infty$  on the equatorial plane.

The discriminant spin $a_{*1}$ determines the classes of \textbf{NSs} according to the location of the inversion points  with respect to the ergosurface.
(Note that spins  $\{a_{*1},a_{*2}\}$ are defined according to the  geodetic properties on the  equatorial plane.).
 For  $a>a_{*1}$, the inversion point is out of the ergoregion ($r>r_{\epsilon}^+$),  viceversa for  $a<a_{*1}$, the inversion point is inside the ergoregion.
Some tori orbiting  geometries $a<a_{*1}$ are co-rotating   (with $\La<0$ and $\Em<0$), while for   $a\in ]a_{*1},a_{*2}[$, tori are counter-rotating with
 $\La<0$ and  $\Em>0$. For   $a>a_{*2}$  there are co-rotating tori with   $\La>0$ and $\Em>0$ (with no inversion point), and counter-rotating tori with  $\La<0$.
\begin{figure}
\centering
\includegraphics[width=5.75cm]{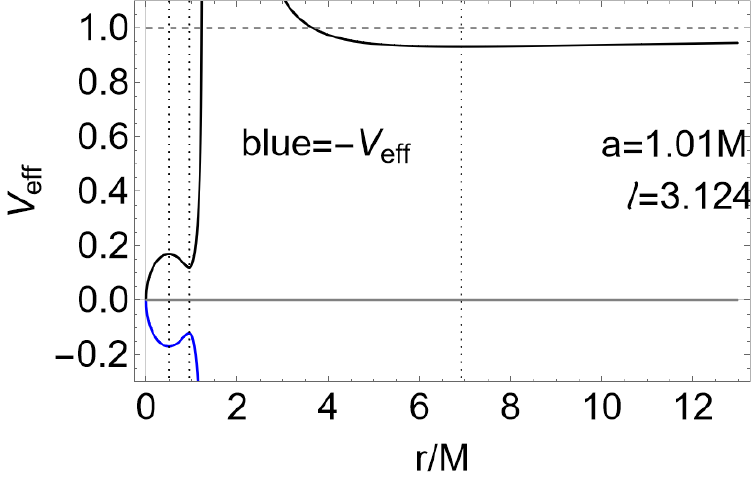}
       \includegraphics[width=5.75cm]{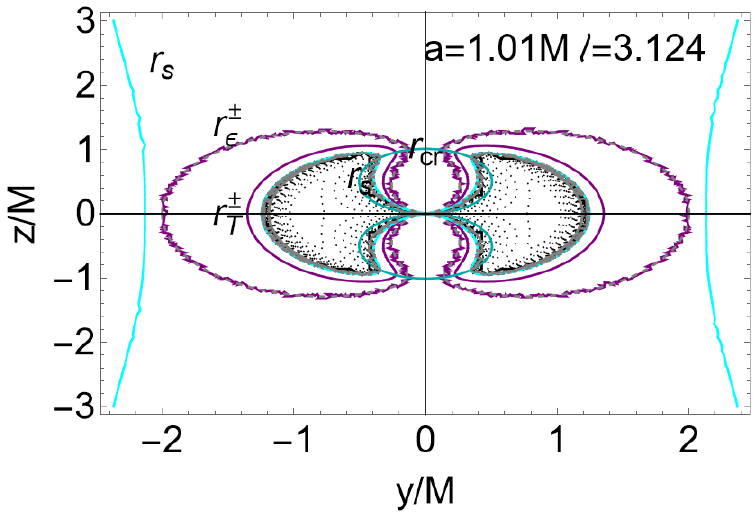}
    \includegraphics[width=5.75cm]{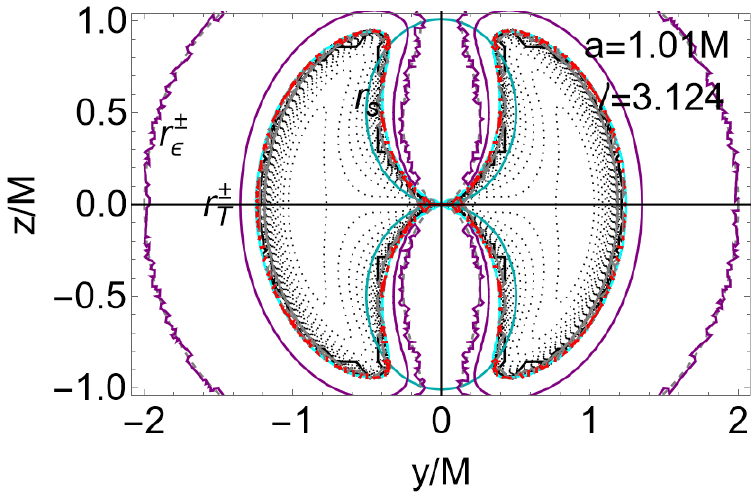}
  \caption{ \textbf{NS}  geometry  with spin $a=1.01M$. The (co-rotating) particle flows  specific angular momentum is  $\ell=3.124$.  The most general solutions $r_\Ta^\pm$ are shown not considering    constraints of Sec.\il(\ref{Sec:fromaccretionflows}). For $\ell^->0$  inversion points are only for space-like particles with $(\Em<0,\La<0)$ in the conditions of Sec.\il(\ref{Sec:co-space-like}). The initial radius is  $r_0<r_\times$, where $r_\times$ is the torus inner edge. The inversion sphere $r_\Ta$ is purple surface.
Left panel show the effective potential for the orbiting torus. Dotted curves show the location of torus cusp $r_{\times}$. Note that at $\ell=3.124$ there are two classes of tori, the inner class admits a cusped torus, the outer class has no cusp. Center panel: equi--density  surfaces are shown, the  cyan curve is the  light surface $r_s$ with rotational frequency $\omega=1/\ell$.  Radius $r_\epsilon^{\pm}$ is the geometry ergosuface.  Radius $r_{cr}=r_\Ta^+=r_\Ta^-$ is defined in Eq.\il(\ref{Eq:sigmatur}). Right panel is a close-up view of the center panel on the  torus.}\label{Fig:PlotcredcatrmarkefNS30}
\end{figure}

From the analysis of Figs\il(\ref{Fig:PlotgluT},\ref{Fig:PlotbluecurveherefNS1p},\ref{Fig:PlotbluecurveherefNS},\ref{Fig:Plotgpass}) it is clear that the inversion point  can be directly related to the outer torus flows (even in the condition $\ell^->0$ with  $\La<0$, for  tachyonic particles),  for $\ell^-<0$,  where there is also the case $\ell^-=\ell^+<0$.  In the determination of the double inversion point, the limiting role represented by the outer horizon for the  \textbf{BH} geometries is played by the radius $r_{cr}$ for the \textbf{NSs}, which also determines the distancing of the inversion point from the central axis of rotation and differentiates slow rotating \textbf{NSs} from faster spinning \textbf{NSs}  --see  Figs\il(\ref{Fig:Plotmoutri}).
\subsubsection{On the double inversion points}\label{Sec:double-inversion-points}
 A distinctive characteristic of the  \textbf{NS}  geometries with respect to the  \textbf{BH} geometries   is the presence of two  inversion point functions $r_\Ta^\pm(\sigma_\Ta)$ for fixed  $\sigma\neq 1$, therefore from  $u^\phi\lessgtr0$ and $u^{\phi}\gtrless0$ at  constant $\ell$ and double inversion points at fixed $\ell$ and  vertical coordinate  $z_\Ta$--Figs\il(\ref{Fig:Plotdacplot2r3}),(\ref{Fig:Plotvinthalloa5p5})--
 Figs\il(\ref{Fig:Plotmoutri},\ref{Fig:Plotmoocalpit1},\ref{Fig:Plotmoocalpit15},\ref{Fig:PlotcredcatrmarkefNS30},\ref{Fig:PlotgluT}).

 In the \textbf{BH} spacetime this characteristic is limited to fast spinning \textbf{BHs}, depending on the fluid specific angular momentum and far from the equatorial plane. A double inversion point in  the \textbf{BH} spacetime occurs in  $]z(+), z_\Ta^{\max}[$ and $]r_+,y(+)[$ where $z_\Ta^{\max}$ is the maximum vertical coordinate of the inversion sphere and  $z(+):r_+$ and $y(+): z=r_+$, where $r_+$ is the \textbf{BH} outer horizon and the coordinates are as follows $r=\sqrt{z^2+y^2}$ and $\sigma=\sin^2\theta=y^2/(y^2+z^2)$  \cite{published}.

The inversion corona  is the same for particles and photons,  but distinguishes co-rotating from counter-rotating fluids  and it is different for accretion-driven and proto-jets driven flows.
Function $r_\Ta^{\pm}(\sigma_\Ta)$ defines a (regular) locus of points surrounding the central singularity, where there can be two inversion points  for  each plane $\sigma\neq1$. The values $(r_\Ta,\sigma_\Ta)$, constrained on the inversion sphere   $r_\Ta^{\pm}(\sigma_\Ta)$, is fixed by the particles (photons) trajectory.
In Figs\il(\ref{Fig:PlotparolA0tun1})  and Figs\il(\ref{Fig:Plotconfsud}) is the analysis of  the relativistic angular velocity $\Omega$, showing  two inversion radii $r_{\Ta}^\pm$ at fixed $\ell$ \emph{and} plane $\sigma_\Ta$. In Sec.\il(\ref{Sec:relative-location})  the relative location of inversion points according $\ell$ was considered.
Defining  the  inner and outer  inversion point (at fixed $\sigma_\Ta$ and $\ell$) as  $r_{\Ta}^{(i)}<r_\Ta^{(o)}$ respectively,  there is $\Omega<0$ for $r<r_{\Ta}^{(i)}\cup r>r_\Ta^{(o)}$, and $\Omega>0$ for $r\in]r_{\Ta}^{(i)},r_{\Ta}^{(o)}[$.
\subsubsection{Proto-jets driven inversion points and inversion points verticality}\label{Sec:inversionpoint-proto-jets}
We conclude this work considering in Sec.\il(\ref{Sec:inversion-proto-jet}), proot-jets driven  inversion points and, in Sec.\il(\ref{Sec:inversion-verti}),  the inversion point location with respect to the singularity rotational axis (verticality of the inversion point).

\textbf{{Inversion points from proto-jets}}\label{Sec:inversion-proto-jet}
 Fluids  with momentum $\ell=\ell^+$ form proto-jets with   $\ell\in \mathbf{L_2^+}\equiv]\ell_{\gamma}^+,\ell^+_{mbo}[$. Fluids with $\ell=\ell^-$ form proto-jets   $\ell\in \mathbf{L_2^-}\equiv ]\ell_{mbo}^-,\ell_\gamma^-[$, where $\ell_{mbo}^->0$ and $\ell_{\gamma}^-=a>0$.

Radii  $r_{mbo}^->r_{\gamma}^-=0$, constraint the  critical points of pressure for  fluids with   $\Em>0$ and $\La>0$  ($\ell=\ell^->0$), with a proto-jet cusp in $r_j\in]0,r_{mbo}^-]$ and center in $r_{center}\in]{r}_{[mbo]},r_{[\gamma]}^-[$, where $r_{[\gamma]}^-=a^2$-- see Figs\il(\ref{Fig:PlotparolA2}). Therefore proto-jets with $\ell=\ell^-$ are always co-rotating\footnote{Open cusped configurations are possible  for example for  $\ell=\ell^-$ at $a<a_2$,  with the formation of axial cusp--\cite{submitted}.} and there are \emph{no} inversion points for  proto-jets driven flows with $\ell=\ell^-$.

Proto-jets  driven inversion points exist only for  fluids with  $\ell=\ell^+<0$ with a cusp out of the ergoregion at $r_j\in ]r_\gamma^+,r_{mbo}^+[$,
and centers in\footnote{There is  a "mixed" region, where  there can be  co-rotating and counter-rotating  proto-jets  structures for  $a>a_7$. From  Eqs\il(\ref{Eq:amsoamno-avv}) it is clear that for  $a>a_{mbo}$ the proto-jets with $\ell=\ell^-$ are out of the ergoregion.} $r\in[r_{[mbo]}^+, r_{[\gamma]}^+]$)--Figs\il(\ref{Fig:PlotparolA2}).
Proto-jets  driven inversion coronas are out of the ergoregion and  bounded by the values
 $]\ell^+_{mbo},\ell^+_{\gamma}[$.
Proto-jets  inversion points are distinguishable from accretion inversion points. Compared with  accretion  driven coronas, proto-jets driven coronas are closer to the ergoregion, approaching  the equatorial plane, and smaller in extension with   larger  thickness.
Increasing the spin, the corona maximum point (verticality) moves far from the attractor (the inversion radius always decreases with increasing $\ell^+<0$  in magnitude).
\textbf{{Inversion points verticality}}\label{Sec:inversion-verti}
Inversion point verticality (vertical coordinate $z_\Ta$, elongation  on the central rotational axis of the inversion point) is particularly relevant for proto-jet driven configurations.

\begin{figure}
\centering
   \includegraphics[width=5.75cm]{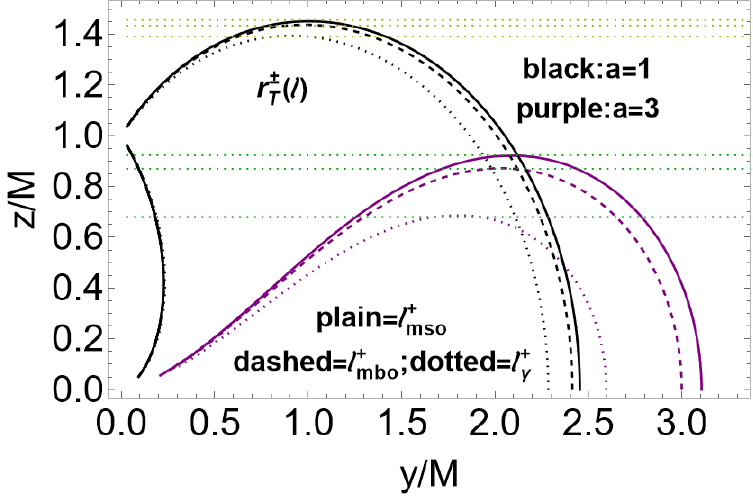}
    \includegraphics[width=5.75cm]{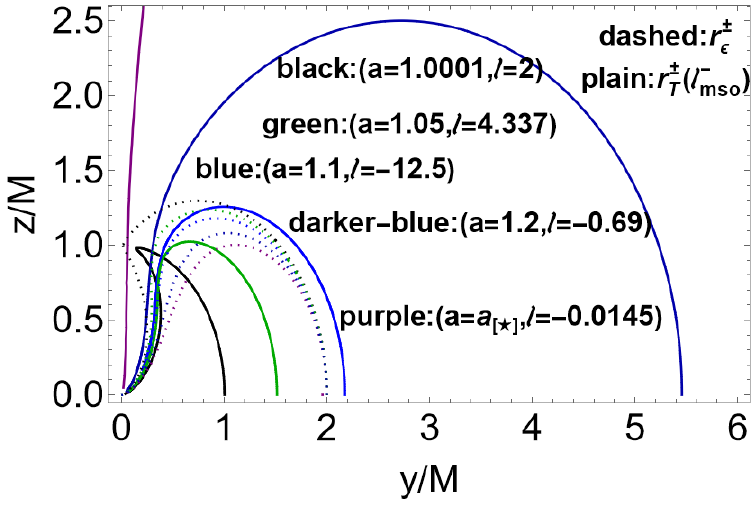}
    \includegraphics[width=5.75cm]{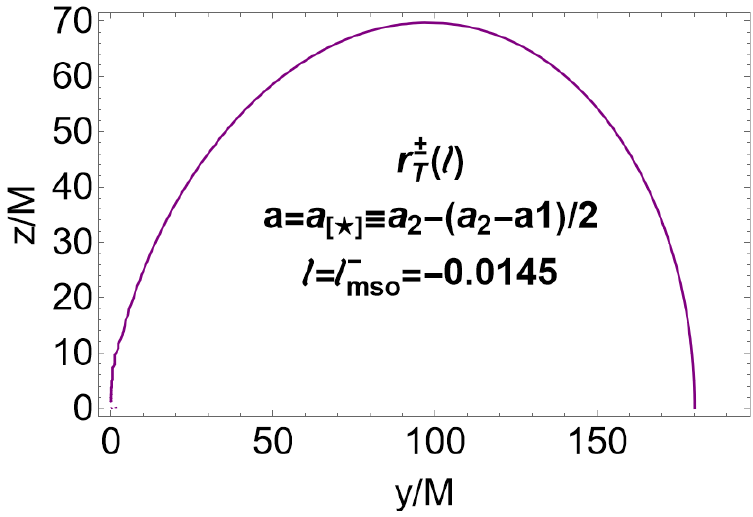}
  \caption{Analysis of the inversion point verticality for fluids with specific angular momentum $\ell=\ell^{+}<0$  (left panel) and $\ell=\ell^-$ (center and right panels).  Here the most general solutions $r_\Ta^\pm$ are shown not considering    constraints of Sec.\il(\ref{Sec:fromaccretionflows}).  For $\ell^->0$  inversion points are only for space-like particles with $(\Em<0,\La<0)$ in the conditions of Sec.\il(\ref{Sec:co-space-like}).  Inversion radius $r^{\pm}_\Ta(\ell)$ is plotted for selected values of $\ell$ and \textbf{NS} spins $a/M$, signed on the panel, in the plane $(y,z)$ where $r=\sqrt{z^2+y^2}$ and $\sigma\equiv\sin^2\theta=y^2/(z^2+y^2)$. Spin $a=M$ is the extreme Kerr \textbf{BH}. Radii $r_\epsilon^\pm$ are the outer and inner ergosurfaces. Notation $(mso, mbo,\gamma)$ refers the marginally stable, marginally bounded and last circular orbit respectively. Purple surface in the center panel is focused on the right panel.}\label{Fig:PlotQCDHEP}
\end{figure}

In Figs\il(\ref{Fig:PlotQCDHEP})  inversion points  location with respect to  the central  rotational axis is shown.
 Radius $r_{cr}$, defined in Eqs\il(\ref{Eq:sigmatur}), provides an indication on the  inversion radius maximum extension.
 Radius $r_{cr}$,    is  related to the  $r_\Ta^\pm$ definition, and it is a  background property dependent on the spin $a$ only, and therefore describes both   fluid particles  and photons. (The inversion radius  does not depend on the particles  radial velocity and  therefore is relevant  also for  outgoing particles--see for example Figs\il(\ref{Fig:PlotbluecurveherefNS})).
\begin{figure}
\centering
    \includegraphics[width=5cm]{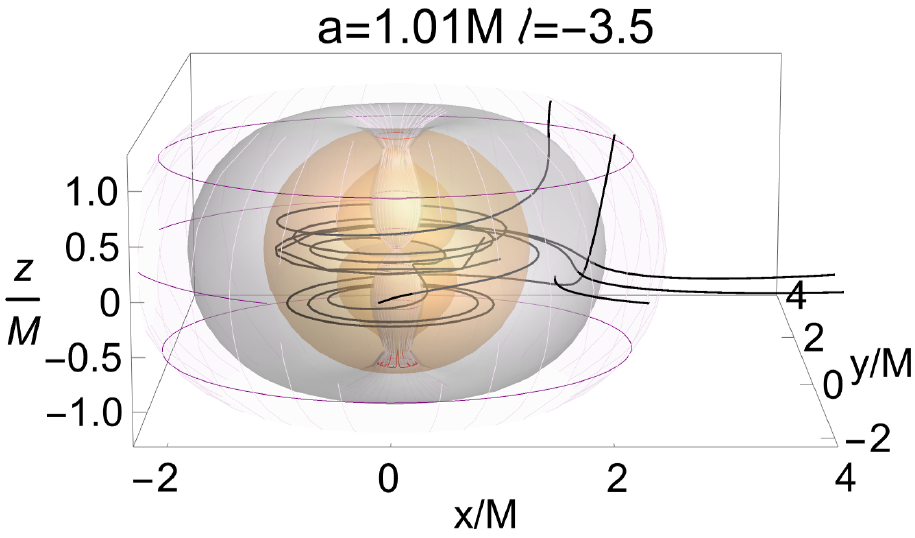}
 \includegraphics[width=5cm]{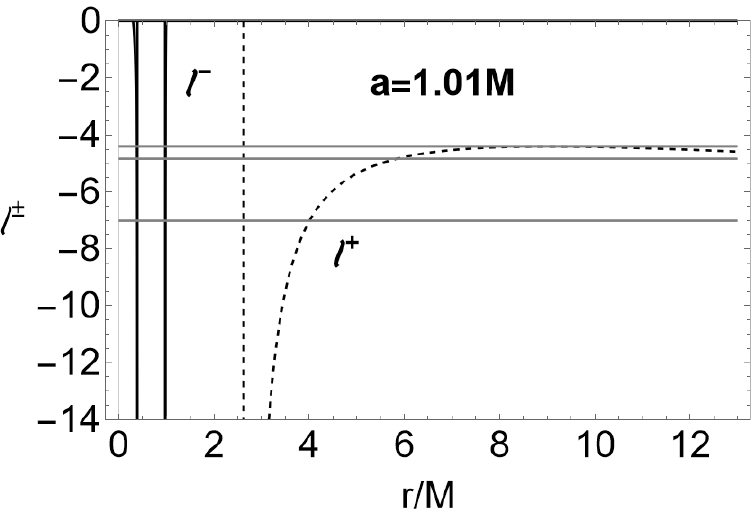}
               \includegraphics[width=5cm]{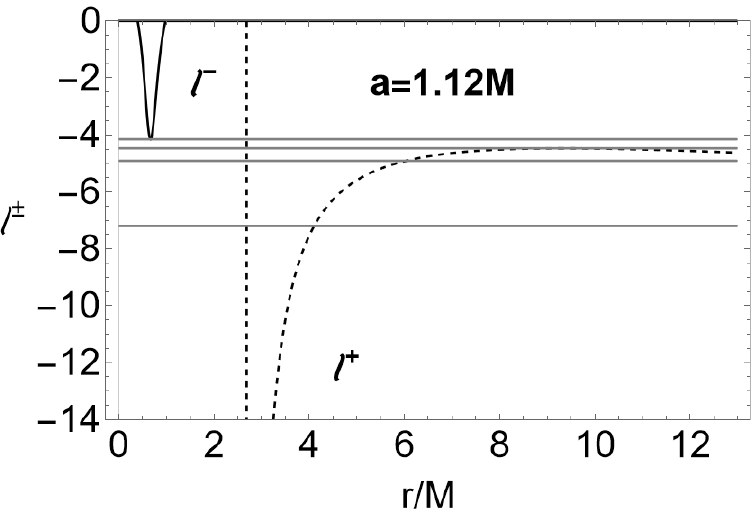}
                \includegraphics[width=5cm]{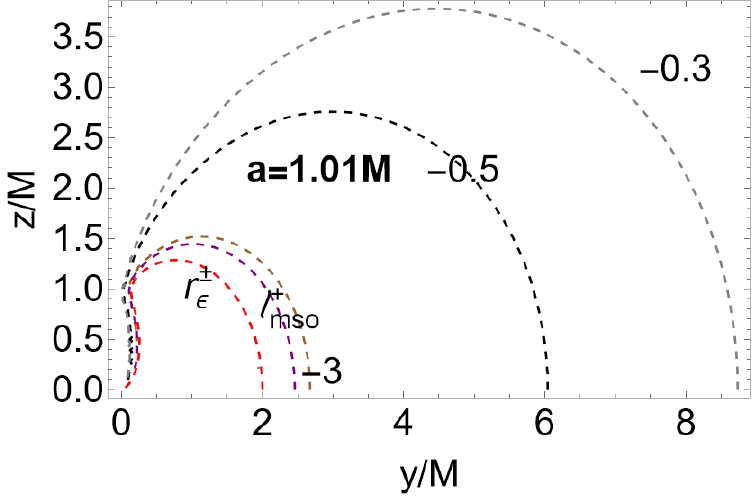}
      \includegraphics[width=5cm]{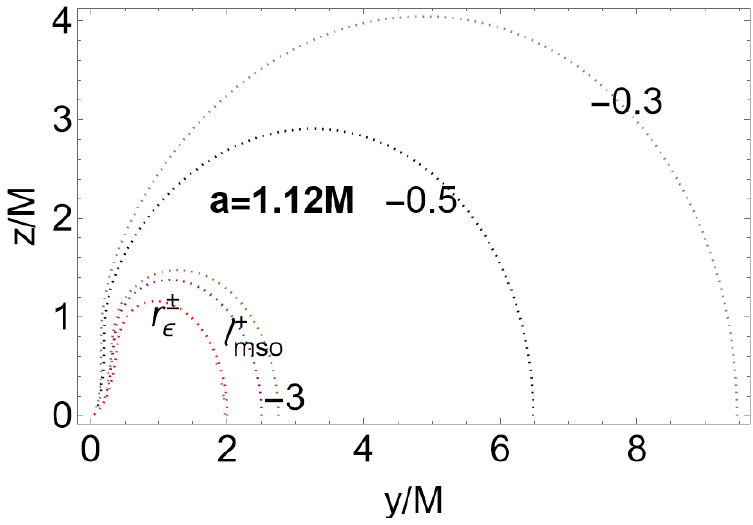}
             \includegraphics[width=5cm]{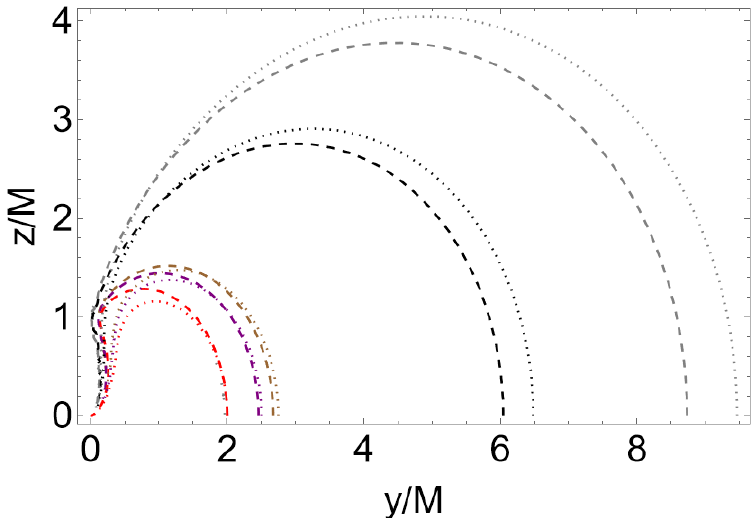}
       \caption{Analysis of the inversion sphere for low angular momenta counter-rotating  tori  $\ell^-\in]\ell_{mso}^+,0[$ in the ergoregion of \textbf{NSs} with $a\in[0,a_2]$. There is $r=\sqrt{z^2+y^2}$ and $\sigma\equiv\sin^2\theta=y^2/(z^2+y^2)$. I Radii $r_\epsilon^\pm$ are the outer and inner ergosurfaces. Notation $mso$ refers the marginally stable circular orbit, $mbo$ to marginally bounded orbit and $(\gamma)$ to the marginally circular orbit. Upper left panel: purple surface is the inversion sphere at specific angular momentum $\ell^-=-3.5>\ell_{mso}^-$. Black lines are test particles ingoing trajectories crossing the inversion spheres and out--going trajectory crossing the inversion sphere. Gray region is the ergosurfaces orange region is the  counter-rotating torus in the ergoregion. Upper center and right panel: specific angular momenta $\ell^\pm<0$ as functions of $r/M$ for selected  spins  signed on the panel. Gray lines are $\ell_{mso}^\pm$ and $(\ell_{mbo}^+,\ell_{\gamma}^+)$. Below  left and center panel, the ergosurfaces and  inversion spheres for selected values of $\ell<0$ signed on the curves and for the \textbf{NSs} spins reported on the panels. Bottom right panel shows super-imposition of the left and center plots. }\label{Fig:PlotSSCR1}
\end{figure}

Inversion corona approaches the ergoregion as   $\ell$  increases in magnitude.

The situation for flows with  $\ell=\ell^+$ is clear:  the maximum of the inversion point verticality $z_\Ta^{\max}$, occurs for $\ell=\ell_{mso}^-$ and $a\approx M$,and  there is  (for proto-jets and accretion driven flows) $z_\Ta^{\max}<1.4M$--Figs\il(\ref{Fig:Plotmoocalpit1}).

 For proto-jets and accretion driven flows, the corona  is larger the smaller the momenta magnitude is, and   the smaller  \textbf{NS} spin   is.
Increasing the spin and decreasing in magnitude $\ell$,  the inversion coronas  extend on the equatorial plane. Furthermore, at $y<y_\Ta^{max}$, the proto-jets and accretion driven  coronas are very close and  with  narrow thickness, reducing  approximately to an orbit.

Let us consider now the more general solution $\ell_\Ta=\ell^-$.
For  $\ell=\ell^-$  the situation is  more complex. Inversion points exist only for accretion driven flows in \textbf{NSs} with $a\in]M,a_2[$. In $a=a_2$, with  $\La<0$,  there is a limiting situation, where for $a\leq a_2$ the inversion sphere for $\ell=\ell^-<0$ (and $\La<0$) increases in extension on the equatorial plane and on the vertical direction becoming, in the limit $a=a_2$ (where $\ell=0$), asymptotically large and therefore  on the \textbf{NS} poles--Figs\il(\ref{Fig:Plotconfsud}).

The inversion sphere is inside or outside the ergoregion according to $\ell=\ell^->0$ (space-like particles) or $\ell=\ell^-<0$--see discussion in Eqs\il(\ref{Eq:spond-sasastar}).

  For  $\ell=\ell_{mso}^\pm$, increasing the spin, the verticality decreases.

   For  $\ell=\ell^+$  the inversion sphere becomes slender and on the equatorial plane moving far from the attractor.

   At $\ell=\ell_{mso}^-<0$, increasing  the spin, the spheres (and the verticality) always increase. (Note that there are tori with  $\ell^-\neq\ell^+<0$.)
 (Inversion spheres for  $\ell^-<0$ are larger  than the  inversion spheres for $\ell^->0$ (spacelike solutions for  tachyons), as the counter-rotating ones  contain  in the ergoregion,  while the co-rotating spheres are  inside the  ergoregion-- Figs\il(\ref{Fig:Plotgusrleona42},\ref{Fig:Plotgusrleona41}, \ref{Fig:Plotgusrleona43})).

\begin{figure}
\centering
   \includegraphics[width=7cm]{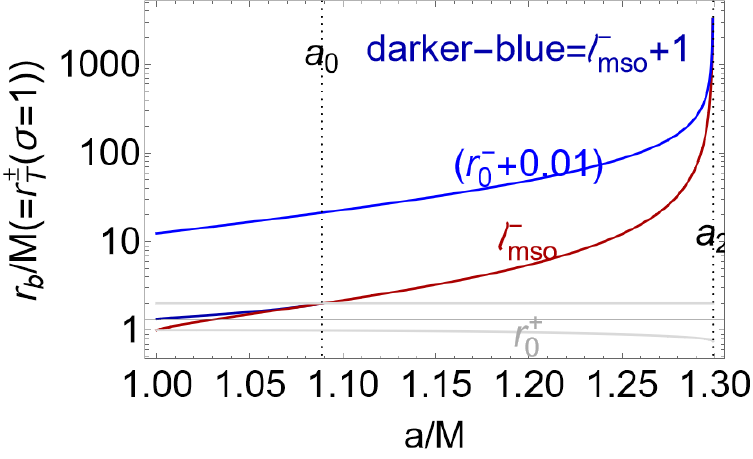}
      \includegraphics[width=7cm]{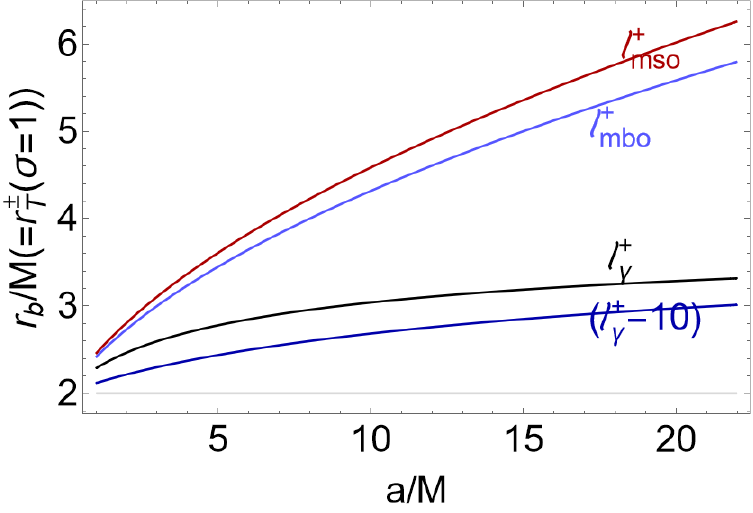}
  \caption{Maximal extension on the equatorial plane of the inversion radius $r_\Ta^\pm=r_b$ for fluids with specific angular momentum $\ell^-$ (left panel) and $\ell^+<0$ (right panel). Notation $(mso)$ refers to marginally stable orbit, $(mbo)$ to marginally bounded orbit, $\gamma$ to the last circular orbit, at $r_0^\pm$  there is $\ell=0$. For $\ell^->0$  inversion points are only for space-like particles with $(\Em<0,\La<0)$ in the conditions of Sec.\il(\ref{Sec:co-space-like}).
  Spins $\{a_0,a_1\}$ are defined in Table\il(\ref{Table:corotating-counter-rotatingspin}).}\label{Fig:Plotprextax}
\end{figure}

 (The inversion sphere maximum for  $\ell^->0$
is provided by the ergoregion, to which the inversion  sphere tends  for large $a$ (i.e. $a\leq a_0$) and  for  $\ell^-\to+ \infty$, occurring for   critical points of pressure close to  $r_\delta^\pm$. These tori are cusped and located inside the ergoregion. They belong to a double toroidal  system with equal $\ell^->0$ (where $\ell_{mso}^->\ell_{\gamma}^-$). The outer co-rotating torus has  $\ell\in\mathbf {L_3^-}$, it is quiescent and with center in  $r>r_{[mso]}^->r_{[\gamma]}^->r_{[mbo]}^-$ and geometrically separated from the inner torus--see Figs\il(\ref{Fig:Plotprextax}).

\subsubsection{Inversion coronas thickness and slow counter-rotating spheres
}\label{Sec:slow-Invers}

We conclude this section with some final  comments  on the coronas thickness  and on  slow counter-rotating spheres  with $\ell^-\in]\ell_{mso}^+,0[$.
 Let us consider Figs\il(\ref{Fig:PlotQCDHEP}) and (\ref{Fig:PlotSSCR1}).

For $\ell^+$,  the accretion coronas thickness is always smaller then the proto-jets coronas thickness   and decreases, approaching the singularity and the rotational axis, increasing on the equatorial plane. The inversion corona is therefore a particular active region close to the \textbf{NS}.  The thickness increases with the \textbf{NSs} spins, increasing also the thickness  differences between the accretion driven and proto-jets driven coronas.

The slow momenta inversion spheres, located out of  the ergoregion,  are not immediately related to flows   coming from   the  counter-rotating tori  in the ergoregion of \textbf{NSs} with
$a\in[M,a_2]$ (especially on the equatorial plane). There is no  inversion corona for these, quiescent or cusped, tori as
 the specific momentum is in $\mathbf{L_1^-}=]\ell_{mso}^+,0[$.

As expected,  the inversion sphere increases, decreasing   the momentum in magnitude and it is always closed (there is always a inversion point on the equatorial plane).

 It should be noted that a solution $r_\Ta^\pm$ (not related to tori) also exists  for  $a>a_2$ (with  $\Em>0$ and  $\La<0$).
It is interesting to note that there are always inversion points for very large distances from the \textbf{NS}, not linked to the frame dragging of the ergoregion, and there are also solutions for \textbf{BH} spacetimes.

The inversion spheres  verticality is very high for very small $\ell$ in magnitude, however here we do not consider this case as it is not related to the orbiting tori.

Interestingly in general, increasing the \textbf{NS} spin,  the inversion sphere increases, contrary to  the case of  high momenta  in magnitude,  where  it decreases with the \textbf{NS} spin,  as for the case $\ell=\ell^+<\ell_{mso}^+$.
\subsubsection{Flow inversion points from orbiting tori}\label{Sec:deta-tori-integr}
Inversion  spheres are fixed by the   tori specific angular momentum.
We  assume    tori    fix  the  initial conditions on the (free) fluid particles trajectories, in terms of energy (for timelike particles inherited by the torus parameter $K$) and momentum ($\ell=$constant),     considering the   (unstable) circular orbit at  $r_\times$ (or $r_0\leq r_\times$)  as torus inner edge, and  time--like particle energy as $K$ (at $r_\times$)  fixed by the torus. The reliability of this assumption   is confronted by the  small variation  of the  (narrow)   inversion coronas  thickness with the initial data (parameter $\ell=$constant)  variation and by the inversion points  independence on  flow initial data (other than the specific angular momentum)-- Figs\il(\ref{Fig:Plotvinthalloa5p5}).

\begin{figure}
\centering
\includegraphics[width=5.75cm]{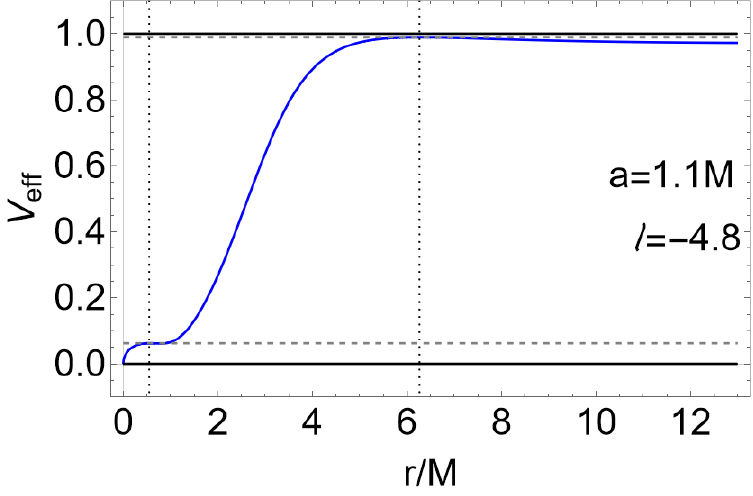}
       \includegraphics[width=5.75cm]{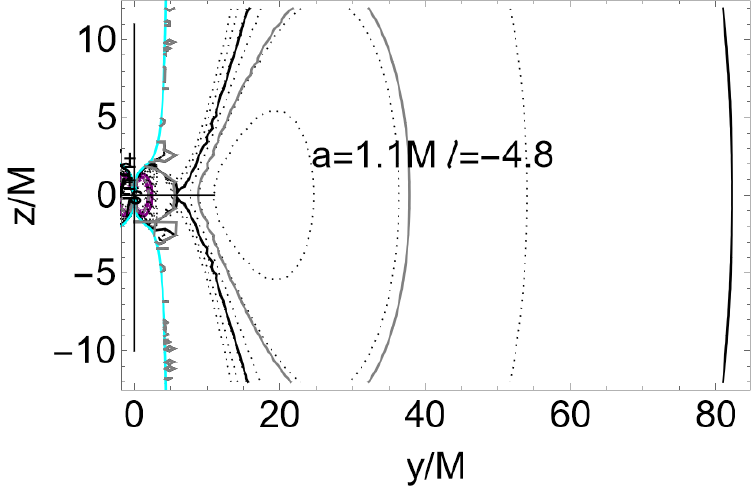}
    \includegraphics[width=5.75cm]{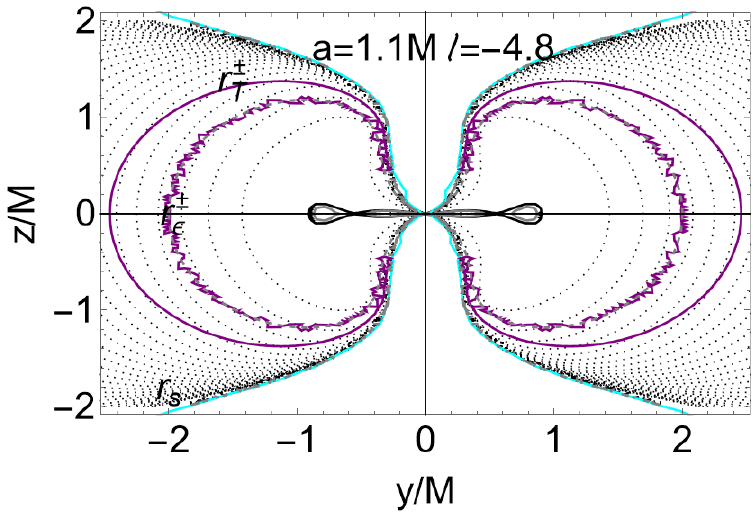}
       \includegraphics[width=5.cm]{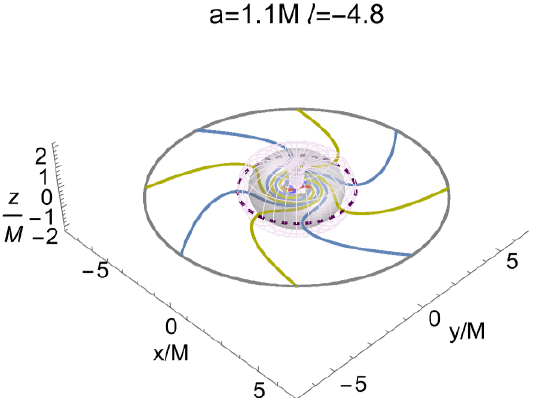}
    \includegraphics[width=4.5cm]{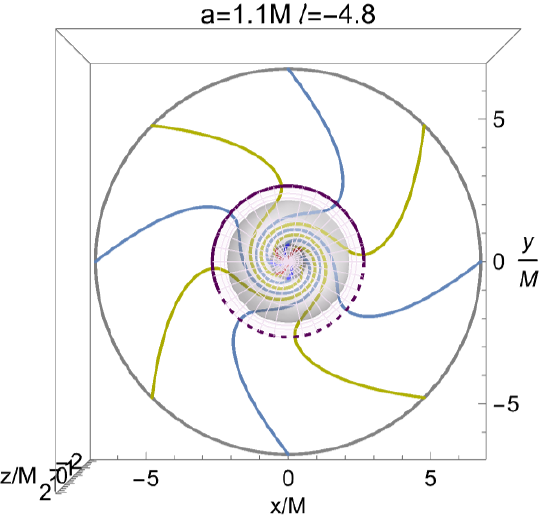}
    \includegraphics[width=4.5cm]{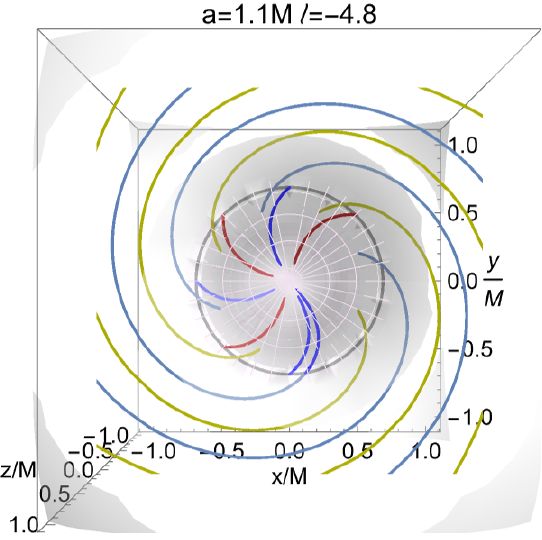}
      \includegraphics[width=4.5cm]{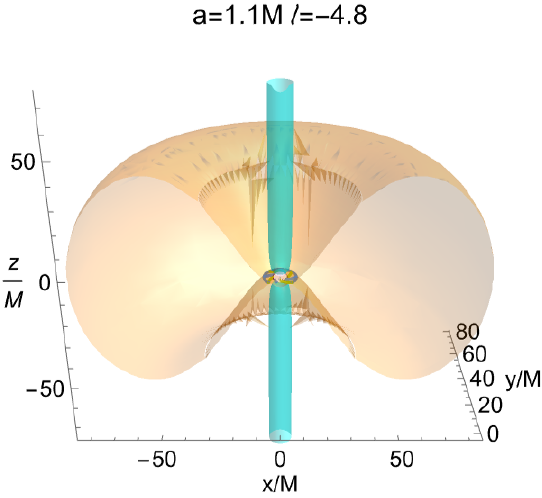}
     \includegraphics[width=4.5cm]{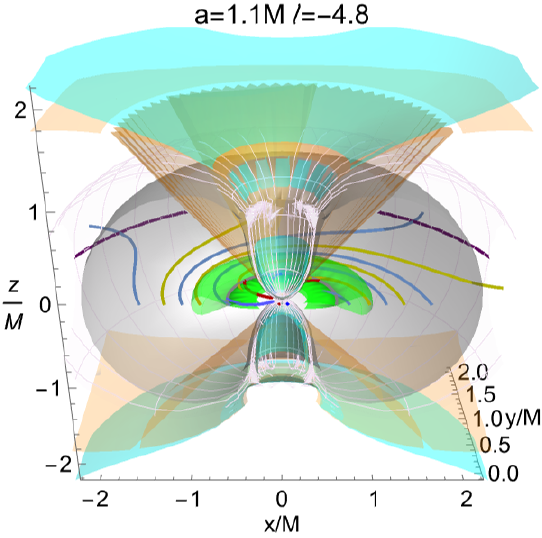}
  \caption{Photon   orbits with trajectories  inversion points  in the \textbf{NS}  geometry  with spin $a=1.1M$. The (counter-rotating) particle flows  specific angular momentum is  $\ell=-4.8$. There are  two orbiting counter-rotating tori  with $\ell^-=\ell^+=-4.8$. In Figs\il(\ref{Fig:PlotbluecurveherefNS1p}) there is the analysis for the test particles   trajectories. The initial radius is  $r_0<r_\times$, where $r_\times$ is the torus inner edge (cusp). The inversion sphere  $r_\Ta$ is the light-purple surface. On the equatorial plane radius $r_\Ta^{\pm}$ is the purple curve.
  Upper line: left panel shows the effective potential for the system of two  counter--rotating tori. Dotted curves show the location of tori cusps $r_{\times}$. Center panel: equi--density  surfaces an outer torus, cyan curve is the  light surface with rotational frequency $\omega=1/\ell$. Right panel is a close-up view of the center panel on the inner torus. Radius $r_\epsilon^{\pm}$ is the geometry ergosuface. Center line:   Gray surface is the geometry inner and outer ergosurfaces.  Center panel is the above view of the left  panel.   Right panel is a close-up view of the center panel, where trajectories from the inner counter-rotating torus flow are shown.  Bottom panels show the outer  torus (orange surface) and inner torus (green surface), while the cyan surface is the light--surface with rotational frequency $\omega=1/\ell$. Right panel is a close-up view of the left panel.}\label{Fig:PlotgluT}
\end{figure}

\begin{figure}
\centering
\includegraphics[width=5.75cm]{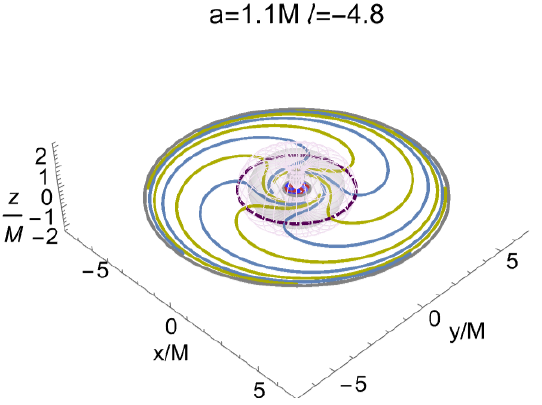}
       \includegraphics[width=5.75cm]{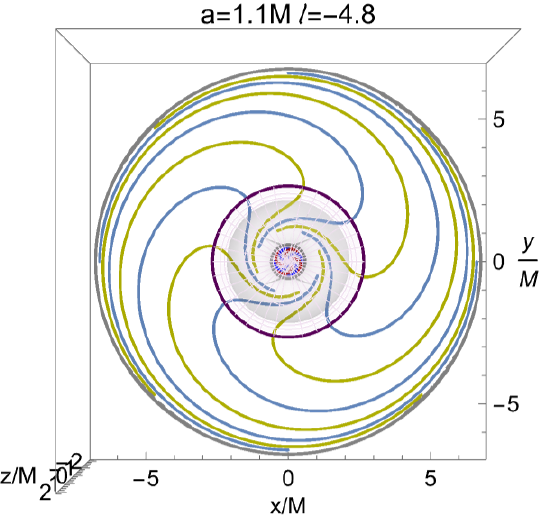}
    \includegraphics[width=5.75cm]{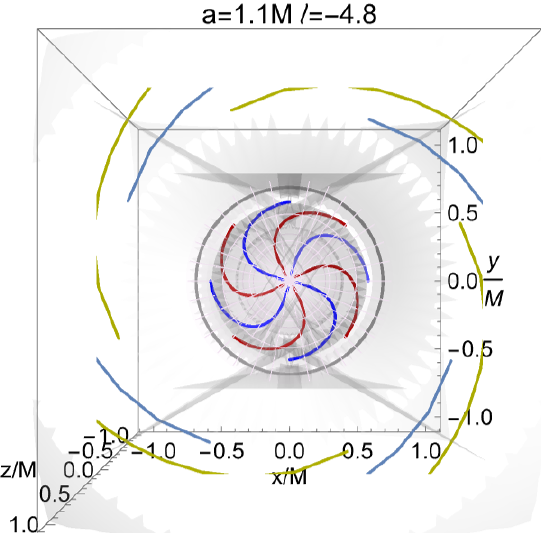}
  \caption{Test particle   orbits with trajectories  inversion points in the \textbf{NS}  geometry  with spin $a=1.1M$. The (counter-rotating) particle flows  specific angular momentum is  $\ell=-4.8$.  There are  two orbiting counter-rotating tori. In Figs\il(\ref{Fig:PlotgluT}) there is the analysis for the photons  trajectories and description of the toroidal models. The initial radius is  $r_0<r_\times$, where $r_\times$ is the torus inner edge. The inversion sphere $r_\Ta$ is the  light-purple surface. On the equatorial plane, radius $r_\Ta^{\pm}$ is the purple curve. Gray surface is the geometry inner and outer ergosurfaces.  Center panel is the above view of the left  panel.   Right panel is a close-up view of the center panel, where trajectories from the inner counter--rotating torus flow are shown.}\label{Fig:PlotbluecurveherefNS1p}
\end{figure}

In the analysis of Figs\il(\ref{Fig:Plotmoocalpit1},\ref{Fig:Plotmoocalpit15mis},\ref{Fig:Plotmoocalpit15}) and Figs\il(\ref{Fig:PlotgluT})    double tori were considered.
Flows  with  $\ell=\ell^-<0$  can be  interpretable, for the outer torus of the pair at  $\ell^-=\ell^+<0$, as accretion or proto-jets driven inversion points.
A distinctive property of Kerr  \textbf{NS} geometries  is the presence of two inversion radii at planes different from the equatorial-- discussed in more details in Sec.\il(\ref{Sec:double-inversion-points}). The inversion point  is uniquely defined by  the $\ell$ and, at fixed  $\ell$, there can be two  (sphere) radii $r_\Ta^\pm(\sigma)$,  defining a   closed inversion sphere. Viceversa, functions   $r_\Ta^{\pm}(\sigma)$ correspond to  one $\ell$. The inversion point does not explicitly depend on the orbital energy of the particle (or photon) or on the torus topology (if quiescent or cusped), which is  fixed  by the  $K$ parameter (related to the energy definition  $\Em$).

In Figs\il(\ref{Fig:Plotmoocalpit15}),
 a limiting closed configuration is shown bounding  the inner torus of a couple with  momentum $\ell^-$, orbiting in the \textbf{NS} ergoregion,  identified with the light surfaces $r_s$ solutions for the  null-like
circular orbits with  rotational frequencies     $\omega_\pm=1/\ell:{g}(\mathcal{L},{\mathcal{L}})=0$, defined from  the  Killing vector $\mathcal{L}=\xi^t +\omega \xi^{\phi}$.
 The  limiting   frequencies $ \omega_\pm$
 bound the
 (time-like) stationary observers   four-velocity, (where, in the \textbf{BH} spacetimes,
  $\omega_\pm(r_{\pm})=\omega_H^{\pm}$ are the frequencies of the outer and inner Kerr horizons respectively). There is  $\omega_{\pm}(r\to0)=1/a$, where $r_\gamma^-=0$ and $\ell_\gamma^-=0$.
\section{Final Remarks}\label{Sec:discussion}
The inversion points have been  analyzed  for accretion driven and proto-jets driven flow in the Kerr super-spinar background. Defined by the condition $u^\phi=0$, the  flow inversion points are related  to the orbiting structures defining accretion driven and proto-jets driven inversion coronas.

The location of inversion points has been considered also in relation to the ergoregion,  characterizing  the    dragging  effects on the accretion flows and jet emission in  Kerr super-spinars and  proving that
\textbf{NSs} are distinguished from the \textbf{BHs} by the presence of  double inversion points on planes  different from the equatorial.

In  accretion  and jet driven  flows,  possible strong observational signatures of these attractors, are highlighted. Our results prove that inversion points can constitute an observational aspect capable of distinguishing the super-spinars.

The flow carries in fact information about the accretion structures around the central attractor. These structures are governed by the balance of the hydrodynamic and centrifugal  forces  and also effects of super-spinars repulsive gravity.
In this paper we addressed in particular  the special case of double tori with equal specific angular momentum and  slow counter-rotating  inversion spheres,
tori and inversion points for  $\ell=\pm
a$  and
excretion driven inversion points.
Finally proto-jets driven inversion points and inversion points verticality have been discussed focused in relation to
 the inversion coronas thickness. (On the other hand, for  slow  tori with $\ell^-\in]\ell_{mso}^+,0[$,  inversion spheres  verticality is very high for very small $\ell$ in magnitude,  this case however is not related to the orbiting tori.).
 With  narrow thickness and small  extension on the equatorial plane and rotational central axis,  the inversion surface  varies   little with the   fluid initial conditions and the details  of the  emission processes. The inversion coronas, including all the possible inversion surfaces,  are a geometric property  of the background. Therefore the information provided by the analysis of  the inversion points  has    a remarkable   observational significance  and
  applicability  to  various  orbiting toroidal  models.

In these different scenarios  our analysis  provides distinguishable physical characteristics of jets and accretion tori  for  possible strong  observational
signature of the primordial Kerr superspinars.
Therefore this work  is part of the research on the possible observational evidences of Kerr naked singularity solutions, and the astrophysical tracers
distinguishing black holes from their super-spinar counterparts having cosmological relevance (as predicted for example by string theory \cite{Horava}).
However  the  observational properties  in  the inversion coronas we expect    depend strongly on the  processes timescales i.e.  the  time flow reaches the inversion points,  depending  on the initial data.

 The accretion  and proto-jets flows are constituted by particles and photons, coming  from   toroids orbiting a central Kerr super-spinar.  Constraints on the accretion or proto-jets  driven flows are grounded on    the assumption  that the  accretion disk   "inner edge"  is located  between marginally bounded orbit and marginally stable orbit for the accretion driven flow and marginally stable and marginally circular orbit for the proto-jets driven flows (with the exception for   inner tori in the ergoregion with $a\in]M,a_2[$), translated into the  range of values  of the   $\ell$ parameter defining the coronas.
 The closed  surfaces, defining    {inversion coronas} are spherical shells, fixed by the fluids specific angular momentum,  where the  particles  and photons flow toroidal  velocity is null. The inversion corona  surround  the central singularity,  distinguishing    co-rotating and counter-rotating flows from proto-jets and accretion driven flows.
(Co-rotating flows ($a\ell>0$) have no (timelike or photon like particle) inversion  points. However, in Sec.\il(\ref{Sec:co-space-like}) we also focused on the $\ell>0$ case  with $\Em<0$ and  $\La<0$,  where spacelike (tachyonic particles) inversion points are possible inside the ergoregion, in relation to tori  orbiting  a very small region of the ergoregion of certain  slowly spinning \textbf{NSs}. Further notes on the co-rotating   flows inversion points in relation to the \textbf{NS} causal structure are in Sec.\il(\ref{Sec:co-space-like}).)
  We deepen the analysis in  Sec.\il(\ref{Sec:relative-location}) discussing in detail the
relative location of the inversion points in accreting flows.

{The results of this analysis are also based on the possibility to relate the  flow inversion points to the orbiting configurations, differentiated in  accretion  and proto-jets structures, defining the driven and proto-jets driven flows respectively. In each case we  distinguish the definition and properties of the  inversion coronas.}

{The special case of double tori with equal specific angular momentum is a notable example of a  trace for  the presence of  a center super-spinar in contrast with a central \textbf{BH} attractor.
A further  remarkable feature is the existence of
tori for specific angular parameter   $\ell=\pm
a$ and the   inversion points of their flows.
We also discussed the existence of  excretion driven inversion points due to the repulsive gravity effects typical of the ergoregions of a super-spinars class.
More in general we focused on the location of inversion points in relation to the ergoregion and  with the respect to the  torus \emph{outer} edge, relevant for  the double orbiting systems at equal $\ell$ in the \textbf{NSs} backgrounds.
\textbf{NSs} are distinguished in fact  from the \textbf{BHs} also by the presence of  double inversion points on planes  different from the equatorial, we discuss this aspect extensively.
Proto-jets driven inversion points and inversion points verticality are focused  a part
addressing the presence of inversion points in the jet emission. Consequently
we characterised  several  inversion coronas morphological features as  thickness, relating to the dimensionless  spin of the central attractor. }

{Slow counter-rotating  inversion spheres,
 due to orbiting tori with small $\ell$ magnitude,  have been pointed out as  a tracer of certain classes of super-spinars.}

{In conclusion properties characterizing these geometries have been  highlighted in  both accretion  and jet flows. As the flow carries information about the accretion structures around the central attractor, we  established that  the inversion points can constitute an observational aspect capable of distinguishing the super-spinars, proving  that the  closed inversion  region is capable to  distinguish proto-jets from  accretion driven flows,  and co-rotating from  counter-rotating flows, providing a  signature of these attractors. As  the  inversion coronas are   small bounded  orbital regions located   out of  ergoregion,   we expect it   be a  remarkably active  part of the accreting flux of matter and photons,   particularly   close  to the central singularity and the rotational axis, where this region may be   characterized  by an increase of the flow luminosity and temperature.
 We plan to apply  in the future the  analysis  using observational data, searching for  signatures of the  Kerr
super-spinars in the   distinguishable physical characteristics of jets and accretion
tori highlighted here. An application that may  detect interesting scenarios, both in the super-spinars and  \textbf{BHs} fields,  could be constituted by the turning points evidence possibly imprinted by counter-rotating  photons in the shadows cast by the central singularity and now  capable  to be observed with the recent advances in observational astronomy\footnote{(\textbf{EHT}: \textbf{E}vent  \textbf{H}orizon \textbf{T}elescope 	 eventhorizontelescope.org).}.}

\section*{Acknowledgements}
The authors would like to express their acknowledgments for the Research Centre for Theoretical Physics and Astrophysics, Institute of Physics, Silesian University in Opava.

\appendix

\section{Tori  model}\label{Sec:fluid-effe-pot}
 Toroidal surfaces are the closed, and closed cusped Polish doughnut (P-D) solutions. The accretion in these geometrically thick tori is driven by the Paczynski  mechanism of violation of  force balance  \cite{Pac-Wii}.
 The toroids  rotation law, $\ell=\ell(\Omega)$ is   independent of  the  details of the  fluids equation of state, and   provides  the   integrability condition of   the Euler equation in the case of  barotropic fluids--\cite{abrafra}.
PD are constructed as
constant pressure  surfaces, constructed in the axis--symmetric spacetimes   applying the  von Zeipel theorem. (The surfaces of constant angular velocity $\Omega$ and of constant specific angular momentum $\ell$ coincide.).

Tori are described by the conditions   $\partial_t \mathbf{q}=0$ and
$\partial_{\varphi} \mathbf{q}=0$,  with $\mathbf{q}$ being a generic spacetime tensor and
assuming  a barotropic equation of state  with    $u^{\theta}=0$ and
$u^r=0$ on the fluid particles velocities. 
The  condition   $\ell=$constant  is assumed for each toroids.  The  continuity equation is therefore
is  identically satisfied. The  fluid dynamics  is  governed by the Euler equation only. The pressure gradients are are regulated by  an effective potential function, $V_{eff}(r;\ell,a)$,  encoding the centrifugal and gravitational components of the force balance:
\bea\label{Eq:K-Veff-fluids}
V_{eff}^2={\frac{g_{t\phi}^2-g_{\phi\phi} g_{tt}}{g_{\phi\phi}+2 g_{t\phi} \ell +g_{tt} \ell ^2}}.
\eea
(There is  $\ell^\pm=\ell(r;a):\partial_r V_{eff}=0$).
In this frame the  torus  edges are solutions of $V_{eff}^2=K^2$   and there is   $K=K_\times$ for the  cusped tori  edges. (We adopt the notation $q_{\bullet}\equiv q(r_{\bullet})$ for any quantity $q$ evaluated on a radius $r_{\bullet}$.). Function $K(r)=V_{eff}(\ell(r))$, characterizes proto-jets cusps and accretion disks cusps as
cusped tori have parameter $K=K_\times\equiv K(r_{\times})\in]K_{center}, 1[\subset ]K_{mso}, 1[$.
 Centers and cusps correspond to the  minimum and maximum of the  fluid effective potential respectively.
Toroidal configurations are defined  by a maximum of the pressure and density (torus center $r_{center}$) and, eventually, a vanishing  of pressure point (torus  cusp $r_{\times}$, or proto-jets cusp $r_j$).

 The matter outflow (at $r\leq r_\times$) occurs  as consequence of the violation of mechanical equilibrium  in the balance of the gravitational and inertial forces and the pressure gradients in the tori  (hydro-gravitational instability  Paczy\'nski  mechanism  \cite{Pac-Wii})): at   ($r\leq r_\times$), the fluid  is  pressure-free  we shortly indicate the toroids   cusp as  the inner edge of accreting torus.
P-D models  are essentially defined only by their radial extension on the equatorial plane, by  the radial  pressure (density) gradients (maximum and minimum points) on the equatorial  plane,  therefore  the main  functions describing tori properties    are projected on the equatorial plane.
\section{Relative location of the inversion points}\label{Sec:relative-location}
A inversion radius $r^\pm_\Ta(\sigma)$ is associated to one and only one $\ell$ value, but it may be associated to a double  toroidal orbital system composed by  two counter-rotating  tori orbiting the central singularities for $\ell<0$.
At plane different from the equatorial $(\sigma\neq0)$,  fluids particles with a momentum $\ell$ have two inversion point radii $r_\Ta^\pm$ and, at fixed $\ell$ and vertical coordinate  $z_\Ta$ two inversion points $y_\Ta$ for planes different  from the equatorial (as well as two $z_	\Ta$ at fixed $y_\Ta$), which constitutes a major difference with the \textbf{BH} case.
 \textbf{NS} geometries are characterized by double inversion points at fixed $\ell$, on a vertical coordinate. (There are also two inversion points at fixed coordinate $y$ on the inversion sphere, below and above the equatorial planes).

We here  consider two  fluids with specific angular momentum $(\ell, \kappa_{\ell} \ell)$  on the  inversion plane $\sigma_\Ta$,  with  $\kappa_{\ell}\in \mathds{R}$--Eq\il(\ref{Eq:lLE}).
 We discuss the relative location of flows inversion points for fluids characterized by a specific relation among their specific momentum.
Relatively counter--rotating/co-rotating fluids are defined  by the condition  $\kappa_{\ell}\lessgtr0$, as \emph{$\ell$counter-rotating}  and \emph{$\ell$co-rotating} respectively. The special case,   $\kappa_{\ell}=1$ considers    inversion points for elements of  same flow (we prove the existence of $\kappa_r\neq1$)\footnote{At  fixed $\sigma_\Ta$ ($\theta_\Ta)$ we study the radii $r_\Ta=\sqrt{z_\Ta^2+y_\Ta^2}$, while in the analysis of the double inversion points  we fixed $z_\Ta$, discussing  coordinates $y_\Ta$.} or from double toroids systems with  with equal specific angular momentum.

Inversion point definition depends only on the specific fluid angular momentum $\ell$,  and it is independent from $\Em$ or $K$. The explicit independence from $K$ ensures the independence   on the tori geometrical thickness, tori topology (cusped or quiescent) and the precise location of the particles at initial data.
Therefore we solve the  equation $\ell (r_\Ta) =\kappa_{\ell} \ell (\kappa_r r_\Ta)$, where $\ell(r_\Ta)$ is in Eq\il(\ref{Eq:lLE}).
It is clear that for $\kappa_\ell<0$, $\ell$counter-rotating flows,
(time-like and photon-like) inversion points are only for one fluid, as co-rotating fluids have no inversion points.

\medskip

\textbf{On the equatorial plane $\sigma=1$}

On the equatorial plane, $\sigma=1$, the situation is as follows:
\bea&&\nonumber
\sigma =1:  \quad r_\Ta=r_{\Ta (\kappa_{\ell}\kappa_r)}\quad\mbox{for}\quad (\kappa_{\ell}<0: \kappa_r\neq1);\quad
  (k\in]0,1[, \kappa_{\ell}<\kappa_r, \kappa_r\neq1),\quad (\kappa_{\ell}=1, \kappa_r=1, r_\Ta\neq 2),\\
&&   (\kappa_{\ell}>1: \kappa_r<k,\kappa_r\neq1),\quad \mbox{where}\quad r_{\Ta (\kappa_{\ell}\kappa_r)}\equiv \frac{\kappa_r-1}{\kappa_{\ell}-\kappa_r}+\sqrt{\frac{(\kappa_{\ell}-1)^2}{(\kappa_{\ell}-\kappa_r)^2}}+1.
\eea
 We also included the limiting case $\kappa_{\ell}=\kappa_r=1$.
It is clear that the $\ell$co-rotating case is much more complex then the  $\ell$counter-rotating case.

\medskip

\textbf{The general case $\sigma\neq1$}

In the general case ($\sigma\neq1$)
we  analyze the sub-cases
$\{(\kappa_{\ell}<0, \kappa_r>1),(\kappa_{\ell}>0, \kappa_r>1),(\kappa_{\ell}>0, \kappa_r<1),(\kappa_{\ell}<0,\kappa_r<1)\}$.

For $ \kappa_r=1$, there are solutions only for $\kappa_{\ell}=1$, (a inversion point $r_\Ta(\sigma_\Ta)$ is uniquely identified by a value of $\ell$) and $r\neq r_\epsilon^\pm$.
The cases $(\kappa_{\ell}<0,\kappa_r\lessgtr1)$   and  $(\kappa_{\ell}>0, \kappa_r\lessgtr 1)$   can be studied considering the problem symmetries.

For  $ (\kappa_{\ell}<0,\kappa_r>1$), $\ell$-counter-rotating fluids, the situation can be summarized as  follows
\bea&&
\mbox{For}\quad(\kappa_{\ell}<0,  \kappa_r>1):\sigma=\sigma_w,\quad\mbox{for}\quad  r\in]0,r_{(\kappa_r)}[,\quad r\neq r_{\kappa_r},\quad\mbox{where}
\\\nonumber
&&
\sigma_w\equiv\frac{r\kappa_r  [\kappa_{\ell} (r-2)-r\kappa_r +2]}{a^2 [\kappa_{\ell} \kappa_r-1]}+1,\quad r_{\kappa_r}\equiv \frac{2}{\kappa_r+1},\quad r_{(\kappa_r)}\equiv\frac{2 (\kappa_{\ell}-1)}{\kappa_{\ell}-\kappa_r}.
\eea
Or, alternately, in terms of the  inversion point  radius (with $\sigma\neq\{0,1\}$) there is
\bea&&
\mbox{For}\quad \kappa_{\ell}<-1:\quad \kappa_r>1, (\sigma_{(\tau)}, r_{(\tau)}^{-}), (\sigma\in]\sigma_{(\tau)},1[, r_{(\tau)}^{\pm}), (\sigma_{(l)}, r_{(\tau)}^{+}).
\\\nonumber
&& \mbox{For}\quad \kappa_{\ell}=-1:\quad \kappa_r>1, (\sigma\in]\sigma_{(l)},1[, r_{(\tau)}^{\pm}).
\\\nonumber
&& \mbox{For}\quad \kappa_{\ell}\in]-1,0[:\quad \kappa_r>1, (\sigma_{(\tau)}, \sigma_{(l)},r_{(\tau)}^{-}), (\sigma>\sigma_{(\tau)}, r_{(\tau)}^{\pm}),
\eea
where
\bea
&&\nonumber
\sigma_{(l)}\equiv\frac{a^2 (\kappa_r+1)^2-4 \kappa_r}{a^2 (\kappa_r+1)^2},\quad \sigma_{(\tau)}\equiv 1-\frac{(\kappa_{\ell}-1)^2 \kappa_r}{a^2\kappa_e (\kappa_{\ell} \kappa_r-1)},\quad r_{(\tau)}^\mp\equiv\frac{\kappa_{\ell}-1}{\kappa_e}\mp\sqrt{\frac{a^2 (\sigma -1) \kappa_e (\kappa_{\ell} \kappa_r-1)+(\kappa_{\ell}-1)^2 \kappa_r}{\kappa_r \kappa_e^2}},
\eea
with ${\kappa_e}\equiv\kappa_{\ell}-\kappa_r$.

We now consider the conditions
$(\kappa_{\ell}>0,\kappa_r>1)$ for $\ell$-co-rotating fluids,
where there is
\bea&&\nonumber
\mbox{For}\quad \kappa_{\ell}\in]0,1[: \left[\kappa_r\in]1,\kappa_{(r)}^+[, (\sigma_{(\tau)},\sigma_{(l)}, r_{(\tau)}^{-}), (\sigma>\sigma_{(\tau)},r_{(\tau)}^{\pm})\right],
 \quad \left[\kappa_{(r)}^+, (\sigma>\sigma_{(\tau)},r_{(\tau)}^{\pm}), (\sigma_{(l)}, r_{(\tau)}^{-})\right],
\\\nonumber
  &&\qquad \qquad \left[\kappa_r\in]\kappa_{(r)}^+,{1}/{k}[, (\sigma>0, r_{(\tau)}^{\pm}), (\sigma_{(l)}, r_{(\tau)}^{-})\right],\quad \left[\kappa_r\geq {1}/{\kappa_{\ell}}: (\sigma\neq\{0,1\}, r_{(\tau)}^{+})\right].
\\\nonumber
&&\mbox{For}\quad \kappa_{\ell}=1: \left[\kappa_r>1, (\sigma\neq\{0,1,\sigma_{(l)}\}, r_{(\tau)}^{+})\right].
\\\nonumber
&&\mbox{For}\quad \kappa_{\ell}>1:\quad\left[\kappa_r\in]1,\kappa_{(r)}^+[, (\sigma_{(\tau)}, r_{(\tau)}^{-}),(\sigma_{(l)}, r_{(\tau)}^{+}), (\sigma>\sigma_{(\tau)},  r_{(\tau)}^{\pm})\right],
\quad \left[\kappa_r\in[\kappa_{(r)}^+,\kappa[, (\sigma>0, r_{(\tau)}^{\pm}), (\sigma_{(l)}, r_{(\tau)}^{+})\right],
\\
&& \qquad \qquad \left[\kappa_r=\kappa_{\ell}, (\sigma>0,\sigma\neq \sigma_{(l)}[, r_{f})\right], \quad \left[\kappa_r>\kappa_{\ell}, (\sigma>0,\sigma\neq \sigma_{(l)}, r_{(\tau)}^{+})\right].
\eea
with
\bea
&&
\kappa_{(r)}^\mp\equiv\frac{a^2 \left[\mp \kappa_{\ell} \sqrt{\frac{\left(a^2-1\right) (\kappa_{\ell}-1)^2 [(a-1) \kappa_{\ell}+a+1] [a(\kappa_{\ell}+1)+\kappa_{\ell}-1]}{a^4 \kappa_{\ell}^2}}+\kappa_{\ell}^2+1\right]-(\kappa_{\ell}-1)^2}{2 a^2 \kappa_{\ell}};\quad
r_{f}\equiv-\frac{a^2 (\sigma -1) (\kappa_{\ell} \kappa_r-1)}{2 (\kappa_{\ell}-1) \kappa_r}.
\eea
The case  $(\kappa_{\ell}>0,\kappa_r<1)$ can be studied from the case $(\kappa_{\ell}>0,\kappa_r>1)$, and considering the problem symmetries. However it can be useful to see explicitly this case as follows
\bea&&
\mbox{For}\quad \kappa_{\ell}\in]0,1[:\quad \left[\kappa_r\in]0,\kappa_{\ell}[, \sigma\in]0,1[,\sigma\neq\sigma_{(l)}, r_{(\tau)}^{+}\right],\quad
\left[\kappa_r=\kappa_{\ell}: \sigma\in]0,1[,\sigma\neq \sigma_{(l)}, r_{f}\right],
\\\nonumber
&&\left[\kappa_r\in]\kappa_{\ell}, \kappa_{(r)}^-], (\sigma\in]0,1[, r_{(\tau)}^{\pm}), (\sigma_{(l)}, r_{(\tau)}^{+})\right],
\quad \left[\kappa_r\in]\kappa_{(r)}^-,1[, (\sigma_{(\tau)}, r_{(\tau)}^{-}),(\sigma_{(l)}, r_{(\tau)}^{+}),  (\sigma>\sigma_{(\tau)}, r_{(\tau)}^{\pm})\right].
\\\nonumber
&&\mbox{For}\quad \kappa_{\ell}=1: \quad\left[\kappa_r\in]0,1[, (\sigma\in]0,1[, r_{(\tau)}^{+}), \sigma\neq\sigma_{(l)}\right].
\\\nonumber
&&\mbox{For}\quad \kappa_{\ell}>1:\quad\left[\kappa_r\in]0,1/\kappa_{\ell}], \sigma\in]0,1[,r_{(\tau)}^{+}, \sigma\neq\sigma_{(l)}\right],\quad
\left[\kappa_r\in]{1}/{\kappa_{\ell}},\kappa_{(r)}^-], (\sigma\in]0,1[, r_{(\tau)}^{\pm}), (\sigma_{(l)}, r_{(\tau)}^{-})\right],
\\\nonumber
&&\left[\kappa_r\in]\kappa_{(r)}^-,1[, (\sigma_{(\tau)}, \sigma_{(l)}, r_{(\tau)}^{-}), (\sigma>\sigma_{(\tau)},r_{(\tau)}^{\pm})\right].
\eea
\section{Notes on the co-rotating flows inversion points}\label{Sec:co-space-like}
\subsection{On the fluids parameters in NSs spacetime}
We can   analyze,  in terms of past/future   directed  and   spacelike particles,  co-rotating and counter-rotating motion    with   negative energy orbits  $\Em<0$ or with  $\La<0$ in the ergoregion.
In order to do that, it is convenient  to  re-define  the energy $
{\Em}\to s \Em$  with   $s\equiv\pm1$, having:
\bea&&\label{Eq:EmLdefsS}
\Em=s (g_{t\phi} \dot{\phi}+g_{tt} \dot{t}),\quad \La=g_{\phi\phi} \dot{\phi}+g_{t\phi} \dot{t},\quad  g_{ab}u^a u^b=\kappa \mu^2,
\eea
$\kappa=(\pm,0)$ is the  normalization constant. For time-like particles,
from the  condition on the energy   $\mathcal{E}/\mu=1$,  as seen at infinity ($r\rightarrow +\infty$)
we set  $s=-1$.

The angular velocity $\Omega$ can be parametrized with momentum ${\ell}\to s \ell$.  Tori orbiting  \textbf{NSs} can be described in terms of  $s{\ell}$ and $s{\Em}$, therefore the following relations hold
\bea\label{Eq:usegnosS}&&
 \left(u^t,u^\phi)\right|_{s=-1}=- \left(u^t,u^\phi)\right|_{s=1;\La=-\La}=\left(u^t,u^\phi)\right|_{s=1;\Em=-\Em}=-\left(u^t,u^\phi)\right|_{s=-1;\Em=-\Em;\La=-\La}
 \\&&
 \left.\Omega\right|_{\ell>0;s=\pm1}=\left.\Omega\right|_{\ell<0;s=\mp1},
\eea
Note, $(\sigma_\Ta,r^{\pm}_\Ta)$ can be expressed in terms of the variable $\tilde{\ell}=s\ell$:
\bea&&\label{Eq:rtur-econ-bi-nign} r_\Ta^\mp(s)\equiv\frac{\mp\sqrt{a^2 \sigma_\Ta^2+(s\ell)^2 \left[a^2 (\sigma_\Ta -1)+1\right]+2 a \sigma_\Ta \text{s$\ell $}}+a \sigma_\Ta +\text{s$\ell $}}{\text{s$\ell $}},
\quad\mbox{and}\quad
\sigma_\Ta(s)\equiv\frac{\text{s$\ell $}\Delta_	\Ta}{a (a \text{s$\ell $}+2 r_\Ta)}.
\eea
Introducing the $s$ sign there is $\ell=s \left.g_{t\phi}/g_{tt}\right|_\Ta$, and $\Em_\Ta \equiv s g_ {tt}(\Ta) \dot {t}_\Ta $.
Co-rotating  particles and fluids with   $s=-1$ (i.e. $({\La}<0$ or ${\ell}<0)$ respectively)  can be studied as co-rotating fluids  (i.e. $({\La}>0$ or ${\ell}>0)$ respectively)  with $s=+1$  and viceversa.

With the  energy re-parametrization $\tilde{\Em}\equiv s \Em$,  accounting for   on the energy sign $(s=\pm 1)$, the fluid effective potential can be parametrized in terms of $\tilde{\ell}=s\ell$ (which can be interpreted as sign of fluid (and particle) rotation). The condition at radial infinity is independent from  $s$ i.e.  $V_{eff}^2=
-\kappa$ for
$
{r\rightarrow+\infty}$.
The condition $V_{eff}^2\geq0$ is fixed by the normalization condition, determined by  the $\kappa$-sign.

\subsection{Inversion radius $r_\Ta^{\pm}$ of the  co-rotating ($\ell>0$) flows}\label{Sec:inversion-radius-coro}
In this section we complete the analysis considering  also the general sign $s=\pm1$, where $\tilde{\ell}\equiv s\ell$ and $\tilde{\Em}\equiv s\Em$.  There are no inversion points for $\Em>0$ and $\La>0$.  A necessary condition for the existence of a inversion point, from definition Eq.\il(\ref{Eq:lLE}) of $\La$ at $(r_\Ta,\sigma_\Ta)$, is $\La<0$ and   therefore it occurs in the co-rotating case only for $\La<0$ and $\Em<0$.
(However we shall prove that these solutions are for spacelike (tachyonic) particles.).
 Conditions discussed  in this article are in fact a necessary but not sufficient condition for the existence of a inversion point for accreting flows.

 Considering co-rotating flows, for $s=-1$, there are only space-like inversion points from co-rotating flows with $\Em<0$ and $\La<0$.
We then  consider the condition $u^\phi=0$ with  $\La \Em>0$ and $(\ell,\La,\Em)$ constants.

There are co-rotating inversion points  for
\bea&&\label{Eq:cap-sen-cam}
 (\ell>0)\quad u^\phi=0, \quad(\La\leq0, \Em\leq0):
\\
&&\nonumber
\mbox{for}\quad s<0: (\sigma_\Ta\in]\sigma_\epsilon^+,1],r_\Ta\in ]r_\epsilon^-,r_\epsilon^+[),
\\\nonumber
&&\mbox{for}\quad  s>0: (\sigma_\Ta\in]0,\sigma_\epsilon^+[, r_\Ta>0), (\sigma_\Ta =\sigma_\epsilon^+, r_\Ta\in ]0,r_\epsilon^-[\cup r_\Ta>r_\epsilon^-), (\sigma_\Ta\in]\sigma_\epsilon^+,1], r_\Ta\in ]0,r_\epsilon^-[\cup r_\Ta>r_\epsilon^+),
\eea
(note that in this case we considered also the case $s>0$) where the test particle energy $\Em$ and the angular momenta $(\La,\ell)$ are
\bea
&& \Em_{[g]}\equiv s u^t \left[\frac{2 r_\Ta}{\Sigma_\Ta}-1\right],\quad\La_{[g]}\equiv-\frac{2 a r_\Ta \sigma_\Ta  u^t}{\Sigma_\Ta}, \quad \ell_{[g]}\equiv-\frac{2 a r_\Ta \sigma_\Ta }{s[2 r_\Ta-\Sigma_\Ta]}.
\eea
Considering the normalization and stand still  conditions  $(u^\phi=0,\La\leq0, \Em\leq0)$, with
$(s<0, k>0)$, there is $(\sigma\in]\sigma_\epsilon^+,1],r\in ]r_\epsilon^-,r_\epsilon^+[, u^t_{(y)})$. For  $ (s>0, k<0)$,  there is  $(\sigma\in]0,\sigma_\epsilon^+[, u^t_{(y)})$, $(\sigma =\sigma_\epsilon^+,r\neq r_\epsilon^-,u^t_{(y)})$, $(\sigma\in]\sigma_\epsilon^+,1], (r\in ]0,r_\epsilon^-[\cup r>r_\epsilon^+,u^t_{(y)})$, where $\Em=k s/u^t, \La=-{2 a r \sigma u^t}/\Sigma,u^t_{(y)}=\sqrt{k ({2 r}/{\delta_0}-1)}$, and $\delta_0\equiv 2 r-\Sigma $.
(%
Tori within these conditions are rather small and are located in the \textbf{NS} ergoregion.).

\begin{figure}
  \includegraphics[width=6.75cm]{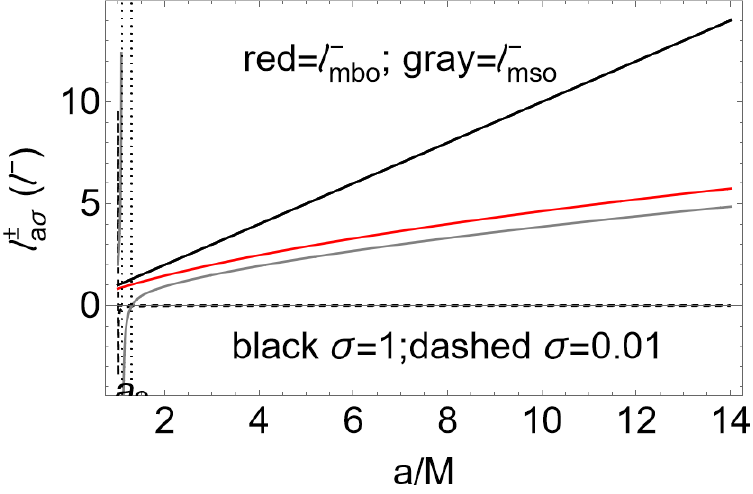}
   \includegraphics[width=6.75cm]{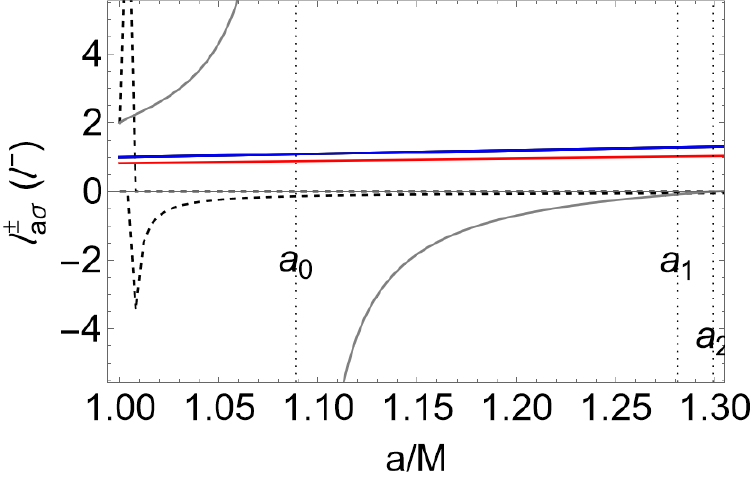}
 \caption{Fluid specific angular momentum $\ell^{\pm}_{a\sigma}$ of Eqs\il(\ref{Eq:sas}) as function the \textbf{NS} spin--mass ratio $a/M$ for different planes $\sigma\equiv\sin^2\theta$.  Right panel is a close-up view of the left panel.  Spins $\{a_0,a_1,a_2\}$ are defined in Table\il(\ref{Table:corotating-counter-rotatingspin}).   }\label{Fig:Plotlaslasso1}
\end{figure}

It is useful to introduce the  following angular momenta and  plane:
\bea&&
\label{Eq:sas}
\ell_{a\sigma}^{\mp}\equiv\frac{a \sigma_\Ta }{1-r_{cr}^2}\mp\sqrt{\frac{a^2 r_{cr}^2 \sigma_\Ta ^2}{\left[1-r_{cr}^2\right]^2}},
\quad \sigma_{a\sigma}\equiv\frac{1}{2} \left[\sqrt{\ell ^2 \left(-\frac{4 \ell }{a}+\ell ^2+4\right)}-\ell ^2\right]+\frac{\ell }{a},
 \eea
--Figs\il(\ref{Fig:Plotlaslasso1}).

There are no inversion points on the poles $\sigma=0$.
More in general inversion radii are for
\bea&&\nonumber
\ell>0:
\\&&\sigma_\Ta\in]\sigma_\epsilon^+,1[:(\ell =\ell_{a\sigma}^+, r_\Ta=r_\Ta^-); (\ell >\ell_{a\sigma}^+, r_\Ta=r_\Ta^\pm), (\sigma =1, \ell >\ell_\gamma^-, r_\Ta=r_\Ta^+),
\eea
or, in other words
\bea\label{Eq:mon-semb}
\ell >\ell_\gamma^-: (\sigma_\Ta\ =\sigma_{a\sigma}, r_\Ta=r_\Ta^-);\quad  \sigma_\Ta\in](\sigma_{a\sigma},1[, r_\Ta=r_\Ta^\pm); (\sigma_\Ta\ =1, r_\Ta=r_\Ta^+).
\eea
Value  $\ell^-(r=0)=\ell_{\gamma}^-=a>\ell_{mbo}^-$  is  the limit for the existence of proto-jets. In the geometries $a>a_2$, for $\ell>\ell_\gamma^-$,  there are only  quiescent tori  centered in $r>r_{[\gamma]}^-$, in this region there is $\La>0$ and there are no inversion points.
For $a\in ]M, a_0[$, $\ell_{mbo}^-<\ell_\gamma^-<\ell_{mso}^-$ in $]r_{\delta}^-, r_{\delta}^+[$, and  therefore   the case $\ell>\ell_\gamma^-$ includes only inversion points from the inner tori for $a\in ]M, a_0[$ (tori are for any $\ell>\ell_{mso}^-$) and quiescent tori at  $r>r_{[\gamma]}^-$. For $a\in]a_0,a_2[$, at $\ell>a$ there are quiescent tori.

Contrary to the counter-rotating case,  the inversion point with  $\ell >
   0$ exists \emph{only} in the ergoregion $r\in ] r_\epsilon^-, r_\epsilon^+[$--see also Figs\il(\ref{Fig:Plotlaslasso1}).
   There is a double inversion point radius $r_\Ta^\pm$, while on the equatorial plane there is there is one only inversion point, within  the condition $\ell>\ell_\gamma^->M$, that is
$
\mbox{for}\quad\sigma=1: \ell >\ell_\gamma^-, r=r_\Ta^+.
$
The co-rotating inversion point is confined in the ergoregion  $]r_\epsilon^-, r_\epsilon^+[$, in fact there is for  $\ell>0$  and  $s=-1$
\bea&&\label{Eq:sigmatacoro-ecc-mili}
 \sigma =\sigma_\Ta:\quad
(\ell \geq \ell_\beta,r\in ]0,2M[)\quad \mbox{or}\quad (\ell >\ell_\gamma^-, \in]0,r_b]),\quad \ell_\beta\equiv  \frac{2 a}{2-r},\quad r_b\equiv \frac{2(\ell- a)}{\ell }
\\&&\label{Eq:rblbm}
\mbox{or alternatively}\quad \sigma=\sigma_\Ta:\ell\geq\ell_\beta >\ell_\gamma^-, \quad 0<r_\Ta\leq r_b<2M.
 \eea
\begin{figure}
\centering
 \includegraphics[width=6.5cm]{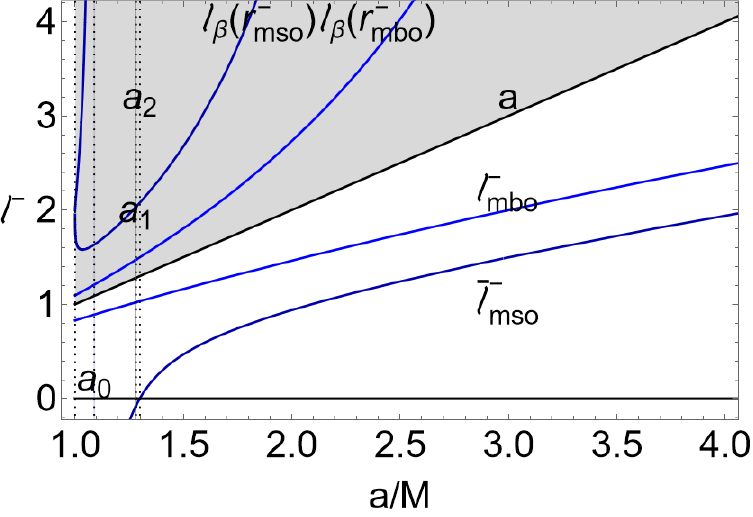}
  \includegraphics[width=6.5cm]{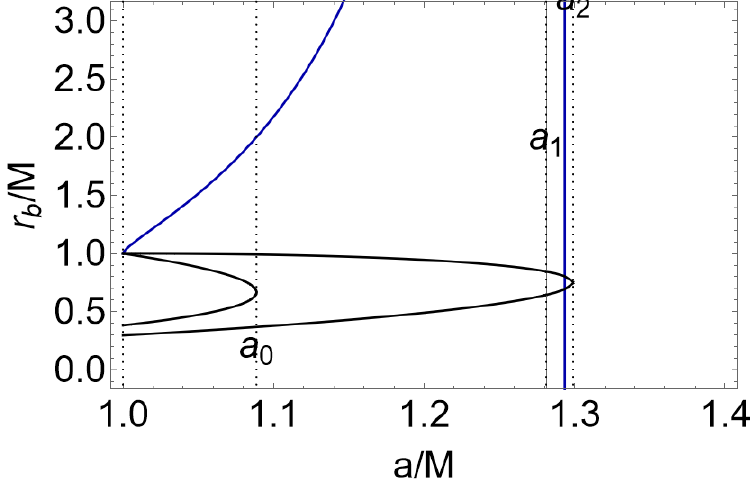}
    \caption{Analysis   of inversion plane $\sigma_\Ta$ with fluid specific angular momentum $\ell^-$, including co-rotating motion $\ell^->0$ and counter-rotating fluids with  $\ell=\ell^-<0$  for spin $a\in]M,a_2[$ and orbital range for the definition of $\ell$ in $]r_0^-,r_0^+[-]r_\delta^-,r_\delta^+[$. Left panel: momentum $\ell^-$ on the geodesic radii are shown as function of the \textbf{NS} spin-mass ratios. Reference value $\ell=\ell_\gamma^-=a$ is shown, function $\ell_\beta$ of Eqs.\il(\ref{Eq:sigmatacoro-ecc-mili}) on the geodesic  structure (relative to fluid momentum $\ell^-$ is also shown. Right panel:  Radius $r_b$ is plotted on different momenta  according to the colors definitions of the right left panel.  Spins $\{a_0,a_1,a_2\}$ are defined in Table\il(\ref{Table:corotating-counter-rotatingspin}).  Black curves are radii $r_0^-<r_{\delta}^-<r_\delta^+<r_0^+$ as functions of the \textbf{NS} spin $a/M$. }\label{Fig:Plotsigmaonlinversion}
\end{figure}
Here however we consider also the case $s=+1$.
The choice of $s$ sign is related to the sing of $(u^t,u^\phi)$ and $(\Em,\La)$. In the following  we drop the notation  $\Ta$ while still intending all the quantities being evaluated at the inversion point.
We consider the normalization condition  with   $\kappa=\{0,-1\}$ and ($u^\phi=0$, $u^t>0$), introducing the following quantities
\bea&&\nonumber
\dot{t}_{[u]}\equiv \sqrt{\frac{\Sigma}{\Sigma-2r}};\quad \dot{\theta}^2_{[u]}\equiv \frac{r \left[(r-2) \dot{t}^2-r\right]-a^2 (\sigma -1) \left(\dot{t}^2-1\right)}{\Sigma^2},\quad \dot{\theta}^2_{[y]}\equiv\frac{(r-2) \dot{t}^2}{r^3},\quad  \dot{\theta}^2_{[g]}\equiv\frac{\dot{t}^2 \left(\Sigma-2r\right)}{\Sigma ^2};\\
&&\dot{r}_{[u]}\equiv-\frac{\Delta \left[a^4 (\sigma -1)^2 \dot{\theta}^2+a^2 (\sigma -1) \left[\dot{t}^2-2 r^2 \dot{\theta}^2-1\right]+r^4 \dot{\theta}^2+r^2-(r-2) r \dot{t}^2\right]}{\Sigma^2};
\\\nonumber
&& \dot{r}^2_{[x]}\equiv-\frac{\Delta \left[a^4 (\sigma -1)^2 \dot{\theta}^2+a^2 (\sigma -1) \left[\dot{t}^2-2 r^2 \dot{\theta}^2\right]+r^4 \dot{\theta}^2-(r-2) r \dot{t}^2\right]}{\Sigma^2}.
\eea
For photon-like particles there is $\dot{r}^2=\dot{r}^2_{[x]}, \La=\La_{[g]}, \Em=\Em_{[g]}$ with
\bea&&
s<0, (\sigma =\sigma_\epsilon^+, r=r_\epsilon^-, \dot{\theta}^2=0), (\sigma\in]\sigma_\epsilon^+,1[, (r=r_\epsilon^\pm, \dot{\theta}^2=0), (\sigma =1, r=2,  \dot{\theta}^2=0);
\\\nonumber
&&s\geq 0: (\sigma\in[0,\sigma_\epsilon^+[,\dot{\theta}^2\in[0,\dot{\theta}^2_{[g]}]), (\sigma =\sigma_\epsilon^+, (r\in ]0,r_\epsilon^-[\cup r>r_\epsilon^-, \dot{\theta}^2\in[0,\dot{\theta}^2_{[g]}]), (r=r_\epsilon^-,  \dot{\theta}^2=0);
\\\nonumber
&&(\sigma\in]\sigma_\epsilon^+,1[, (r\in ]0,r_\epsilon^-[\cup r>r_\epsilon^+,\dot{\theta}^2\in[0,\dot{\theta}^2_{[g]}]), (r=r_\epsilon^\pm,  \dot{\theta}^2=0);
\\
&&(\sigma =1, (r=2, \dot{\theta}^2=0); (r>2,\dot{\theta}^2\in[0,\dot{\theta}^2_{[y]}])).
\eea
For time-like particles there is $ \dot{r}^2=\dot{r}_{[u]}, \La=\La_{[g]}, \Em=\Em_{[g]}$ with
\bea&&\nonumber
s\geq 0:
\\
&&\nonumber (\sigma\in[0,\sigma_\epsilon^+[, (\dot{t}=\dot{t}_{[u]}, \dot{\theta}^2=0);
\\
&& (\dot{t}>\dot{t}_{[u]}, \dot{\theta}^2\in[0,\dot{\theta}^2_{[u]}]);
\\&&\nonumber
 (\sigma =\sigma_\epsilon^+, (r\neq r_\epsilon^-, (\dot{t}=\dot{t}_{[u]}, \dot{\theta}^2=0); \\
 &&\nonumber
 (\dot{t}>\dot{t}_{[u]}, \dot{\theta}^2\in[0,\dot{\theta}^2_{[u]}]); \\\nonumber
&&\sigma\in]\sigma_\epsilon^+, 1], ((r\in ]0,r_\epsilon^-[\cup r>r_\epsilon^+,
\\
&&(\dot{t}=\dot{t}_{[u]}, \dot{\theta}^2=0); (\dot{t}>\dot{t}_{[u]}, \dot{\theta}^2\in[0,\dot{\theta}^2_{[u]}]).
\eea

The flow inversion points at $s=+1$ and $s=-1$ are in separated orbital regions and therefore distinguishable. Considering also the normalization condition, with  $u^{\phi}=u^r=u^{\theta}=0$ (\emph{stand still} condition), there is
$(\Em\geq 0, \La\leq0, \ell\leq0, u^\phi=0)$ with  $(\Em_{(x)},\La_{(x)}, \ell_{(x)}=\La_{(x)}/\Em_{(x)},u_{(x)}^t)$ for: $(s<0, k<0, a>1): (\sigma\in]0,\sigma_\epsilon^+[, r>0), (\sigma =\sigma_\epsilon^+, r\neq r_\epsilon^-),
(\sigma\in]\sigma_\epsilon^+,1], (r\in ]0,r_\epsilon^-[\cup r>r_\epsilon^+)$. There is  $(s>0, k>0, a>1): \sigma\in]\sigma_\epsilon^+,1], r\in] r_\epsilon^-,r_\epsilon^+[$.
Where
$u_{(x)}^t\equiv
\sqrt{k \left({2 r}/({2r-\Sigma})-1\right)},\Em_{(x)}={k s}/u^t
$, $\La_{(x)}\equiv-2 a r \sigma u^t/\Sigma$.
The case  $\ell=0$  is possible  only for $\La=0$ and  $\sigma=0$ (on the \textbf{BH} axis) where
$
\Em={ u^t \Delta}/({a^2+r^2})
$
Considering also the normalization  and  the stand still condition, the case $\ell=0$  is possible for any  $s$ and $\kappa$ only  for $\La=0$ and  $\sigma=0$ (on the rotational axis),
where
$
\Em={\kappa s}/u^t$, $ u^t=\sqrt{-\kappa (a^2+r_\Ta^2)/\Delta_\Ta}$.

Let us define %
:
 \bea&&
\Em_{\Gamma}\equiv \dot{t_{\Ta}}\left[1-\frac{2 r_{\Ta}}{\Sigma_{\Ta}}\right],\quad\La_{\Gamma}=-\frac{2 a r_{\Ta} \sigma_{\Ta}  \dot{t}_{\Ta}}{\Sigma_{\Ta}},
 \eea

  For  $s=-1$ there are \emph{no} inversion for points for time-like or photon-like particles with  $(\La<0,\Em<0)$. There are however spacelike solutions (at $u^t>0$), in the ergoregion  with $(\dot{r}_{\Ta}^2=\dot{r}_J^2,\La=\La_{\Gamma} ,\Em=\Em_{\Gamma})$ for
\bea&&
\sigma_{\Ta}\in]\sigma_\epsilon^+,  1], r_{\Ta}\in]r_\epsilon^-,r_\epsilon^+[,(\dot{t}_{\Ta}\in]0,\dot{t}_{\Gamma}^-[, \dot{\theta}_{\Ta}^2\in[0,\dot{\theta}^2_J]); (\dot{t}_{\Ta}=\dot{t}_{\Gamma}^-,\dot{\theta}^2=0),
\eea
where
\bea&&
\dot{t}_{\Gamma}^-\equiv\sqrt{\frac{2 r_{\Ta}}{2r_{\Ta}-\Sigma_{\Ta}^2}-1};\quad \dot{\theta}^2_J\equiv\frac{r_{\Ta} \left[(r_{\Ta}-2) \dot{t}_{\Ta}^2+r_{\Ta}\right]-a^2 (\sigma_{\Ta} -1) \left(\dot{t}_{\Ta}^2+1\right)}{\Sigma_{\Ta}^2};
\\\nonumber
&&\dot{r}_J^2\equiv-\frac{\Delta_{\Ta} \left[a^4 (\sigma_{\Ta} -1)^2 \dot{\theta}_{\Ta}^2+a^2 (\sigma_{\Ta} -1) \left(\dot{t}_{\Ta}^2+1-2 r_{\Ta}^2 \dot{\theta}_{\Ta}^2\right)+r_{\Ta}^4 \dot{\theta}_{\Ta}^2-r_{\Ta} \left[(r_{\Ta}-2) \dot{t}_{\Ta}^2+r_{\Ta}\right]\right]}{\Sigma_{\Ta}^2}.
\eea

\end{document}